\documentclass[12pt]{article}
\usepackage{mathtools,amssymb,mathrsfs,microtype}
\usepackage{graphics} 
\usepackage[hidelinks,colorlinks=true,linkcolor=blue,citecolor=red]{hyperref}

\usepackage{subcaption,comment,multirow,draft}

\usepackage[framemethod=TikZ]{mdframed}
\definecolor{darkcerulean}{rgb}{0.03, 0.27, 0.49}

\usepackage[normalem]{ulem}

\numberwithin{equation}{section}

\usetikzlibrary{decorations.pathmorphing,calc}

\def\CC{{\mathcal C}}
\def\CL{{\mathcal L}}
\def\CN{{\mathcal N}}

\def\CZ{{\mathcal Z}}
\def\bZ{{\mathbb{Z}}}

\usepackage{mdframed}
\newenvironment{claim}{  \begin{mdframed}[linecolor=black!0,backgroundcolor=black!10]\noindent\itshape\ignorespaces}{\end{mdframed}}

\usepackage{cite}

\begin{document}

\begin{titlepage}

\title{Non-Local Conserved Currents and Continuous Non-Invertible Symmetries}
\author{Diego Delmastro$^{1}$, Adar Sharon$^{1}$, and Yunqin Zheng$^{2,3}$}

\address{${}^{1}$Simons Center for Geometry and Physics, \\Stony Brook University, Stony Brook, NY, 11790, United States\\\bigskip 
${}^{2}$Kavli Institute for Theoretical Sciences, \\University of Chinese Academy of Sciences, Beijing, 100190, China\\\bigskip 
${}^{3}$C. N. Yang Institute for Theoretical Physics, \\Stony Brook University, Stony Brook, NY, 11790, United States}

\bigskip 
\abstract

We embark on a systematic study of continuous non-invertible symmetries, focusing on 1+1d CFTs. We describe a generalized version of Noether's theorem, where continuous non-invertible symmetries are associated to \emph{non-local} conserved currents: point-like operators attached to extended topological defects. The generalized Noether's theorem unifies several constructions of continuous non-invertible symmetries in the literature, and allows us to exhibit many more examples in diverse theories of interest. We first review known examples which are non-intrinsic (i.e., invertible up to gauging), and then describe $\textit{new}$ examples in Wess-Zumino-Witten models and products of minimal models.  
For some of these new examples, we show that these continuous non-invertible symmetries are intrinsic if we demand that a certain global symmetry is preserved. The continuous non-invertible symmetries in products of minimal models also allow us to construct new examples of defect conformal manifolds in a single copy of a minimal model. Finally, we comment on continuous non-invertible symmetries in higher dimensions. 

\vfill

\begin{flushleft}
\small
\texttt{diego.delmastro@scgp.stonybrook.edu} \\
\texttt{adar.sharon@scgp.stonybrook.edu} \\
\texttt{zhengyunqin@ucas.ac.cn}
\end{flushleft}

\end{titlepage}

{\hypersetup{linkcolor=black}
\setcounter{tocdepth}{3}
\tableofcontents
}

\newpage

\section{Introduction}

A key breakthrough in the program of generalized global symmetries was the identification of global symmetries with topological operators, which led to numerous generalizations \cite{Gaiotto:2014kfa}. Let us restrict ourselves to linear, internal symmetries. It is believed that the most general structure that describes finite symmetries is that of (higher) fusion categories. In 1+1d, the finite fusion category symmetries are extensively discussed in various references, see for instance~\cite{Chang:2018iay,Komargodski:2020mxz, Thorngren:2019iar, Thorngren:2021yso, Bhardwaj:2017xup, Kong:2020cie,Lin:2022dhv,Huang:2021zvu,Lu:2022ver,Aasen:2020jwb,Lin:2023uvm,Huang:2021nvb,Burbano:2021loy,Gaiotto:2020iye}. In higher dimensions, the finite higher fusion category symmetries have been investigated recently in e.g.~\cite{Koide:2021zxj,Bhardwaj:2023ayw,Kaidi:2021xfk,Choi:2021kmx,Choi:2022fgx,Choi:2022zal,Inamura:2023qzl,Bhardwaj:2023wzd,Bhardwaj:2022kot,Bhardwaj:2022yxj,Kaidi:2022cpf,Bartsch:2022ytj,Antinucci:2022vyk,Antinucci:2023ezl,Cordova:2023bja,Bhardwaj:2024xcx,Bullimore:2024khm}.

While finite generalized symmetries have been extensively investigated, their continuous counterpart has not been systematically explored when the topological operators are allowed to be non-invertible. 
This is partially because the related mathematical framework, i.e., the continuous avatar of (higher) fusion categories, is still under development. Yet, a few examples of continuous non-invertible symmetries have been studied in the recent literature~\cite{Thorngren:2021yso,Heidenreich:2021xpr,Bhardwaj:2022yxj,Jacobson:2024muj,Hsin:2024aqb,Hsin:2025ria,Nguyen:2021yld,Antinucci:2022eat,Fuchs:2007tx,GarciaEtxebarria:2022jky,Damia:2023gtc,Antinucci:2025uvj,Choi:2025ebk,Seifnashri:2025fgd}, which we review below. These examples are either gauge-related to invertible symmetries, or infinitesimally close to the symmetries of the above sort, in a sense we review below. As a result, these are examples of \textit{non-intrinsic} non-invertible symmetries\cite{Kaidi:2022uux}.  
In this paper we find new examples of continuous non-invertible symmetries in 1+1d CFTs that are beyond the types mentioned above, and which behave as genuine continuous non-invertible symmetries in a sense we make precise. We also discuss some applications of these new symmetries. 

In the rest of this introduction we review some known examples of continuous non-invertible symmetries,
and then discuss the merit of understanding continuous non-invertible symmetries from non-local conserved currents. We finally summarize the key examples and applications.

\subsection{Non-Intrinsic Examples of Continuous Non-Invertible Symmetries}\label{sec:triv_examples}

\subsubsection{Non-Invertible Coset Symmetry}\label{sec:introgauging}

One systematic construction of continuous non-invertible symmetries is via gauging a \emph{non-normal} finite subgroup of a continuous symmetry~\cite{Heidenreich:2021xpr,Thorngren:2021yso,Bhardwaj:2022yxj,Jacobson:2024muj,Hsin:2024aqb,Hsin:2025ria,SSW:2021unpublished,Seifnashri:2025fgd}. For instance, consider a $G\rtimes K$ 0-form symmetry, where $G$ is a continuous symmetry and $K$ is a finite group which is assumed to be anomaly free. Importantly, $K$ acts on $G$ non-trivially. Then gauging $K$ renders $G$ non-invertible. This is known as the coset construction~\cite{Hsin:2024aqb,Hsin:2025ria}.\footnote{More precisely, since the non-invertible symmetry is given by a ring of the double cosets $K \backslash G/K$, it may be more appropriate to call it double coset construction \cite{Cao:2025qnc}. We thank Yuan Miao for the discussion.
}

A typical example of this kind is the so-called ``cosine" symmetry. Consider a 1+1d QFT $\mathcal{Q}$ with $G=U(1)$ and $K=\bZ_2$. The $\bZ_2$ is the charge conjugation symmetry that flips the sign of the $U(1)$ charge. The topological defect that implements $U(1)$ is
\begin{equation}
    U_{\alpha}(\mathcal{M})= \exp\biggl(i \alpha \int_{\mathcal{M}} \star J\biggr)\;,\quad \alpha\simeq \alpha+ 2\pi \;,
\end{equation}
and the generator of $\bZ_2$ is $\eta(\mathcal{M})$. Below, to simplify the notation, we often suppress the manifold dependence $\mathcal{M}$. These topological operators obey the fusion rules
\begin{equation}
    \eta\times \eta=1\;, \quad U_\alpha \times U_\beta = U_{\alpha+ \beta}\;, \quad \eta\times  U_\alpha = U_{-\alpha} \times \eta\;.
\end{equation}
After gauging $\bZ_2$, depending on the value of $\alpha$, we get operators of different types:
\begin{itemize}
    \item When $\alpha=0$ mod $2\pi$, i.e.~the identity operator, it obviously survives in the gauged theory $\mathcal{Q}/\bZ_2$. Moreover, since it is stable under $\bZ_2$, it splits into two lines, labeled by the representation of the gauged $\bZ_2$ \cite{Moore:1989yh}. These are the identity and the quantum symmetry line $\widehat{\eta}$ \cite{Vafa:1989ih}. 
    \item When $\alpha=\pi$ mod $2\pi$, $U_{\pi}$ also commutes with $\bZ_2$, hence it also survives in the gauged theory $\mathcal{Q}/\bZ_2$, and splits into two operators related by fusing $\widehat{\eta}$. 
    \item When $\alpha\notin \pi\bZ$, only the linear combination 
    \begin{eqnarray}\label{eq:Vcosine}
        V_{|\alpha|}  := U_\alpha + U_{-\alpha}= 2\cos\biggl(\alpha \int_{\mathcal{M}} \star J\biggr)
    \end{eqnarray}
    is gauge invariant. Because the right hand side is a cosine function of the current, such symmetry is termed the ``cosine symmetry''. A crucial feature of the cosine symmetry is that it obeys the non-invertible fusion rule
    \begin{eqnarray}\label{eq:cosinefusion}
        V_{|\alpha|}\times 
        V_{|\beta|} = V_{|\alpha+ \beta|}+ V_{|\alpha- \beta|}\;.
    \end{eqnarray}
\end{itemize}

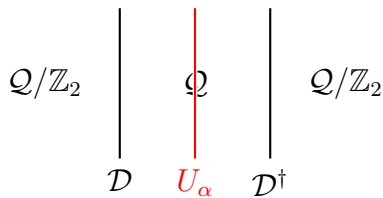
\begin{figure}
    \centering
    \begin{tikzpicture}
        \draw[thick] (0,0) -- (0,2);
        \draw[thick] (2,0) -- (2,2);
        \node[] at (1,1) {$\mathcal{Q}$};
        \draw[line width=1.8pt, white] (1,0) -- (1,2);
        \draw[thick, red] (1,0) -- (1,2);
        \node[] at (-1,1) {$\mathcal{Q}/\bZ_2$};
        \node[] at (3,1) {$\mathcal{Q}/\bZ_2$};
        \node[below] at (0,0) {$\mathcal{D}$};
        \node[below] at (2,0) {$\mathcal{D}^\dagger$};
        \node[below, red] at (1,0) {$U_\alpha$};
    \end{tikzpicture}
    \caption{Sandwich construction of cosine symmetry. }
    \label{fig:sandwich}
\end{figure}

Recently, it has been pointed out in \cite{Jacobson:2024muj} that the definition \eqref{eq:Vcosine} is not very precise---it only ensures that the operator is invariant under the $\bZ_2$ global charge conjugation, whereas it is demanded to be invariant under $\bZ_2$ \emph{gauge} charge conjugation. This issue was solved by more careful analysis both on the lattice \cite{Jacobson:2024muj} and in the continuum \cite{Hsin:2024aqb,Hsin:2025ria}. 

Let's review the construction in \cite{Hsin:2024aqb,Hsin:2025ria}. After gauging $\bZ_2$, the resulting theory $\mathcal{Q}/\bZ_2$ is separated from $\mathcal{Q}$ by a topological interface $\mathcal{D}$. Starting with a $U(1)$ symmetry operator $U_{\alpha}$ in $\mathcal{Q}$, under gauging $\bZ_2$, the resulting symmetry operator is $U_\alpha$ sandwiched by $\mathcal{D}$ and $\mathcal{D}^\dagger$ from its left and right, respectively. See Figure \ref{fig:sandwich} for an illustration. Note that the resulting line $V_{|\alpha|}:=\mathcal{D} U_\alpha \mathcal{D}^\dagger$ is not necessarily simple. This sandwich construction reproduces and improves the cosine construction \eqref{eq:Vcosine}. The different types of operators are now: 
\begin{itemize}
    \item When $\alpha=0,\pi$ mod $2\pi$, $U_\alpha$ commutes with gauging, hence one can move $U_\alpha$ outside the sandwich. Using $\mathcal{D}\times \mathcal{D}^\dagger = 1+ \widehat{\eta}$, we obtain
    \begin{eqnarray}\label{eq:Vnpi}
        V_{n\pi} := \mathcal{D} U_{n\pi } \mathcal{D}^\dagger=U_{n\pi }+ U_{n\pi }\widehat{\eta}\;,
    \end{eqnarray}
    which is non-simple. The two simple lines $U_{n\pi}$ and $U_{n\pi}\widehat{\eta}$ differ by the quantum $\bZ_2$ line $\widehat{\eta}$, which is associated with the sign representation of $\bZ_2$ charge conjugation. This reproduces the first two bullet points above. 
    \item When $\alpha \notin \pi \bZ$, we get the operator 
    \begin{eqnarray}
    \label{eq:Valpha}
        V_{|\alpha|} := \mathcal{D} U_\alpha \mathcal{D}^\dagger\;.
    \end{eqnarray}
    $V$ does not depend on the sign of $\alpha$. To see this, we note that $\mathcal{D}$ absorbs $\eta$ from its right, i.e.~$\mathcal{D}\times \eta = \mathcal{D}$. Similarly $\mathcal{D}^\dagger$ absorbs $\eta$ from its left, i.e.~$\eta \times \mathcal{D}^\dagger= \mathcal{D}^\dagger$. Then $V_{|\alpha|} := \mathcal{D} U_\alpha \mathcal{D}^\dagger = \mathcal{D}\eta U_\alpha \mathcal{D}^\dagger= \mathcal{D} U_{-\alpha} \eta\mathcal{D}^\dagger=\mathcal{D} U_{-\alpha} \mathcal{D}^\dagger= V_{|-\alpha|}$. 
    To compute the fusion rule, we simply write
    \begin{equation}
    \begin{aligned}
        V_{|\alpha|}\times V_{|\beta|} &= \mathcal{D} U_\alpha  (\mathcal{D}^\dagger \mathcal{D}) U_\beta \mathcal{D}^\dagger\\
        &= \mathcal{D} U_\alpha (1+ \eta) U_\beta \mathcal{D}^\dagger\\
        &=\mathcal{D} U_{\alpha+\beta} \mathcal{D}^\dagger +  \mathcal{D} U_{\alpha-\beta} \mathcal{D}^\dagger\\
        &= V_{|\alpha+ \beta|}+ V_{|\alpha- \beta|}\;.
    \end{aligned}
    \end{equation}
    This reproduces \eqref{eq:cosinefusion}. 
    When $\alpha\pm \beta \in \pi \bZ$ on the right hand side, the corresponding term should be evaluated using \eqref{eq:Vnpi}. 
\end{itemize}

For the purpose of constraining dynamics along the renormalization group (RG) flow, this construction of continuous non-invertible symmetries is considered uninteresting.  The reason is that gauging a finite symmetry commutes with the RG flow. Any dynamical consequence one draws from the continuous non-invertible symmetry $(G\rtimes K)/K$ in the theory $\mathcal{Q}/K$ can be viewed as a consequence of the invertible symmetry $G\rtimes K$ in the theory $\mathcal{Q}$. Hence, if continuous non-invertible symmetries are going to prove interesting at all for the purpose of studying the RG flow, it must be the case that some of them do \emph{not} originate from gauging a non-normal subgroup of a continuous symmetry. In other words, it is desirable to search for an \emph{intrinsic} continuous non-invertible symmetry, one that remains non-invertible for all global forms of the theory under consideration.

\subsubsection{Continuous Non-Invertible Symmetry in the Compact Boson}
\label{sec:compbosonintro}

Continuous non-invertible symmetries can also arise as we move along a conformal manifold. The quintessential example of this (which will be discussed at length in Section \ref{sec:c=1}) is the $c=1$ compact boson \cite{Ginsparg:1988ui,Ginsparg:1987eb}. It is well-known that this theory has a conformal manifold parametrized by a radius $R$, and at a special point on this manifold, i.e.~$R=1$, the theory becomes the $SU(2)_1$ WZW model whose global symmetry is enhanced from $U(1)\times U(1)$ to $[SU(2)\times SU(2)]/\bZ_2$. One can ask what happens to these enhanced symmetries as we move away from $R=1$.

It turns out that away from this point, the enhanced symmetry becomes non-invertible \cite{Fuchs:2007tx}. One way to understand this is to note that for $R=p/q\in \mathbb{Q}$ the theory is obtained from the $R=1$ theory by gauging a $\bZ_p\times \bZ_q$ (non-normal) subgroup of $[SU(2)\times SU(2)]/\bZ_2$. Since we are gauging a non-normal subgroup, by the coset construction reviewed in Section \ref{sec:introgauging}, the $[SU(2)\times SU(2)]/\bZ_2$ symmetry becomes non-invertible. 
Since any non-rational $R$ can be approached by a sequence of rational $R$'s, the symmetry for non-rational $R$ is 
always infinitesimally away\footnote{We use the Zamolodchikov metric to define distances on the conformal manifold.} from a point on the conformal manifold where it can be made invertible by a discrete gauging. 
We will thus also consider these examples as non-intrinsic.\footnote{Indeed, as we explain, since these symmetries are infinitesimally close to non-intrinsic non-invertible symmetries, their properties are similar.}

Recently, Ho Tat Lam studied many examples of conformal manifolds obtained via current-current deformations, and showed that a dense set of points in these conformal manifolds are actually related by gauging \cite{hotatlam1,hotatlam2}. Therefore, if there is a special point in the conformal manifold where there is an enhanced continuous symmetry, by our previous discussion, the theory at other points of the conformal manifold can have continuous non-invertible symmetries---significantly generalizing the example of the compact boson. However, once again all such examples are infinitesimally close to a point where they are non-intrinsic, and so we regard them as non-intrinsic.

\subsection{Local vs. Non-Local Conserved Currents}

With these considerations in mind, one of the purposes of this paper is to address the following question:
\begin{claim}
\vspace{-1\baselineskip}
\begin{eqnarray}
\label{eq:keyquestion}
    \text{Do intrinsic continuous non-invertible symmetries exist? }
\end{eqnarray}
\end{claim}
In other words, are there continuous non-invertible symmetries which cannot be made invertible by gauging or by moving infinitesimally on a conformal manifold? So far all the examples reviewed above are non-intrinsic, and it would be desirable to search for intrinsic examples which require new insights on continuous non-invertible symmetries.

This new insight comes from conserved currents. 
For conventional continuous global symmetries, Noether's theorem shows that every such symmetry is associated with a local conserved current $J$ satisfying $d\star\! J=0$. In short,
\begin{eqnarray}
    \text{Continuous symmetry} \longleftrightarrow \text{Local conserved current}\;.
\end{eqnarray}
Integrating the conserved current over a spacelike slice defines a conserved charge. Exponentiating the conserved charge gives a topological operator.

For a non-invertible symmetry, is there an analogue of Noether's theorem? Motivated by the known examples reviewed in Section \ref{sec:triv_examples}, a natural  conjecture is that every continuous non-invertible symmetry is associated with a \emph{non-local} conserved current, such that integrating it on a spacelike slice produces the non-invertible topological operator. Indeed, focusing on 1+1d CFTs, we will be able to prove a \textbf{generalized Noether's theorem}:
\begin{claim}
\vspace{-1\baselineskip}
\begin{eqnarray}
\label{eq:key}
\text{Continuous non-invertible symmetry} \longleftrightarrow \text{Non-local conserved current. }
\end{eqnarray}
\end{claim}
For our purposes, a \emph{current} is an operator of dimension $(h,\bar{h})=(1,0)$ (or $(0,1)$), and a non-local conserved current is a point-like current attached to an extended (often discrete) topological line. The main purpose of this paper is to examine the statement \eqref{eq:key} and study it in a wide range of examples.
We emphasize that this proposal has been hinted at in various examples, see e.g.~\cite{Thorngren:2021yso, Antinucci:2025uvj}, and discussed in a talk by Shu-Heng Shao \cite{SSW:2021unpublished2} which is based on unpublished work \cite{SSW:2021unpublished}. In \cite{Ambrosino:2025myh}, non-local currents are used to construct translation invariant (but not necessarily topological) defects.

As a first check, the cosine symmetry in 1+1d reviewed in Section \ref{sec:introgauging} is consistent with \eqref{eq:key}. In the theory $\mathcal{Q}$, there is a conserved current $J$, generating a $U(1)$ symmetry. Charge conjugation symmetry $\bZ_2$ acts on the current via $J\to -J$, meaning that the current is $\bZ_2$ odd. Once we gauge the $\bZ_2$ symmetry, it is well-known that every $\bZ_2$ odd local operator in $\mathcal{Q}$ becomes a $\widehat{\bZ_2}$ even \emph{non-local} operator attached to the topological line $\widehat{\eta}$ associated with the quantum $\bZ_2$ symmetry in $\mathcal{Q}/\bZ_2$. As we explicitly show in the main text, integrating such a non-local current on a circle precisely gives rise to non-invertible topological operators \eqref{eq:Vnpi} and \eqref{eq:Valpha}.  This construction extends to all the coset symmetries. 

We will see that understanding continuous non-invertible symmetries in terms of non-local conserved currents offers an opportunity to solving \eqref{eq:keyquestion}, i.e.~searching for an intrinsic non-invertible symmetry. As we will see, in 1+1d CFTs it is easier to look for non-local currents than topological operators. If a non-local conserved current can be made local by gauging, the non-invertible symmetry can be made invertible, and hence is non-intrinsic. Conversely, looking for intrinsic continuous non-invertible symmetries amounts to searching for a non-local conserved current which \emph{cannot} be made local upon gauging.\footnote{One additional way to construct a continuous non-invertible symmetry not mentioned above is to take a continuous invertible symmetry generator $U_\alpha$ and stack it with a discrete non-invertible defect $\mathcal{N}$. But since this symmetry has a local conserved current we can also call this type of symmetry non-intrinsic.} 
 In the main text, we will provide a series of examples of currents which cannot be made local if we demand some additional symmetry to be preserved.\footnote{It is difficult to prove that a non-local current can never be made local by gauging, since there are very few theories for which all topological operators, and hence all possible gaugings, are known. As a result we must reduce ourselves to proving that a symmetry is intrinsic under some simplifying assumptions. }

\subsection{Summary of Examples and Results}

Our discussions focus on basic and well-known rational CFTs in 1+1d, and we will show that non-local currents are surprisingly ubiquitous. The examples discussed in detail here are:

\paragraph{$c=1$ CFTs.} We review the continuous non-invertible symmetries in the circle and orbifold branches, originally discussed in \cite{Fuchs:2007tx,Thorngren:2021yso}, and emphasize the role of non-local currents.  

\paragraph{The minimal models $\mathcal{M}_m$.} We show that the minimal models do not have non-local currents in Appendix \ref{app:min_mod}. This is to be expected, since the minimal models are highly constrained theories.

\paragraph{Products of two  minimal models $\mathcal{M}_m\times\overline{\mathcal{M}}_m$.} We show that for an infinitely large family of $m$'s, these examples \textit{do} have non-local currents at the end of Verlinde lines. As a result, these theories possess continuous non-invertible symmetries. As an application, we show that these symmetries imply that $\mathcal{M}_m$ has a conformal manifold of defects for these values of $m$, following the arguments of \cite{Antinucci:2025uvj}.

\paragraph{$G_k$ WZW models.} We mostly focus on $G=SU(2)$, in which case we prove that there is an infinite family of $k$'s where the theories contain non-local currents at the end of Verlinde lines (and so continuous non-invertible symmetries). We discuss the corresponding charges that one can construct and their action on local operators. We also show that these non-invertible symmetries cannot be made invertible by any topological manipulation that preserves the $SU(2)$ chiral algebra. In other words, we show that these continuous non-invertible symmetries are intrinsic if we demand that the $SU(2)$ chiral algebra is preserved. Finally, we show that non-local currents are common in general $G_k$ WZW models, and appear for various combinations of target spaces $G$ and levels $k$.

\subsection{Organization}

The organization of this paper is as follows. In Section \ref{sec:Noether} we prove the generalized Noether's theorem, i.e.~we prove that one can construct a continuum of non-invertible defects from a non-local current and vice versa, in 1+1d CFTs.
In Section \ref{sec:c=1} we apply this formalism to the well-known non-intrinsic examples of non-invertible symmetries in the case of the $c=1$ CFT as a warm-up. In Section \ref{sec:currentWZW} we encounter our first new example of a continuous non-invertible symmetry in the case of $G_k$ WZW models for specific target manifolds $G$ and levels $k$. We mostly focus on the case $G=SU(2)$ and discuss the action of these new charges on local operators. Then in Section \ref{sec:phantom_syms} we discuss another family of nontrivial examples appearing in specific products of minimal models. We also discuss an application of these continuous non-invertible symmetries, which allows us to find a continuum of conformal defects in some of the original minimal models. In Section \ref{sec:higher_d} we make some preliminary comments on higher-dimensional examples, and briefly comment on the non-invertible ABJ symmetry. Finally in Section \ref{sec:conclusions} we conclude and discuss some open questions.

\section{A Generalized Noether's Theorem for Continuous Non-Invertible Symmetries}\label{sec:Noether}

We focus on continuous non-invertible symmetries in conformal field theories in 1+1d, and discuss the proposal \eqref{eq:key} in both directions.

\subsection{Continuous TDL $\rightarrow$ Non-Local Conserved Current}

We review the discussion in \cite{Thorngren:2021yso}. Suppose $\CN$ is a topological defect line (TDL) labeled by a continuous parameter (which we suppress). There is an exactly marginal operator $J$ living on the TDL that generates deformations of this continuous parameter. Because $J$ is marginal, its conformal dimension is
\begin{eqnarray}
	\Delta_J= h_J + \bar{h}_J =1\;.
\end{eqnarray}
Moreover, $\CN$ being topological implies the conservation equation $d\star\! J=0$, which further implies that $J$ has spin 1, 
\begin{eqnarray}
	S_J = h_J - \bar{h}_J = \pm 1\;.
\end{eqnarray}
Combining the above conditions together, we find 
\begin{eqnarray}
	(h_J, \bar{h}_J) = (1,0) \text{ or } (0,1)\;.
\end{eqnarray}
So there is a (anti-)holomorphic operator $J$ with conformal weight $(1,0)$ (or $(0,1)$) living on the line $\CN$.

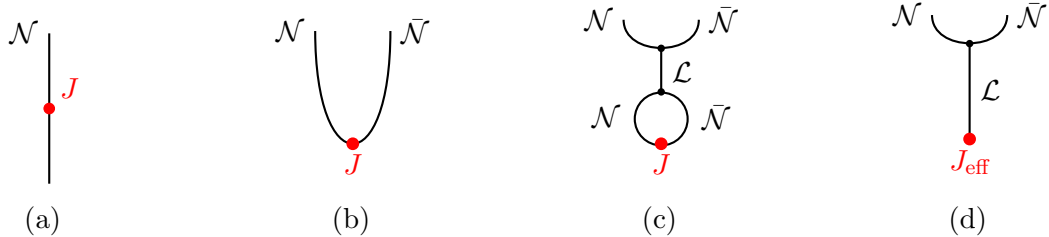
\begin{figure}
      \centering
     \begin{subfigure}[b]{0.24\textwidth}
         \centering
    \begin{tikzpicture}[scale=1, every node/.style={font=\small}]
    \begin{scope}[shift={(0,0)}]
        \draw[thick] (0,-1) -- (0,1) node[pos=1, left] {$\mathcal N$};
        \filldraw[red] (0,0) circle (2pt) node[above right] {$J$};
    \end{scope}
    \end{tikzpicture}
         \caption{}
     \end{subfigure}
     \begin{subfigure}[b]{0.24\textwidth}
         \centering
         \begin{tikzpicture}[scale=1, every node/.style={font=\small}]
    \begin{scope}[shift={(0,0)}]
        \draw[thick] (-0.5,2) .. controls (-0.5,0) and (0.5,0) .. (0.5,2);
        \filldraw [thick,red] (0,0.5) circle (2pt);
        \node[right] at (0.5,2) {$\bar{\mathcal{N}}$};
        \node[left] at (-0.5,2) {$\mathcal{N}$};
        \node[below, red] at (0,0.5) {$J$};
    \end{scope}
    \end{tikzpicture}
    \caption{}
    \label{fig:conserved_current_b}
     \end{subfigure}
     \begin{subfigure}[b]{0.24\textwidth}
         \centering
           \begin{tikzpicture}[scale=1, every node/.style={font=\small}]
    \begin{scope}[shift={(0,0)}]
        \draw[thick] (-0.5,2) .. controls (-0.5,1.5) and (0.5,1.5) .. (0.5,2);
        \draw[thick] (0,1) -- (0,1.65);
        \draw [thick] (0,0.68) circle (10pt);
        \filldraw [thick] (0,1.04) circle (1pt);
        \filldraw [thick] (0,1.62) circle (1pt);
        \filldraw [thick,red] (0,0.35) circle (2pt);
        \node[right] at (0.5,2) {$\bar{\mathcal{N}}$};
        \node[left] at (-0.5,2) {${\mathcal{N}}$};
        \node[right] at (0.4,0.7) {$\bar{\mathcal{N}}$};
        \node[left] at (-0.4,0.7) {${\mathcal{N}}$};
        \node[right] at (0,1.3) {$\mathcal{L}$};
        \node[below, red] at (0,0.35) {$J$};
    \end{scope}
    \end{tikzpicture}
         \caption{}
         \label{fig:conserved_current_c}
     \end{subfigure}
     \begin{subfigure}[b]{0.24\textwidth}
         \centering
           \begin{tikzpicture}[scale=1, every node/.style={font=\small}]
    \begin{scope}[shift={(0,0)}]
        \draw[thick] (-0.5,2) .. controls (-0.5,1.5) and (0.5,1.5) .. (0.5,2);
        \draw[thick] (0,0.35) -- (0,1.65);
        \filldraw [thick] (0,1.62) circle (1pt);
        \filldraw [thick,red] (0,0.35) circle (2pt);
        \node[left] at (-0.5,2) {${\mathcal{N}}$};
        \node[right] at (0.5,2) {$\bar{\mathcal{N}}$};
        \node[right] at (0,1) {$\mathcal{L}$};
        \node[below, red] at (0,0.35) {$J_{\text{eff}}$};
    \end{scope}
    \end{tikzpicture}
         \caption{}
     \end{subfigure}
    
        \caption{
        Folding the topological line $\mathcal{N}$ along a conserved current $J$. We suppress all arrows denoting orientation of the lines. (a) A conserved current living on a topological defect line $\mathcal{N}$. (b) Folding $\mathcal{N}$ along $J$. (c) Fusing the $\mathcal{N}$ with $\bar{\mathcal{N}}$ produces the the intermediate line $\mathcal{L}$. (d) Shrinking the bubble on which $J$ lives yields an effective current $J_{\text{eff}}$ living at the end of $\mathcal{L}$. 
        }
        \label{fig:conserved_current}
\end{figure}

We can then fold $\CN$ at the locus of $J$, as shown in Figure \ref{fig:conserved_current_b}. If we fuse the topological line $\mathcal{N}$ with its orientation reversal $\bar{\mathcal{N}}$, we produce a summation of ``tadpole" diagrams, as in Figure \ref{fig:conserved_current_c}, where $J$ lives on the bubble. The line $\mathcal{L}$ appears in the fusion channel 
\begin{eqnarray}
    \mathcal{N}\times \bar{\mathcal{N}} = \sum_{\mathcal{L}} N_{\mathcal{N}\bar{\mathcal{N}}}^{\mathcal{L}} \mathcal{L}\;.
\end{eqnarray}
Finally, shrinking the bubble produces an effective conserved current $J_{\text{eff}}$ living at the end of the line $\mathcal{L}$. Note that not every $\mathcal{L}$ in the fusion channel of $\mathcal{N}\times \bar{\mathcal{N}}$ is guaranteed to support an effective current $J_{\text{eff}}$. For certain lines $\mathcal{L}$, after shrinking the bubble, one can get a zero operator. 

Let's consider a few special cases. 
\begin{enumerate}
    \item When $\mathcal{N}$ is a topological line defect for an invertible symmetry, then $\mathcal{N}\times \bar{\mathcal{N}} =1$, hence after folding, the bubble decouples from $\mathcal{N}$ on the top in Figure \ref{fig:conserved_current_c}. Shrinking the bubble gives a local conserved current, as expected. 
    \item When $\mathcal{N}$ is a topological line defect obtained by a direct sum of invertible continuous defects, i.e.~$\mathcal{N}$ is a non-simple continuous symmetry line, then after folding and shrinking the bubble, the only $J_{\text{eff}}$ that survives is associated with $\mathcal{L}=1$. 
    \item In the cosine symmetry example from Section \ref{sec:introgauging}, $J_{\text{eff}}$ lives at the end of the line $\mathcal{L}= \widehat{\eta}$ which generates the quantum $\bZ_2$ symmetry. 
\end{enumerate}

For a continuous non-invertible symmetry that is not constructed from a direct sum of invertible lines (as in the second example above), there must be currents $J_{\text{eff}}$ living at the end of non-trivial lines $\mathcal{L}$. Hence a continuous non-invertible symmetry implies a non-local conserved current living at the end of a topological line. Such a topological line can either be a discrete line (as in the example of the cosine symmetry), or can be a continuous line (as in the example of the compact boson, see Section \ref{sec:c=1}).

\subsection{Non-Local Conserved Current $\rightarrow$ Continuous TDL}

For invertible symmetries, identifying currents is often much easier than constructing a topological defect directly. Conveniently, Noether's framework provides the link between the two: given a conserved current $J$, its integral defines a charge $Q=\int_{M_{d-1}}\star J$, and the exponential map produces a topological defect, $U_\alpha=e^{i\alpha Q}$. As we shall see in a few examples below, for non-invertible symmetries it is also often the case that (non-local) currents are easy to find, the obvious question being -- how do we produce non-invertible topological defects out of them? Is there a version of the exponential map for these non-local conserved currents? 

The answer turns out to be \emph{yes}, at least in conformal field theories in $1+1d$. In this section we prove this statement.

\subsubsection{Conserved Current on a Base Line}
\label{sec:baseline}

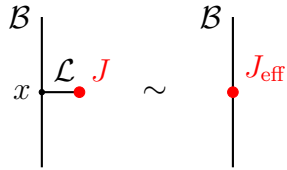
\begin{figure}
    \centering
    \begin{tikzpicture}
    \begin{scope}
        \draw[thick] (0,-1) -- (0,1) node[pos=1, left] {$\mathcal B$};
        \draw[thick] (0,0) -- (0.5,0);
        \filldraw[red] (0.5,0) circle (2pt) node[above right] {$J$};
        \node[above] at (0.3,0) {$\mathcal{L}$};
        \filldraw[black] (0,0) circle (1pt) node[left] {$x$};
    \end{scope}
    \node[] at (1.5,0) {$\sim$};
    \begin{scope}[xshift=1in]
        \draw[thick] (0,-1) -- (0,1) node[pos=1, left] {$\mathcal B$};
        \filldraw[red] (0,0) circle (2pt) node[above right] {$J_{\text{eff}}$};
    \end{scope}
    \end{tikzpicture}
    \caption{$J$ living at the end of a topological line $\mathcal{L}$ (left) can be used to define a $J_{\text{eff}}$ on the base loop $\mathcal{B}$ (right). }
    \label{fig:JonB}
\end{figure}

Our starting point is a non-local current $J$, of conformal weight $(h,\bar{h})= (1,0)$, living at the end of a (discrete) topological line $\mathcal{L}$. To construct a conserved charge and topological defect, one needs to worry about the other endpoint of $\mathcal{L}$. One way to deal with the other end is to let it \emph{topologically} terminate on a (possibly non-simple) \emph{base topological line} $\mathcal{B}$ at the topological junction $x$, See Figure \ref{fig:JonB}. The existence of a topological junction $x$ between $\mathcal{B}$ and $\mathcal{L}$ means that
\begin{eqnarray}\label{eq:BBL}
    \mathcal{B}\times \mathcal{L} \supset \mathcal{B}\;.
\end{eqnarray}
By shrinking $\mathcal{L}$, we find a conserved current $J_{\text{eff}}$ living along the base line $\mathcal{B}$.  Below, we will suppress the subscript ``eff'' and the label for the topological junction $x$ for simplicity. 

One immediate consequence of \eqref{eq:BBL} is that the topological base line $\mathcal{B}$ must be non-invertible, if $\mathcal{L}$ is not an identity line. 
Also note that a (not necessarily simple) line $\mathcal B$ satisfying~\eqref{eq:BBL} always exists when $\mathcal{L}$ is self-dual, since the choice $\mathcal B=1+\mathcal L$ is a valid one. The construction in Figure~\ref{fig:JonB}, where we allow for a general choice of $\mathcal B$, affords us flexibility in defining a topological defect. For example, it is often convenient to choose $\mathcal B$ to be simple, whenever possible.

When $\mathcal B$ exists, it is very much not unique. For example, given one such baseline, we can always construct other valid baselines by combining it with other line defects: if $\mathcal B$ satisfies~\eqref{eq:BBL}, then so do $\mathcal B'=\mathcal B\times\mathcal M$ and $\mathcal B'=\mathcal B+\mathcal M$, for any line $\mathcal M$. We will comment on this freedom in our choice of $\mathcal B$, and its effect on the corresponding continuous non-invertible defect, in the next subsection, after we explain how to build such a defect by integrating the non-local current along $\mathcal B$.

\subsubsection{Construction of the Defect}
\label{sec:defectconstruction}

We now consider an operator $J$ of dimensions $(h,\bar h)=(1,0)$ living along a topological base line $\mathcal{B}$. There is a natural construction of a corresponding topological defect: we start with the line $\mathcal{B}$ supported on a closed loop $\gamma$ in the path integral, and then deform this line by the local perturbation $J$ integrated along $\gamma$, by including $-i\alpha\oint_{\gamma} \frac{dz}{2\pi i} J(z)$.\footnote{This general procedure is known as a ``pinning field deformation'' in the defect literature. Here we are performing this deformation in the specific case of a conformal defect which is also topological.} Here we are using radial quantization for convenience. The partition function is thus modified to 
\begin{eqnarray}
    \int \mathcal{D}\phi\ e^{- S[\phi] + i \alpha \oint_{\gamma} \frac{dz}{2\pi i} J(z)} \mathcal{B}(\gamma) \;,
\end{eqnarray}
and so the schematic form of the defect is 
\begin{eqnarray}\label{eq:schematic}
    \mathcal{N}_\alpha(\gamma) = ``\ \mathcal{B}(\gamma) e^{i\alpha \int_{\gamma} \frac{dz}{2\pi i} J(z)}\ "\;.
\end{eqnarray}
However, \eqref{eq:schematic} does not mean the tensor product of $\mathcal{B}(\gamma)$ and $e^{i\alpha \int_{\gamma} \frac{dz}{2\pi i} J(z)}$ -- for one thing, the exponential is not well-defined by itself, the current only exists along $\mathcal B$. To define \eqref{eq:schematic} more precisely, we use the Taylor series
\begin{equation}\label{eq:def_of_U}
\begin{aligned}
\mathcal{N}_\alpha(\gamma)\ &:=\ 
\begin{tikzpicture}[scale=1.5, baseline=-3pt]
    \draw[thick] (1.2,0) circle [radius=0.7];
\end{tikzpicture}
+i\alpha\oint_\gamma \frac{dz}{2\pi i}\ 
\begin{tikzpicture}[scale=1.5, baseline=-3pt]
    \draw[thick] (4.1,0) circle [radius=0.7];
    \fill[red] (4.8,0) circle [radius=1pt];
    \node[red, left] at (4.8,0) {$J(z)$};
    \end{tikzpicture}\\
    &\ +\frac{(i\alpha)^2}{2!}\displaystyle\oint_{\gamma} \frac{dz_1}{2\pi i}\,\oint_{\gamma} \frac{dz_2}{2\pi i}\ 
\begin{tikzpicture}[scale=1.5, baseline=-3pt, every node/.style={font=\normalsize}]
    \draw[thick] (8.4,0) circle [radius=0.7];
    \fill[red] ($(8.4,0)+(45:0.7)$) circle [radius=1pt];
    \node[red, below left] at ($(8.4,0)+(45:0.7)$) {$J(z_1)$};
    \fill[red] ($(8.4,0)+(-135:0.7)$) circle [radius=1pt];
    \node[red, above right] at ($(8.4,0)+(-135:0.7)$) {$J(z_2)$};
\end{tikzpicture}\ +\ \cdots
\end{aligned}
\end{equation}
where the black line represents the base loop $\mathcal{B}$. The zero-th order term is the topological base loop $\mathcal{B}$. The first term $i\alpha Q$ is proportional to the conserved charge $Q= \oint \frac{dz}{2\pi i} 
\begin{tikzpicture}[scale=0.4, baseline=-3pt]
    \draw[thick] (4.1,0) circle [radius=0.7];
    \fill[red] (4.8,0) circle [radius=3pt];
    \end{tikzpicture}
$. For higher order terms, one needs to specify what happens when multiple $J$'s collide, which potentially would yield divergences. Therefore it is important to define a regularization procedure.

First we discuss how this works for a standard invertible symmetry (where $J$ is a local operator and the base line $\mathcal{B}$ is the trivial line). In this case, the topological line is $\mathcal{N}_\alpha = e^{i\alpha \oint_{\gamma} \frac{dz}{2\pi i} J(z)}$, which can be written as
\begin{equation}
\mathcal{N}_\alpha(\gamma) = 1 +i\alpha\oint_{\gamma} \frac{dz}{2\pi i}\ 
\begin{tikzpicture}[scale=1.5,baseline=-3pt]
  \draw[dashed] (4.1,0) circle [radius=0.7];
  \fill[red] (4.8,0) circle [radius=1pt];
  \node[red, left] at (4.8,0) {$J(z)$};
\end{tikzpicture}\ +\frac{(i\alpha)^2}{2!}\displaystyle\oint_{\gamma} \frac{dz_1}{2\pi i}\,\oint_{\gamma} \frac{dz_2}{2\pi i}\ 
  \begin{tikzpicture}[scale=1.5,baseline=-3pt]
  \draw[dashed] (8.4,0) circle [radius=0.7];
  \fill[red] ($(8.4,0)+(45:0.7)$) circle [radius=1pt];
  \node[red, above right] at ($(8.4,0)+(45:0.7)$) {$J(z_1)$};
  \fill[red] ($(8.4,0)+(-135:0.7)$) circle [radius=1pt];
  \node[red, above right] at ($(8.4,0)+(-135:0.7)$) {$J(z_2)$};
\end{tikzpicture}\! +\cdots
\end{equation}
where we have used dashed lines to denote integration contours, since now there are no topological lines which need to be attached. Then radial ordering consists of separating the contours of integration for the various $J$'s by slightly moving them in the transverse direction. For example, the term of order $\alpha^2$ will now take the form
\begin{equation}
\frac{(i\alpha)^2}{2!}\oint_{\gamma_1} \frac{dz_1}{2\pi i}\,\oint_{\gamma_2} \frac{dz_2}{2\pi i}\ 
\begin{tikzpicture}[scale=1.5, baseline=-3pt]
  \draw[dashed] (8.4,0) circle [radius=0.7];
  \draw[dashed] (8.4,0) circle [radius=0.6];
  \fill[red] ($(8.4,0)+(45:0.7)$) circle [radius=1pt];
  \node[red, above right] at ($(8.4,0)+(45:0.7)$) {$J(z_1)$};
  \fill[red] ($(8.4,0)+(-135:0.6)$) circle [radius=1pt];
  \node[red, above right] at ($(8.4,0)+(-135:0.6)$) {$J(z_2)$};
\end{tikzpicture}\;,
\end{equation}
where $\gamma_1$ is of larger radius than $\gamma_2$,
and so on. The integration can now be done without worrying about short-distance singularities, and we take the contours to coincide at the end. This procedure is especially convenient for proving that these defects are topological since no singularities will appear.

Now we can discuss the generalization to non-local currents. Radial ordering of the integration contour of non-local currents requires us to split the integration contour from the location of the topological base line. This is made possible by the separation as in Figure \ref{fig:JonB}: we separate a current living on $\mathcal{B}$ from $\mathcal{B}$, at the cost of connecting $J$ to $\mathcal{B}$ by a topological line $\mathcal{L}$. For instance, at the linear order we find
\begin{equation}\label{eq:charge_reg}
i\alpha\oint \frac{dz}{2\pi i}\; 
\begin{tikzpicture}[baseline=10pt]
    \draw[thick] (0,0.5) circle (.7cm);
\filldraw[red] (0,1.2) circle (1.5pt);
\node[below, red] at (0,1.2) {$J(z)$};
\end{tikzpicture}\longrightarrow
i\alpha\oint \frac{dz}{2\pi i}\; 
\begin{tikzpicture}[baseline=10pt]
    \draw[thick] (0,0.5) circle (1cm);
\draw[dashed] (0,0.5) circle (.6cm);
\draw[thick] (0,1.5) -- (0,1.1);
\filldraw[red] (0,1.1) circle (1.5pt);
\node[below, red] at (0,1.1) {$J(z)$};
\end{tikzpicture}\;.
\end{equation}
We emphasize that solid lines correspond to topological lines (which are not necessarily identical), while dashed lines correspond to integration contours. 
We are assuming one-dimensional junction spaces for simplicity, otherwise we need to keep track of various choices of vectors at vertices.

When considering higher order terms, an additional subtlety arises compared with local currents. Suppose there are two currents $J(z_1)$ and $J(z_2)$. If we integrate $z_2$ for fixed $z_1$, we need to consider both configurations where $z_2$ is to the left of $z_1$ and to the right of $z_1$. For self-consistency, we would like to demand that the regularization process is compatible between the two configurations. Concretely, starting with the configuration where $z_2$ is to the left of $z_1$, then performing the regularization, then moving $z_2$ to the right of $z_1$, the net result is consistent with starting with $z_2$ to the right of $z_1$, then performing the regularization. Graphically, this means
\begin{eqnarray}\label{eq:comm}
\begin{tikzpicture}[scale=1, baseline=(current bounding box.center), every node/.style={font=\normalsize}]
    \draw[thick] (3,0) -- (7,0);
    \draw[dashed] (3,0.5) -- (7,0.5);
    \draw[dashed] (3,1) -- (7,1);
    \draw[thick] (4,0) .. controls (4,1.5)  and (6,0)  .. (6,1);
    \draw[thick] (5,0) -- (5,0.5);
    \fill[red] (5,0.5) circle [radius=1.5pt];
    \fill[red] (6,1) circle [radius=1.5pt];

    \node[red, above] at (5,0.5) {$J(z_1)$};
    \node[red, above] at (6,1) {$J(z_2)$};
\end{tikzpicture}
\quad \overset?= \quad
\begin{tikzpicture}[scale=1, baseline=(current bounding box.center), every node/.style={font=\normalsize}]
    \draw[thick] (3,0) -- (7,0);
    \draw[dashed] (3,0.5) -- (7,0.5);
    \draw[dashed] (3,1) -- (7,1);
    \draw[thick] (5,0) -- (5,0.5);
    \draw[thick] (6,0) -- (6,1);
    \fill[red] (5,0.5) circle [radius=1.5pt];
    \fill[red] (6,1) circle [radius=1.5pt];

    \node[red, above] at (5,0.5) {$J(z_1)$};
    \node[red, above] at (6,1) {$J(z_2)$};
\end{tikzpicture}\;,
\end{eqnarray}
where the bottom horizontal line is $\mathcal{B}$, and the vertical lines are $\mathcal{L}$. This equation is not satisfied for every $\mathcal{L}$ that allows a conserved current on its end. When $\mathcal{L}$ is an invertible line, it is easy to show that \eqref{eq:comm} holds, by using the property that the lines have spin 1 which ensures the absence of an anomaly by the spin selection rules\cite{Lin:2019kpn,Lin:2021udi}. But for $\mathcal{L}$ non-invertible, \eqref{eq:comm} generically does not hold. To consistently define an operator for arbitrary $\mathcal{L}$, we need to prevent $z_2$ passing through $z_1$, i.e.~we must perform path ordering along the angular direction. More precisely, the path ordering along the angle direction is defined to be\footnote{One may wonder whether the path ordering is compatible with the fact that the $\arg(z_i)$ is circle valued. One can avoid this problem by picking an arbitrary point on the circle as a base point, and fixing the integration domain of $\arg(z_i)$ to be $[0,2\pi)$, as is standard in the definition of the path ordered Wilson loops.}
\begin{eqnarray}
    \mathcal{P}[J(z_1) J(z_2)] := 
    \begin{cases}
        \begin{tikzpicture}[baseline=(current bounding box.center)]
            \draw[thick] (0,0) -- (3,0);
            \draw[thick, dashed] (0,0.5) -- (3,0.5);
            \draw[thick, dashed] (0,1) -- (3,1);
            \draw[thick] (1,0) -- (1,0.5);
            \fill[red] (1,0.5) circle [radius=1.5pt];
            \draw[thick] (2,0) -- (2,1);
            \fill[red] (2,1) circle [radius=1.5pt];
            \node[above, red] at (1,0.5) {$J(z_1)$}; 
            \node[above, red] at (2,1) {$J(z_2)$}; 
        \end{tikzpicture}\;, & \text{ if } \arg(z_1) \leq \arg(z_2) \\~\\
        \begin{tikzpicture}[baseline=(current bounding box.center)]
            \draw[thick] (0,0) -- (3,0);
            \draw[thick, dashed] (0,0.5) -- (3,0.5);
            \draw[thick, dashed] (0,1) -- (3,1);
            \draw[thick] (1,0) -- (1,1);
            \fill[red] (1,1) circle [radius=1.5pt];
            \draw[thick] (2,0) -- (2,0.5);
            \fill[red] (2,0.5) circle [radius=1.5pt];
            \node[above, red] at (1,1) {$J(z_2)$}; 
            \node[above, red] at (2,0.5) {$J(z_1)$}; 
        \end{tikzpicture}\;, & \text{ if } \arg(z_2) \leq \arg(z_1) \\
    \end{cases}
\end{eqnarray}
We note that in the case of a local current, the path ordering along the angle direction is immaterial, because \eqref{eq:comm} is trivially satisfied. 

Implementing the radial ordering and path ordering, we can regularize the $n$'th order term in the definition \eqref{eq:def_of_U}. This term is 
\begin{equation}
\def\cs{0.7}
\begin{tikzpicture}[scale=1.5, baseline=(current bounding box.center), every node/.style={font=\normalsize}]
  \node at (5.9,0) {$\displaystyle \frac{(i\alpha)^n}{n!}\oint \prod_i \frac{dz_i}{2\pi i}\mathcal{P} \bigg[$ };
  \node at (9.6,0) {$\displaystyle \bigg]$};
  \draw[thick] (8.0,0) circle [radius=\cs];
  \fill[red] ($(8,0)+(45:\cs)$) circle [radius=1pt];
  \node[red, above right] at ($(8,0)+(45:\cs)$) {$J(z_n)$};
  \fill[red] ($(8,0)+(-0:\cs)$) circle [radius=1pt];
  \node[red,  right] at ($(8,0)+(-0:\cs)$) {$J(z_1)$};
  \fill[red] ($(8,0)+(-45:\cs)$) circle [radius=1pt];
  \node[red, below right] at ($(8,0)+(-45:\cs)$) {$J(z_2)$};
  
  \fill[red] ($(8,0)+(-90:\cs)$) circle [radius=1pt];
  \fill[red] ($(8,0)+(-135:\cs)$) circle [radius=1pt];
  \fill[red] ($(8,0)+(-180:\cs)$) circle [radius=1pt];
  \fill[red] ($(8,0)+(135:\cs)$) circle [radius=1pt];
  \node[above] at ($(8,0)+(90:\cs)$) {$\cdots$};
\end{tikzpicture}\;.
\end{equation}
The regularized version is
\begin{equation}
\def\cs{1}
\begin{tikzpicture}[scale=1.5, baseline=(current bounding box.center), every node/.style={font=\normalsize}]
  \node at (5.7,0) {$\displaystyle \frac{(i\alpha)^n}{n!}\oint \prod_i \frac{dz_i}{2\pi i} \mathcal{P} \bigg[$};
  \node at (9.9,0) {$\displaystyle \bigg]$};
  \draw[thick] (8.0,0) circle [radius=\cs];
  \draw[dashed] (8.0,0) circle [radius=\cs-0.1];
  \draw[dashed] (8.0,0) circle [radius=\cs-0.2];
  \draw[dashed] (8.0,0) circle [radius=\cs-0.3];
  \draw[dashed] (8.0,0) circle [radius=\cs-0.4];
  \draw[dashed] (8.0,0) circle [radius=\cs-0.5];
  \draw[dashed] (8.0,0) circle [radius=\cs-0.6];
  \draw[dashed] (8.0,0) circle [radius=\cs-0.7];

  \draw[thick] ($(8,0)+(45:\cs-0.7)$) -- ($(8,0)+(45:\cs)$);
  \draw[thick] ($(8,0)+(135:\cs-0.6)$) -- ($(8,0)+(135:\cs)$);
  \draw[thick] ($(8,0)+(180:\cs-0.5)$) -- ($(8,0)+(180:\cs)$);
  \draw[thick] ($(8,0)+(-135:\cs-0.4)$) -- ($(8,0)+(-135:\cs)$);
  \draw[thick] ($(8,0)+(-90:\cs-0.3)$) -- ($(8,0)+(-90:\cs)$);
  \draw[thick] ($(8,0)+(-45:\cs-0.2)$) -- ($(8,0)+(-45:\cs)$);
  \draw[thick] ($(8,0)+(-0:\cs-0.1)$) -- ($(8,0)+(0:\cs)$);
  
  \fill[red] ($(8,0)+(45:\cs-0.7)$) circle [radius=1pt];
  \fill[red] ($(8,0)+(-0:\cs-0.1)$) circle [radius=1pt];
  \fill[red] ($(8,0)+(-45:\cs-0.2)$) circle [radius=1pt];  
  \fill[red] ($(8,0)+(-90:\cs-0.3)$) circle [radius=1pt];
  \fill[red] ($(8,0)+(-135:\cs-0.4)$) circle [radius=1pt];
  \fill[red] ($(8,0)+(-180:\cs-0.5)$) circle [radius=1pt];
  \fill[red] ($(8,0)+(135:\cs-0.6)$) circle [radius=1pt];

  \node[red, above right] at ($(8.1,0)+(45:\cs)$) {$\!J(z_n)$};
  \node[red,  right] at ($(8,0)+(0:\cs-0.1)$) {$\ J(z_1)$};
  \node[red, below right] at ($(8,0)+(-45:\cs-0.2)$) {$\ J(z_2)$};
  \node[above] at ($(8,0)+(90:\cs)$) {$\cdots$};

\end{tikzpicture}\;.
\end{equation}

In summary, we define the symmetry defect from the non-local current via
\begin{eqnarray}\label{eq:topdef}
\def\cs{1}
    \begin{tikzpicture}[scale=1.5, baseline=(current bounding box.center), every node/.style={font=\normalsize}]
  \node at (5.1,0) {$\displaystyle \mathcal{N}_\alpha = \sum_{n=0}^{\infty}\frac{(i\alpha)^n}{n!}\oint \prod_i \frac{dz_i}{2\pi i} \mathcal{P} \bigg[$};
  \node at (9.9,0) {$\displaystyle \bigg]$};
  \draw[thick] (8.0,0) circle [radius=\cs];
  \draw[dashed] (8.0,0) circle [radius=\cs-0.1];
  \draw[dashed] (8.0,0) circle [radius=\cs-0.2];
  \draw[dashed] (8.0,0) circle [radius=\cs-0.3];
  \draw[dashed] (8.0,0) circle [radius=\cs-0.4];
  \draw[dashed] (8.0,0) circle [radius=\cs-0.5];
  \draw[dashed] (8.0,0) circle [radius=\cs-0.6];
  \draw[dashed] (8.0,0) circle [radius=\cs-0.7];

  \draw[thick] ($(8,0)+(45:\cs-0.7)$) -- ($(8,0)+(45:\cs)$);
  \draw[thick] ($(8,0)+(135:\cs-0.6)$) -- ($(8,0)+(135:\cs)$);
  \draw[thick] ($(8,0)+(180:\cs-0.5)$) -- ($(8,0)+(180:\cs)$);
  \draw[thick] ($(8,0)+(-135:\cs-0.4)$) -- ($(8,0)+(-135:\cs)$);
  \draw[thick] ($(8,0)+(-90:\cs-0.3)$) -- ($(8,0)+(-90:\cs)$);
  \draw[thick] ($(8,0)+(-45:\cs-0.2)$) -- ($(8,0)+(-45:\cs)$);
  \draw[thick] ($(8,0)+(-0:\cs-0.1)$) -- ($(8,0)+(0:\cs)$);
    
  \fill[red] ($(8,0)+(45:\cs-0.7)$) circle [radius=1pt];
  \fill[red] ($(8,0)+(-0:\cs-0.1)$) circle [radius=1pt];
  \fill[red] ($(8,0)+(-45:\cs-0.2)$) circle [radius=1pt];  
  \fill[red] ($(8,0)+(-90:\cs-0.3)$) circle [radius=1pt];
  \fill[red] ($(8,0)+(-135:\cs-0.4)$) circle [radius=1pt];
  \fill[red] ($(8,0)+(-180:\cs-0.5)$) circle [radius=1pt];
  \fill[red] ($(8,0)+(135:\cs-0.6)$) circle [radius=1pt];

  \node[red, above right] at ($(8.1,0)+(45:\cs)$) {$\!J(z_n)$};
  \node[red,  right] at ($(8,0)+(0:\cs-0.1)$) {$\ J(z_1)$};
  \node[red, below right] at ($(8,0)+(-45:\cs-0.2)$) {$\ J(z_2)$};
  \node[above] at ($(8,0)+(90:\cs)$) {$\cdots$};

\end{tikzpicture}
\end{eqnarray}

Our next task is to show that this is topological. Before doing that, we make a couple of remarks about the object we just constructed:
\begin{itemize}
\item The defect $\mathcal N_\alpha$ is not necessarily identical to the continuous defect $\mathcal N$ from the previous subsection. This situation is no different from the invertible case: for example, one can begin with a continuous $U_\alpha$, and stack it with an invertible discrete $C$. The defect $U_\alpha C$ is still continuous, and therefore it still hosts a current $J$. If we integrate this current, and exponentiate it, we recover $U_\alpha$ instead of $U_\alpha C$. The same happens in the non-invertible case: the defect~\eqref{eq:topdef} need not coincide on the nose with the initial defect that hosted the non-local current $J$.
\item The defect $\mathcal N_\alpha$ depends on the choice of baseline $\mathcal B$. We already mentioned in the previous subsection that there is no unique choice of $\mathcal B$. For example, given any valid choice, we can always combine it with other topological lines to yield admissible baselines: if $\mathcal B$ hosts a current, then so do $\mathcal B\times\mathcal M$ and $\mathcal B+\mathcal M$. This ambiguity has a simple effect on~\eqref{eq:topdef}:
\begin{equation}
\mathcal N_\alpha\bigr|_{\mathcal B\times\mathcal M}=\mathcal N_\alpha\bigr|_{\mathcal B}\times\mathcal M \;,\qquad \mathcal N_\alpha\bigr|_{\mathcal B+\mathcal M}=\mathcal N_\alpha\bigr|_{\mathcal B}+\mathcal M\;.
\end{equation}
The ambiguity in our choice of baseline is even larger than this: there are in general multiple valid choices of $\mathcal B$ that are not related via the two operations above. The full set of line defects includes the original discrete symmetries, plus the continuous defects $\mathcal N_\alpha\bigr|_{\mathcal B}$ for all valid choices of $\mathcal B$, modulo all the redundancies inherent to this choice. It would be nice to understand the proper mathematical structure that describes this set of objects.
\item For a given choice of $\mathcal B$, the defect $\mathcal N_\alpha$ also depends on the choice of junction $x\in\operatorname{Hom}(\mathcal B\times\bar{\mathcal B},\mathcal L)$. In this work we only consider one-dimensional junction spaces so this will play no role, but it would be interesting to understand better how this choice affects the defect $\mathcal N_\alpha$.
\end{itemize}

\subsubsection{Topological Invariance}

We need to show that the operator \eqref{eq:topdef} is topological, hence qualifying as a symmetry operator. Colloquially, an operator is topological if it is transparent to the stress-energy tensor. In 1+1d CFT, this can be made more precise: the operator $\mathcal{N}_\alpha$ is topological if 
\begin{eqnarray}\label{eq:NLLbar}
    [\mathcal{N}_\alpha, L_n]=[\mathcal{N}_\alpha, \bar{L}_n]=0, \quad \forall \ n\in \bZ\;,
\end{eqnarray}
where the $L_n$'s are the usual Virasoro generators \cite{Fuchs:2007tx}.

Since $\mathcal{N}_\alpha$ is constructed from holomorphic operators, it automatically commutes with $\bar{L}_n$. Hence the only non-trivial task is to show $[\mathcal{N}_\alpha, L_n]=0$ for $\forall \ n\in \bZ$. We will prove that they commute order by order in $\alpha$.

\paragraph{Zero-th Order:}
The zero-th order term in $\alpha$ is simply the topological base loop $\mathcal{B}$, which by assumption commutes with $L_n$.

\paragraph{First Order:}
The first order term in $\alpha$ is $i\alpha Q$, where $Q$ is the charge 
\begin{equation}\label{eq:charge}
Q=\oint_{\gamma_b} \frac{dz}{2\pi i}\, J(z):=\ \oint_{\gamma_b}\frac{dz}{2\pi i}\,
\tikz[every node/.style={font=\small},baseline=10pt]{
\draw[thick] (0,0.5) circle (1cm);
\draw[dashed] (0,0.5) circle (.6cm);
\draw[thick] (0,1.5) -- (0,1.1);
\fill[red] (0,1.1) circle (1.5pt);
}\;,
\end{equation}
where we adopt the regularization in \eqref{eq:charge_reg}. The red dot represents $J(z)$, and the black dotted line is the integration contour $\gamma_b$.  On the other hand, the Virasoro generator $L_n$ is defined by
\begin{eqnarray}
    L_n=\oint_{\gamma_g} \frac{dw}{2\pi i}\ w^{n+1}T(w):=\ 
\oint_{\gamma_g} \frac{dw}{2\pi i}\ w^{n+1}\ \tikz[every node/.style={font=\small},baseline=10pt]{
\draw[dashed, blue!40!green] (0,0.5) circle (.8cm);
\filldraw[blue!40!green] (.8,.5) circle (1.5pt);
}\;,
\end{eqnarray}
where the green dot represents $T(w)$, and the green dotted line represents the integration contour $\gamma_g$ of $T(w)$.

The commutator is 
\begin{equation}\label{eq:QLn}
\begin{aligned}
[Q,L_n]&=\oint_{\gamma_{b}}\frac{dz}{2\pi i} \oint_{\gamma_g} \frac{dw}{2\pi i} w^{n+1} \bigg[\ \begin{tikzpicture}[every node/.style={font=\small},baseline=10pt]
\draw[thick] (0,0.5) circle (1cm);
\draw[dashed, blue!40!green] (0,0.5) circle (.6cm);
\draw[thick] (0,1.5) -- (0,1.3);
\fill[red] (0,1.3) circle (1.5pt);
\draw[dashed] (0,0.5) circle (.8cm);
\filldraw[blue!40!green] (.6,.5) circle (1.5pt);
\end{tikzpicture}
- 
\begin{tikzpicture}[every node/.style={font=\small},baseline=10pt]
\draw[thick] (0,0.5) circle (1cm);
\draw[dashed] (0,0.5) circle (.6cm);
\draw[thick] (0,1.5) -- (0,1.1);
\fill[red] (0,1.1) circle (1.5pt);
\draw[dashed, blue!40!green] (0,0.5) circle (.8cm);
\filldraw[blue!40!green] (.8,.5) circle (1.5pt);
\end{tikzpicture}
\bigg]
\\[+2ex]
&=\oint_{\gamma_g} \frac{dw}{2\pi i}  \oint_{z\circlearrowleft w}\frac{dz}{2\pi i} w^{n+1} \ \tikz[every node/.style={font=\small},baseline=10pt]{
\draw[thick] (0,0.5) circle (1cm);
\draw[dashed] (0.5,0.5) circle (.3cm);
\draw[thick] (0.8,0.5) -- (1,0.5);
\fill[red] (0.8,0.5) circle (1.5pt);
\draw[dashed, blue!40!green] (0,0.5) circle (.5cm);
\filldraw[blue!40!green] (0.5,0.5) circle (1.5pt);
}
\end{aligned}
\end{equation}
In the first line, we use radial ordering so that in $QL_n$, the current $J(z)$ is on a larger circle than $T(w)$, while in $L_n Q$ the current $J(z)$ is on a smaller circle than $T(w)$. When $J(z)$ and $T(w)$ are sufficiently far away, the two terms in the bracket cancel against each other. The only non-trivial contribution to the integral in the first line occurs when $J(z)$ and $T(w)$ are close to each other, which reduces to integrating $J(z)$ around $T(w)$. Because $J(z)$ and $T(w)$ are close to each other, we can take the operator product expansion (OPE) between them, and integrating over $z$ around $w$ amounts to extracting the residue, i.e.~the coefficient proportional to $(z-w)^{-1}$. The standard OPE between $T(w)$ and $J(z)$ is\footnote{This OPE follows from the fact that $J$ is a primary and from holomorphicity. The presence of the topological line $\mathcal{L}$ attached to $J(z)$ does not change the form of OPE. }
\begin{eqnarray}
    T(w) J(z) \stackrel{w\to z}{\sim} \frac{J(z)}{(w-z)^2} + \frac{\partial_z J(z)}{w-z} + \text{reg}\;.
\end{eqnarray}
Taking the OPE in the opposite ordering, we get
\begin{eqnarray}\label{eq:JT}
    J(z) T(w) \stackrel{z\to w}{\sim} \frac{J(w)}{(z-w)^2} + \text{reg}\;.
\end{eqnarray}
Hence, the residue is zero, implying that the second line in \eqref{eq:QLn}  vanishes after integrating over $z$. In summary we proved that $[\mathcal{N}_\alpha, L_n]$ vanishes to the first order in $\alpha$, or equivalently, the charge $Q$ is conserved. 

We comment that at this order the proof is pretty much the same as the proof for invertible symmetries, see e.g.~\cite[Chapter 2.6]{Polchinski:1998rq}. The only difference is the presence of topological lines $\mathcal{L}$ and $\mathcal{B}$. But since $T(z)$ is transparent to topological lines, the computation is essentially the same as the proof for $[Q, L_n]=0$ for an ordinary symmetry.

\paragraph{Second Order:}
The second order term in $\alpha$ is $\frac{(i\alpha)^2}{2!}$ multiplied by 
\begin{equation}
\def\cs{1}
\oint_{\gamma_b} \prod_{i=1}^{2}\frac{dz_i}{2\pi i} \mathcal{P}\bigg[~
\begin{tikzpicture}[scale=1.5, baseline=(current bounding box.center), every node/.style={font=\normalsize}]
  \draw (8.4,0) circle [radius=0.7];
  \draw[dashed] (8.4,0) circle [radius=0.6];
  \draw[dashed] (8.4,0) circle [radius=0.5];

  \draw[thick] ($(8.4,0)+(45:0.6)$) -- ($(8.4,0)+(45:0.7)$);
  \draw[thick] ($(8.4,0)+(-135:0.5)$) -- ($(8.4,0)+(-135:0.7)$);

  \fill[red] ($(8.4,0)+(45:0.6)$) circle [radius=1pt];
  \fill[red] ($(8.4,0)+(-135:0.5)$) circle [radius=1pt];
\end{tikzpicture}
~\bigg]\;.
\end{equation}
Computing the commutator with $L_n$, 
\begin{equation}
\oint_{\gamma_b} \prod_{i=1}^2\frac{dz_i}{2\pi i} \oint_{\gamma_g} \frac{dw}{2\pi i} w^{n+1}\mathcal{P}\bigg[
\begin{tikzpicture}[scale=1.5, baseline=(current bounding box.center), every node/.style={font=\normalsize}]
  \draw (8.4,0) circle [radius=0.7];
  \draw[dashed] (8.4,0) circle [radius=0.6];
  \draw[dashed] (8.4,0) circle [radius=0.5];
  \draw[dashed, blue!40!green] (8.4,0) circle [radius=0.4];

  \draw[thick] ($(8.4,0)+(45:0.6)$) -- ($(8.4,0)+(45:0.7)$);
  \draw[thick] ($(8.4,0)+(-135:0.5)$) -- ($(8.4,0)+(-135:0.7)$);

  \fill[red] ($(8.4,0)+(45:0.6)$) circle [radius=1pt];
  \fill[red] ($(8.4,0)+(-135:0.5)$) circle [radius=1pt];
  \fill[blue!40!green] ($(8.4,0)+(-45:0.4)$) circle [radius=1pt]; 
\end{tikzpicture}
- 
\begin{tikzpicture}[scale=1.5, baseline=(current bounding box.center), every node/.style={font=\normalsize}]
  \draw (8.4,0) circle [radius=0.7];
  \draw[dashed, blue!40!green] (8.4,0) circle [radius=0.6];
  \draw[dashed] (8.4,0) circle [radius=0.5];
  \draw[dashed] (8.4,0) circle [radius=0.4];

  \draw[thick] ($(8.4,0)+(45:0.5)$) -- ($(8.4,0)+(45:0.7)$);
  \draw[thick] ($(8.4,0)+(-135:0.4)$) -- ($(8.4,0)+(-135:0.7)$);

  \fill[red] ($(8.4,0)+(45:0.5)$) circle [radius=1pt];
  \fill[red] ($(8.4,0)+(-135:0.4)$) circle [radius=1pt];
  \fill[blue!40!green] ($(8.4,0)+(-45:0.6)$) circle [radius=1pt]; 
\end{tikzpicture}
\bigg]\;,
\end{equation}
where the path ordering is only over the $J(z_1)$ and $J(z_2)$. Since $T(w)$ is not attached to any topological line, it does not participate in path ordering. We have also used the fact that topological lines are invisible to $T$.
Now we add and subtract a configuration where $T(w)$ is integrated between the two integration contours for the currents:
\begin{equation}\label{eq:insert}
\oint_{\gamma_b} \prod_{i=1}^2\frac{dz_i}{2\pi i} \oint_{\gamma_g} \frac{dw}{2\pi i} w^{n+1}\mathcal{P}\bigg[
\begin{tikzpicture}[scale=1.5, baseline=(current bounding box.center), every node/.style={font=\normalsize}]
  \draw (8.4,0) circle [radius=0.7];
  \draw[dashed] (8.4,0) circle [radius=0.6];
  \draw[dashed] (8.4,0) circle [radius=0.5];
  \draw[dashed, blue!40!green] (8.4,0) circle [radius=0.4];

  \draw[thick] ($(8.4,0)+(45:0.6)$) -- ($(8.4,0)+(45:0.7)$);
  \draw[thick] ($(8.4,0)+(-135:0.5)$) -- ($(8.4,0)+(-135:0.7)$);

  \fill[red] ($(8.4,0)+(45:0.6)$) circle [radius=1pt];
  \fill[red] ($(8.4,0)+(-135:0.5)$) circle [radius=1pt];
  \fill[blue!40!green] ($(8.4,0)+(-45:0.4)$) circle [radius=1pt]; 
\end{tikzpicture}
- 
\begin{tikzpicture}[scale=1.5, baseline=(current bounding box.center), every node/.style={font=\normalsize}]
  \draw (8.4,0) circle [radius=0.7];
  \draw[dashed] (8.4,0) circle [radius=0.6];
  \draw[dashed,blue!40!green] (8.4,0) circle [radius=0.5];
  \draw[dashed] (8.4,0) circle [radius=0.4];

  \draw[thick] ($(8.4,0)+(45:0.6)$) -- ($(8.4,0)+(45:0.7)$);
  \draw[thick] ($(8.4,0)+(-135:0.4)$) -- ($(8.4,0)+(-135:0.7)$);

  \fill[red] ($(8.4,0)+(45:0.6)$) circle [radius=1pt];
  \fill[red] ($(8.4,0)+(-135:0.4)$) circle [radius=1pt];
  \fill[blue!40!green] ($(8.4,0)+(-45:0.5)$) circle [radius=1pt]; 
\end{tikzpicture}
+ 
\begin{tikzpicture}[scale=1.5, baseline=(current bounding box.center), every node/.style={font=\normalsize}]
  \draw (8.4,0) circle [radius=0.7];
  \draw[dashed] (8.4,0) circle [radius=0.6];
  \draw[dashed,blue!40!green] (8.4,0) circle [radius=0.5];
  \draw[dashed] (8.4,0) circle [radius=0.4];

  \draw[thick] ($(8.4,0)+(45:0.6)$) -- ($(8.4,0)+(45:0.7)$);
  \draw[thick] ($(8.4,0)+(-135:0.4)$) -- ($(8.4,0)+(-135:0.7)$);

  \fill[red] ($(8.4,0)+(45:0.6)$) circle [radius=1pt];
  \fill[red] ($(8.4,0)+(-135:0.4)$) circle [radius=1pt];
  \fill[blue!40!green] ($(8.4,0)+(-45:0.5)$) circle [radius=1pt]; 
\end{tikzpicture}
- 
\begin{tikzpicture}[scale=1.5, baseline=(current bounding box.center), every node/.style={font=\normalsize}]
  \draw (8.4,0) circle [radius=0.7];
  \draw[dashed, blue!40!green] (8.4,0) circle [radius=0.6];
  \draw[dashed] (8.4,0) circle [radius=0.5];
  \draw[dashed] (8.4,0) circle [radius=0.4];

  \draw[thick] ($(8.4,0)+(45:0.5)$) -- ($(8.4,0)+(45:0.7)$);
  \draw[thick] ($(8.4,0)+(-135:0.4)$) -- ($(8.4,0)+(-135:0.7)$);

  \fill[red] ($(8.4,0)+(45:0.5)$) circle [radius=1pt];
  \fill[red] ($(8.4,0)+(-135:0.4)$) circle [radius=1pt];
  \fill[blue!40!green] ($(8.4,0)+(-45:0.6)$) circle [radius=1pt]; 
\end{tikzpicture}
\bigg]\;.
\end{equation}
The first two terms and the last two terms each form a pair where one of the $J(z)$'s is commuted past $T(w)$. In each pair, we apply the discussion above, where the only contribution from the integration of $J(z)$ is when it is close to $T(w)$, and we perform the OPE between them. By \eqref{eq:JT}, the residue vanishes, and so the difference between the first two terms vanishes. A similar equation applies to the second pair. In summary, we proved that $[\mathcal{N}_\alpha, L_n]$ vanishes to the second order in $\alpha$. 

\paragraph{Arbitrary Order:} The above discussion can be straightforwardly generalized to arbitrary higher orders. Like \eqref{eq:insert}, we add and subtract terms, and group the configurations in pairs so that in each pair $T(w)$ is only commuted past the closest $J(z)$, as in \eqref{eq:insert}. By the same argument as above, the $z_i$ integration in each pair vanishes because of the absence of a residue. This completes the proof of \eqref{eq:NLLbar}.

\subsubsection{Comments on Non-Invertible Fusion Rules and Conserved Charges}

We make a few miscellaneous comments. 

First, since we proved that the operator $\mathcal{N}_\alpha$ constructed out of a non-local current is topological, $\mathcal{N}_\alpha$ represents a global symmetry. It is natural to ask for the fusion rule between these topological operators. We relegate the explicit computation of fusion rule, among other aspects, to future works. However, the fusion rule must be non-invertible, because $\mathcal{N}_0$, i.e.~when $\alpha=0$, equals the base line $\mathcal{B}$, which is shown to be non-invertible in Section \ref{sec:baseline}. Hence generically $\mathcal{N}_\alpha$ satisfies a non-invertible fusion rule.

We would also like to comment on the relation between the conserved charge $Q$, as defined in \eqref{eq:charge}, and the topological operator $\mathcal{N}_\alpha$, as defined in \eqref{eq:topdef}. For an invertible symmetry, they are related by the exponential map. However, for a non-invertible symmetry, as we see from the definition \eqref{eq:topdef}, $\mathcal{N}_\alpha$ is not $\exp(i \alpha Q)$ because of the existence of the base line. Alternatively, one may construct a \emph{different} topological operator by
\begin{eqnarray}
    \mathcal{V}_\alpha = e^{i \alpha Q}\;,
\end{eqnarray}
where $Q$ is defined in \eqref{eq:charge}. The topologicalness of $\mathcal{V}_\alpha$ automatically follows from the topologicalness of $Q$ as shown in \eqref{eq:QLn}. 
However, it is not clear why this would define a good defect (in the sense that it has a well-defined defect Hilbert space).

\section{Non-Local Conserved Currents in $c=1$ CFTs}
\label{sec:c=1}

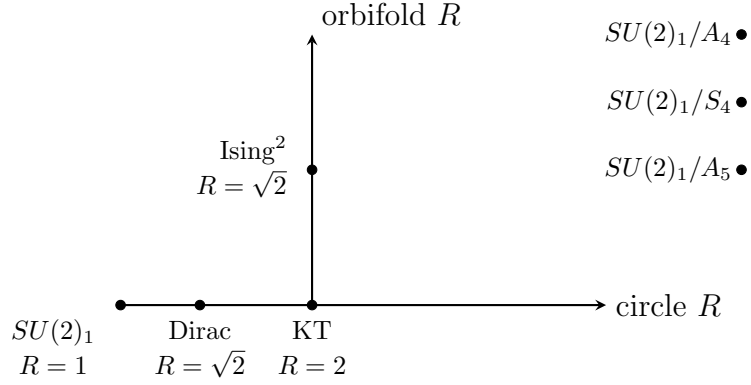
\begin{figure}
    \centering
    \footnotesize{
    \begin{tikzpicture}[scale=1.8]

\def\xSU{1.4142}
\def\xDirac{2.0}
\def\xKT{2.8284}
\def\yIsing{1.0}
\def\ymax{2.0}
\def\xmax{5}

\draw[->, thick,>=stealth] (\xSU, 0) -- (\xmax, 0) node[anchor=west] {\normalsize{circle $R$}};
\draw[->, thick,>=stealth] (\xKT, 0) -- (\xKT, \ymax) node[anchor=south west] {\normalsize{orbifold $R$}};

\filldraw (\xSU, 0) circle (1pt);
\node[anchor=north east] at (\xSU, 0) {
  \begin{tabular}{c}
    $SU(2)_1$ \\
    $R = 1$
  \end{tabular}
};

\filldraw (\xDirac, 0) circle (1pt);
\node[anchor=north] at (\xDirac, 0) {
  \begin{tabular}{c}
    Dirac \\
    $R = \sqrt{2}$
  \end{tabular}
};

\filldraw (\xKT, 0) circle (1pt);
\node[anchor=north] at (\xKT, 0) {
  \begin{tabular}{c}
    KT \\
    $R = 2$
  \end{tabular}
};

\filldraw (\xKT, \yIsing) circle (1pt);
\node[anchor=east] at (\xKT, \yIsing) {
  \begin{tabular}{r}
    $\text{Ising}^2$ \\
    $R = \sqrt{2}$
  \end{tabular}
};

\def\discx{6}
\filldraw (\discx,2) circle (1pt);
\node[anchor=east] at (\discx, 2) {$SU(2)_1/A_4$
};
\filldraw (\discx,1.5) circle (1pt);
\node[anchor=east] at (\discx, 1.5) {$SU(2)_1/S_4$
};
\filldraw (\discx,1) circle (1pt);
\node[anchor=east] at (\discx, 1) {$SU(2)_1/A_5$
};

\end{tikzpicture}
}
    \caption{Space of $c=1$ CFTs. }
    \label{fig:c=1}
\end{figure}

Starting from this section, we provide examples of non-local conserved currents in 1+1d (unitary) CFTs. The simplest CFTs with smallest central charges are the minimal models, with $c<1$. But as we discuss in Appendix \ref{app:min_mod}, there aren't any non-local currents in these theories. The next candidate is therefore the $c=1$ CFTs, see Figure~\ref{fig:c=1} for the space of $c=1$ CFTs. Continuous non-invertible symmetries in this model have already been discussed in \cite[Section 3.5 $\&$ 4.7]{Thorngren:2021yso}. See also \cite{Chang:2020imq,Damia:2023gtc}. 
In this section, we review them from the perspective of non-local currents.

We begin by collecting the basic data of the compact boson on the circle branch. See \cite{Ginsparg:1987eb,Ginsparg:1988ui}  for more details. 
The Lagrangian is 
\begin{eqnarray}\label{eq:compactboson}
    \frac{R^2}{4\pi} \partial_\mu \phi \partial^\mu \phi\;, \quad \phi\sim \phi+2\pi\;.
\end{eqnarray}
The theory has a T-duality, mapping $R\leftrightarrow 1/R$. The scalar $\phi$ is mapped to $\theta$, with the relation $d\phi= -\frac{i}{R^2} \star\!d\theta$. The local primary operators are 
\begin{eqnarray}\label{eq:Vnw}
    V_{n,w}= e^{i n \phi} e^{i w \theta}\;,
\end{eqnarray}
whose conformal weight $(h,\bar{h})$ is
\begin{eqnarray}\label{eq:hhbar}
    h= \frac14 \left(\frac{n}{R}+ w R\right)^2, \qquad \bar{h}= \frac14 \left(\frac{n}{R}- w R\right)^2\;.
\end{eqnarray}

For generic $R$, there is a $U(1)_m\times U(1)_w$ symmetry, associated with the conserved currents 
\begin{eqnarray}
    J^m= i \frac{R^2}{2\pi} d\phi\;, \qquad J^w= \frac{1}{2\pi} \star\!d\phi\;,
\end{eqnarray}
with $\star^2=-1$ acting on a 1-form. The corresponding topological operators are
\begin{eqnarray}
    \mathcal{L}^m_\beta= e^{i \beta\int \star J^m}= e^{-i \frac{\beta}{2\pi} \int d\theta}\;, \qquad \mathcal{L}^w_\alpha = e^{i \alpha \int \star J^w} = e^{- i \frac{\alpha}{2\pi} \int d \phi}\;,
\end{eqnarray}
where we used the relation between $\theta$ and $\phi$. This means that the operator 
\begin{eqnarray}\label{eq:Vab}
    V_{\frac{\alpha}{2\pi}, \frac{\beta}{2\pi}} =e^{i\frac{\alpha}{2\pi} \phi} e^{i\frac{\beta}{2\pi} \theta}\;,
\end{eqnarray}
is attached to $\mathcal{L}^m_\beta \mathcal{L}^w_\alpha$. The conformal dimension of $V_{\frac{\alpha}{2\pi}, \frac{\beta}{2\pi}}$ is obtained by simply replacing $n,w$ with $\frac{\alpha}{2\pi}, \frac{\beta}{2\pi}$: 
\begin{eqnarray}\label{eq:hhbar2}
    h= \frac{1}{4} \left(\frac{\alpha}{2\pi R}+ \frac{\beta R}{2\pi} \right)^2\;, \qquad \bar{h}= \frac14 \left(\frac{\alpha}{2\pi R}- \frac{\beta R}{2\pi} \right)^2\;.
\end{eqnarray}
Indeed, when $\alpha,\beta \in 2\pi \bZ$, the lines are trivial, and $\eqref{eq:Vab}$ reduces to the local operator \eqref{eq:Vnw}. 
The orbifold branch is obtained from gauging the charge conjugation symmetry $\bZ_2^C$ of the compact boson theory \eqref{eq:compactboson}, where $\bZ_2^C$ acts by $\phi\to -\phi$. 

We now search for local and non-local currents in the circle branch and orbifold branch.

\subsection{Circle Branch: Currents Attached to a Continuous Line}

Let's begin by searching for local holomorphic currents of conformal weights $(h,\bar{h}) = (1,0)$. The anti-holomorphic currents of weight $(0,1)$ can be found in a similar way. There are two obvious currents, $J^m$ and $J^w$. To see whether $V_{n,w}$ can be a local current, we require $(h,\bar{h}) = (1,0)$ in \eqref{eq:hhbar}, i.e.~
\begin{eqnarray}
    \frac14 \left(\frac{n}{R}+ w R\right)^2=1\;, \qquad \frac14 \left(\frac{n}{R}- w R\right)^2=0\;.
\end{eqnarray}
This simplifies to 
\begin{eqnarray}
    n= \pm R,\qquad w= \pm \frac{1}{R}\;,
\end{eqnarray}
where the two $\pm$ signs are correlated. Since $n$ and $w$ are both integers, the only possibility is when $R=1$, i.e.~at the self-dual radius. Hence at $R=1$, there are two additional holomorphic conserved currents, denoted by
\begin{eqnarray}\label{eq:Jpm}
    J^1_+= e^{i\phi +i \theta}, \qquad J^1_- = e^{- i \phi - i \theta}\;,
\end{eqnarray}
where the meaning of the superscript $1$ will be transparent soon. 
It is well-known \cite{Ginsparg:1988ui, Ginsparg:1987eb} that these currents generate an $SU(2)$ chiral algebra. 

To search for non-local holomorphic currents, we set $(h,\bar{h})=(1,0)$ in \eqref{eq:hhbar2}, i.e.~
\begin{eqnarray}
    \frac{1}{4} \left(\frac{\alpha}{2\pi R}+ \frac{\beta R}{2\pi} \right)^2=1, \qquad \frac14 \left(\frac{\alpha}{2\pi R}- \frac{\beta R}{2\pi} \right)^2=0\;.
\end{eqnarray}
This simplifies to 
\begin{eqnarray}
    \alpha= \pm 2\pi R, \qquad \beta= \pm \frac{2\pi}{R} \;,
\end{eqnarray}
where the two $\pm$ signs are correlated. Hence there are two non-local currents 
\begin{eqnarray}\label{eq:Jpm2}
    {J}^R_+= e^{i R \phi + i \theta/R}, \qquad {J}^R_- = e^{-i R \phi - i \theta/R}\;.
\end{eqnarray}
${J}^R_+$ is attached to $\mathcal{L}^m_{2\pi/R}\mathcal{L}^w_{2\pi R}$, and ${J}^R_-$ is attached to $\mathcal{L}^m_{-2\pi/R}\mathcal{L}^w_{-2\pi R}$. They generate continuous non-invertible symmetries. When $R=1$, the line $\mathcal{L}^m_{2\pi/R}\mathcal{L}^w_{2\pi R}$ is trivial, and they reduce to local currents \eqref{eq:Jpm}, which explains the superscript $1$. Next we analyze these currents for other values of $R$.

The case $R= p/q\in \mathbb{Q}$ (for $\gcd(p,q)=1$) was discussed in \cite{Thorngren:2021yso, Fuchs:2007tx}. 
Starting with the $R=1$ CFT, if we simultaneously gauge $\bZ_p^m\times \bZ_q^w\subset U(1)^m\times U(1)^w$, we get the $R=\frac{p}{q}$ CFT. As a consequence, the local current $J^1_\pm$ of the $R=1$ CFT in \eqref{eq:Jpm} induces the non-local current ${J}^R_{\pm}$ of the $R=\frac{p}{q}$ CFT in \eqref{eq:Jpm2}. In other words, the continuous non-invertible symmetry obtained by integrating the non-local conserved currents ${J}^R_\pm$ are \emph{non-intrinsic}---they can be made invertible via discrete gauging.

The case $R\notin \mathbb{Q}$ was less explored. The non-local conserved current is now attached to a $U(1)^m\times U(1)^w$ topological line. Integrating such a current as prescribed in Section \ref{sec:defectconstruction} still gives rise to a (possibly non-compact) topological defect line, hence it generates a continuous non-invertible symmetry. We anticipate this continuous non-invertible symmetry to be \emph{non-intrinsic} because ${J}^R_\pm$ in the radius $R$ theory and $J^1_\pm$ in the $R=1$ theory are related by flat $U(1)$ gauging.\footnote{Another perspective of being non-intrinsic is that infinitesimally close to the $R\in \mathbb{Q}$ theory which is non-intrinsic, as reviewed in Section \ref{sec:compbosonintro}.} We will not discuss the flat $U(1)$ gauging in this paper, but we refer to \cite{hotatlam1,hotatlam2, Choi:2025ebk,Argurio:2024ewp} for a more detailed discussion.

\subsection{Orbifold Branch: Non-Local Currents Attached to Non-Invertible Lines}

We proceed to explore the local and non-local currents in the orbifold branch of $c=1$ CFT. The radius $R$ CFT on the orbifold branch is obtained by gauging the $\bZ_2^C$ charge conjugation symmetry of the radius $R$ CFT on the circle branch.

We quickly summarize the results above for the compact boson on the circle branch of radius $R$, where the holomorphic currents are $J^m, J^w, J^R_{+}, J^R_{-}$. When $R=1$, all currents are local; while when $R>1$, the $J^R_{\pm}$ are non-local. The lines attached to non-local currents are $\mathcal{L}^m_{2\pi/R} \mathcal{L}^w_{2\pi R}$. When $R=p/q\in \mathbb{Q}$ is rational, this line generates $\bZ_{pq}$ symmetry. When $R\notin \mathbb{Q}$ is not rational, the line does not generate any finite subgroup of $U(1)$, hence is a TDL of $U(1)$ symmetry. For simplicity we denote 
\begin{eqnarray}
    G_{R}= 
    \begin{cases}
        \bZ_{pq}, & R=p/q\in \mathbb{Q}\\
        U(1), & R\notin \mathbb{Q}\;.
    \end{cases}
\end{eqnarray}
Under $\bZ_2^C$, these currents transform as 
\begin{eqnarray}
\begin{split}
    J^m\to -J^m, \qquad J^w\to -J^w, \qquad J^R_{\pm} \to J^R_{\mp}\;.
\end{split}
\end{eqnarray}

After gauging, we get the orbifold branch $c=1$ CFT with radius $R$. Since $J^m$ is charged under $\bZ_2^C$ before gauging, gauging $\bZ_2^C$ makes $J^m$ a non-local conserved current attached to a $\bZ_2$ line. The associated continuous non-invertible symmetry is the cosine symmetry as reviewed in Section \ref{sec:introgauging}. The same applies to $J^w$. For the $J^R_{\pm}$, after gauging $\bZ_2^C$, the topological line they attach further becomes a non-invertible line corresponding to the Wilson line of 
\begin{eqnarray}
    \mathrm{Rep}(G_R \rtimes \bZ_2^C)\;.
\end{eqnarray}
The end point carries a two dimensional representation of $G_R \rtimes \bZ_2^C$. The non-local current can be made local via gauging $\mathrm{Rep}(G_R \rtimes \bZ_2^C)$ which maps the theory back to the self-dual point on the circle branch, hence the non-invertible symmetry is non-intrinsic.

One can also discuss the non-local currents in the three exceptional points of the moduli space. We will not study them in this work, for some discussions see \cite{Thorngren:2021yso,Yu:2025iqf}.

\section{Conditionally-Intrinsic Non-Local Conserved Currents in WZW Models}
\label{sec:currentWZW}

We proceed to finding non-local conserved currents in $c>1$ CFTs. The classification of these CFTs is far from understood. However, a well-studied class of such CFTs are the diagonal Wess-Zumino-Witten (WZW) models. In this section, we will find that even the simplest family---$SU(2)_k$ WZW models---have interesting non-local conserved currents, and hence continuous non-invertible symmetries, that go beyond the type discussed in Section \ref{sec:c=1}. Namely, certain non-local currents cannot be rendered local by gauging while preserving the $SU(2)$ chiral algebra.
After an extensive discussion of $SU(2)_k$, we also comment on non-local currents of $G_k$ for various other target manifolds at the end of the section. 

\subsection{Preliminaries of $SU(2)_k$ WZW Models}
\label{sec:su2k}

We begin by collecting the basic data of a  diagonal $SU(2)_k$ WZW model \cite{francesco2012conformal,Ginsparg:1988ui,Eberhardt2019WZW}, with more details appearing in Appendix \ref{app:SU2kWZW}. The central charge is $c=3k/(2+k)$. There are $k+1$ local primary operators $\mathcal{O}_\lambda$, labeled by an integrable weight $\lambda$ of the $\mathfrak{su}(2)$ Lie algebra, i.e.~$0\leq \lambda \leq k$. The conformal weight is $(h_\lambda, h_\lambda)$, with 
\begin{eqnarray}\label{eq:conformalweightmain}
    h_\lambda = \frac{\lambda (\lambda+2)}{4(k+2)}\;.
\end{eqnarray}
Let $\chi_\lambda(\tau)$ be the affine character of $\lambda$. The torus partition function of the CFT is 
\begin{eqnarray}\label{eq:Z}
    \mathcal{Z}(\tau) = \sum_{\lambda=0}^k \chi_\lambda(\tau) \bar{\chi}_\lambda(\bar{\tau})\;.
\end{eqnarray}

For arbitrary $k$, there is an $[SU(2)\times SU(2)]/\bZ_2$ global symmetry, where the two $SU(2)$ factors act on holomorphic and anti-holomorphic fields respectively. The torus partition function indeed counts exactly three $(1,0)$ and $(0,1)$ operators
\begin{eqnarray}
    q^{\frac{c}{24}} \bar{q}^{\frac{c}{24}} \mathcal{Z}(\tau) \supset 3 q + 3 \bar{q}\;,
\end{eqnarray}
where $q=e^{2\pi i \tau}$. 
We denote the $(1,0)$ currents as $J^a$, $a=1,2,3$. They satisfy the $\mathfrak{su}(2)_k$ current algebra
\begin{eqnarray}\label{eq:JJOPE}
    J^a(z) J^b(w) \sim  \frac{k \delta^{ab}}{(z-w)^2} + \sum_{c=1}^{3} \frac{i \epsilon^{abc}}{z-w} J^c(w)\;.
\end{eqnarray}
The conserved charge and topological operator are 
\begin{eqnarray}
    Q^a = \oint \frac{dz}{2\pi i} J^a(z)\;, \qquad U_{\alpha, \vec n}= e^{i \alpha n^a Q^a}\;,
\end{eqnarray}
where $\vec{n}= (n^1, n^2, n^3)$ is a unit vector. 

Apart from the $[SU(2)\times SU(2)]/\bZ_2$ continuous symmetry, it is well-known that there are topological line operators which commute with $J^a$ and $\bar{J}^a$. They are termed \emph{Verlinde lines}. There are $k+1$ such lines, and share the same label $\lambda$ as the local primary operator. We denote the lines by $\mathcal{L}_\lambda$. These lines form an $SU(2)_k$ modular tensor category (MTC). The fusion rules are 
\begin{eqnarray}
    \mathcal{L}_\mu \times \mathcal{L}_\nu= \sum_{\lambda=0}^{k} N_{\mu\nu}^{\lambda} \mathcal{L}_{\lambda}\;,
\end{eqnarray}
where 
\begin{eqnarray}\label{eq:N}
    N_{\mu\nu}^\lambda = 
	\begin{cases}
		1, & |\mu-\nu| \leq \lambda \leq \min(\mu+\nu, 2k-\mu-\nu), \mu+\nu+\lambda=0 \text{ mod }2 \\
		0, & \text{otherwise}\;.
	\end{cases}
\end{eqnarray}
The topological spin of the line $\mathcal{L}_\lambda$ is $e^{ 2\pi i h_\lambda}$, which form the diagonal elements of the modular T-matrix. The modular S-matrix is given by
\begin{eqnarray}\label{eq:modularS}
    S_{\lambda \mu} = \sqrt{\frac{2}{k+2}} \sin\biggl(\pi \frac{(\lambda+1)(\mu+1)}{k+2}\biggr)\;,
\end{eqnarray}
which is related to the fusion coefficient \eqref{eq:N} via the Verlinde formula 
\begin{eqnarray}\label{eq:Verlindeformula}
	N_{\mu\nu}^\lambda = \sum_{\sigma=0}^k \frac{S_{\mu \sigma} S_{\nu \sigma} S^*_{\lambda \sigma}}{S_{0 \sigma}}\;. 
\end{eqnarray}

Note that the Verlinde lines commute with the holomorphic $J^a$ which generates the $SU(2)$ chiral algebra, as well as the anti-holomorphic $\bar{J}^a$ which generates the  $SU(2)$ anti-chiral algebra. This automatically ensures that the Verlinde lines commute with the stress energy tensor $T$ and $\bar{T}$, because these are composites of the $SU(2)$ currents via the Sugawara construction. Hence the Verlinde lines are topological. It would be nice to look for further topological lines that commute with $T$ and $\bar{T}$ but do not commute with $J^a$ and $\bar{J}^a$. As far as the authors know, such lines are not classified. For this reason, we will only look for non-local currents at the end of Verlinde lines.

\subsection{Non-Local Conserved Currents}

\subsubsection{Condition for the Existence of Non-Local Conserved Currents}

We now look for conserved currents of weight $(1,0)$ at the end of a Verlinde line $\mathcal{L}_{\lambda}$. Using the state-operator correspondence, this is equivalent to $(1,0)$ states in the $\mathcal{L}_\lambda$ twisted Hilbert space. Such states/operators are counted by the $\mathcal{L}_\lambda$ twisted partition function, 
\begin{eqnarray}\label{eq:Zlambda}
	\CZ_{\lambda}(\tau)=  \sum_{\nu\sigma} N_{\lambda \nu}^\sigma \chi_\nu (\tau) \bar\chi_\sigma(\bar\tau)\;,
\end{eqnarray}
where $N_{\lambda \nu}^\sigma$ is the fusion coefficient \eqref{eq:N}. This means that primary states in the $\mathcal{L}_\lambda$ twisted sector have conformal weight $(h_\nu, h_\sigma)$ provided $N_{\lambda \nu}^\sigma\neq 0$. There are two cases for the $(1,0)$ current: it is either a primary operator or a descendant operator. 

If the current is a primary operator, we demand 
\begin{eqnarray}
    (h_\nu, h_\sigma) = (1,0) \;.
\end{eqnarray}
Using \eqref{eq:conformalweightmain}, we get $\nu= \sqrt{4k+9}-1$ and $\sigma=0$. 
Since $\sigma=0$, $N_{\lambda \nu}^\sigma\neq 0$ if and only if $\lambda = \nu$, hence the line attached to the primary $(1,0)$ conserved current is attached to $\mathcal{L}_\lambda$ with
\begin{eqnarray}
    \lambda =  \sqrt{4k+9}-1\;.
\end{eqnarray}
Since $\lambda$ is an integer and $0\leq \lambda \leq k$, $4k+9$ must be a perfect square of an odd integer $2n +1$ with $n\leq k/2$. Solving this condition, one gets the allowed values are $k\in \mathbb{K}$ where
\begin{equation}\label{eq:k}
\begin{aligned}
	\mathbb{K}& :=\{ (n+2)(n-1) \;| \;n\in\mathbb{Z}_{\geq 2}\} \\
    &\ =\{4, 10, 18, 28, 40,\dots\;\}\;.
\end{aligned}
\end{equation}

If the current is a descendant operator, the only possibility is that it belongs to the descendant of the primary operator of weight $(h_\nu,h_\sigma)=(0,0)$, implying $\nu=\sigma=0$, and hence $\lambda=0$. In other words, these are local conserved currents discussed in Section \ref{sec:su2k}.

We conclude that
\begin{claim}
	When $k\in\mathbb{K}$, the diagonal $SU(2)_k$ WZW model has at least one holomorphic defect current in the $\CL_\lambda$ defect Hilbert space with $\lambda= 2n= \sqrt{4k+9}-1$. 
\end{claim}

Repeating the same computation for an anti-holomorphic conserved current of weight $(0,1)$, we arrive at the same condition.

From the fusion coefficient \eqref{eq:N}, the topological defect line $\CL_\lambda$ is invertible only when $\lambda=0$ or $k$. The former is uninteresting because it means the current is local. So the defect current is attached to an invertible line only when $k=\lambda$. Using $k= (n+2)(n-1)$ and $\lambda = 2n$, the condition $k=\lambda$ means $k=4$. In this case, the topological line $\mathcal{L}_4$ is a $\bZ_2$ line, hence the corresponding continuous symmetry is the cosine symmetry if we choose the base line $\mathcal{B}$ to be $\mathcal{L}_0+\mathcal{L}_4$. For higher $k$ the defect is non-invertible, and the corresponding continuous symmetries are new continuous non-invertible symmetries.

\subsubsection{$SU(2)$ Representation of Non-Local Currents}
\label{sec:su2rep}

The holomorphic non-local currents should form a representation under the left $SU(2)$ symmetry. Since the $SU(2)$ does not admit projective representations, we only need to discuss the linear representation, which is purely determined by the isospin $j$, or equivalently its dimension $2j+1$. This can be read off from the $\mathcal{L}_\lambda$ twisted partition function \eqref{eq:Zlambda},  with chemical potential $z$ turned on for the Cartan of $SU(2)$ (i.e.~we introduce $U(1)$ lines along space and $\CL_\lambda$ line along time). We denote this partition function by $\CZ_\lambda(z;\tau)$. We review the computation in Appendix \ref{app:Z_twist}. For $k\in \mathbb{K}$. the non-local currents can be identified by extracting the terms linear in $q$ and $\bar{q}$:
\begin{eqnarray}
	q^{\frac{c}{24}} \bar{q}^{\frac{c}{24}} \CZ_\lambda(z;\tau) \supset \sum_{n=-\lambda/2}^{\lambda/2} (y^{n} q + \bar{y}^n \bar{q})\;.
\end{eqnarray}
with $y=e^{2\pi i z}$. We thus learn that there are $\lambda+1$ holomorphic non-local currents which form a single $\lambda+1$ dimensional representation (of isospin $j=\lambda/2$), and similarly for the antiholomorphic currents. 

Another way to see this is to note that in the 3d TQFT construction, the Verlinde line $\lambda$ is obtained from the $SU(2)$ Wilson line with isospin-$\lambda/2$ representation, see Figure \ref{fig:SU2k}. Hence the operator living at the end also carries the same $SU(2)$ representation.

\subsubsection{Continuous Non-Invertible Symmetries}

Given a non-local conserved current attached to a topological line $\mathcal{L}_\lambda$, we can further construct a conserved current living on a topological base line $\mathcal{B}$. As discussed in Section \ref{sec:baseline}, there are multiple choices of base lines, as long as there is a topological junction among $(\mathcal{B}, \mathcal{B}, \mathcal{L}_\lambda)$. For instance, if we demand the base line $\mathcal{B}$ to be a simple line $\mathcal{L}_\beta$ for certain $\beta$, then requiring $N_{\beta\beta}^{2n}=1$ gives 
\begin{eqnarray}\label{eq:nbeta}
    n\leq \beta \leq n^2-2\;.
\end{eqnarray}
Since $n\leq n^2-2$ trivially holds for $n\geq 2$, for every $k=(n+2)(n-1)\in \mathbb{K}$, the base line can always be chosen to be a simple line $\mathcal{L}_\beta$ in the window \eqref{eq:nbeta}. 

One can further take the base line $\mathcal{B}$ to be a non-simple line. For instance, one can take $\mathcal{B}= \mathcal{L}_\alpha + \mathcal{L}_\beta$, as long as there is a topological junction among $(\mathcal{L}_\alpha, \mathcal{L}_\beta, \mathcal{L}_{2n})$, i.e.~$N_{\alpha\beta}^{2n}=1$. For instance, one can always take 
\begin{eqnarray}
    \mathcal{B} = \mathcal{L}_0 + \mathcal{L}_{2n}\;.
\end{eqnarray}
When $n=2$ (i.e.~$k=4$), this precisely gives the cosine symmetry. These different choices in principle lead to different non-invertible symmetries, with additional constraints on dynamics.

Upon integrating the non-local conserved current on the base line, we get plenty of continuous non-invertible symmetries in $SU(2)_k$ WZW models.

\subsection{Intrinsic Non-Locality}

We have found non-local currents in $SU(2)_k$ WZW models when $k\in \mathbb{K}$. We would now like to ask whether these currents can be made local by any topological manipulation of the theory (e.g.~discrete gauging). 

Except for $k=4$, our topological line $\mathcal{L}_\lambda$ always has non-integral quantum dimension, which immediately precludes the possibility of a strong gauging \cite{Choi:2023xjw} that liberates the current. Weak gauging, on the other hand, is much harder to rule out: we would need to list all potential algebra objects that contain $\mathcal L_\lambda$, and check whether they are actually gaugeable or not. As a matter of fact, the full list of topological lines in this WZW model is unknown, so we cannot even make such a list to begin with -- we do not know enough about this CFT to prove or disprove the possibility of a weak gauging.

\begin{table}
  \centering
  \small
  \renewcommand{\arraystretch}{1.1}
  \begin{tabular}{|c|c|c|}
  \hline
    level & partition function & diagram \\ \hline \hline
    $k\ge0$
      & $\displaystyle\sum_{\lambda=0}^k\lvert\chi_\lambda\rvert^2$
      & $A_{k+1}$ \\ [1ex] \hline
    $k=4\rho\ge4$
      & $\displaystyle\sum_{j=0}^{\rho-1}\lvert\chi_{2j}+\chi_{4\rho-2j}\rvert^2
         +2\lvert\chi_{2\rho}\rvert^2$
      & $D_{2\rho+2}$ \\[3ex] \hline
    $k=4\rho-2\ge6$
      & $\displaystyle\sum_{j=0}^{2\rho-1}\lvert\chi_{2j}\rvert^2
         +\lvert\chi_{2\rho-1}\rvert^2
         +\sum_{j=0}^{\rho-2}\bigl(\chi_{2j+1}\bar\chi_{4\rho-2j-3}+\text{c.c.}\bigr)$
      & $D_{2\rho+1}$ \\[4ex]\hline
    $k=10$
      & $|\chi_0+\chi_6|^2+|\chi_3+\chi_7|^2+|\chi_4+\chi_{10}|^2$
      & $E_6$ \\[3ex] \hline
    $k=16$
      & $\displaystyle
         \begin{aligned}
           &|\chi_0+\chi_{16}|^2+|\chi_4+\chi_{12}|^2+|\chi_6+\chi_{10}|^2\\
           &\quad+|\chi_8|^2+[(\chi_2+\chi_{14})\bar\chi_8+\text{c.c.}]
         \end{aligned}$
      & $E_7$ \\ [4ex] \hline
    $k=28$
      & $|\chi_0+\chi_{10}+\chi_{18}+\chi_{28}|^2
         +|\chi_6+\chi_{12}+\chi_{16}+\chi_{22}|^2$
      & $E_8$ \\\hline
  \end{tabular}
  \caption{ADE Classification of $SU(2)\times SU(2)$-invariant modular partition functions.}
  \label{tab:WZW}
\end{table}

This means that we simply don't understand $SU(2)_k$ well enough to establish whether our non-invertible symmetry is intrinsic or not: there might exist \emph{some} topological line that, when added to $\mathcal L$, leads to a gaugeable object that makes $J$ local. One set of topological lines that is well understood is the subset that commutes with the current algebra, i.e., the Verlinde lines reviewed in Section \ref{sec:su2k}. So, with this in mind, the best we can do is to rule out gaugings that preserve the $SU(2)$ chiral algebra. In this subsection we show that, apart from a finite set of $k$'s, our non-local currents cannot be made local while preserving the $SU(2)$ chiral algebra.

\subsubsection{Probe from ADE Classification}

All possible discrete (bosonic) gaugings, compatible with the $SU(2)$ chiral algebra, are enumerated by an ADE classification \cite{Cappelli:1986hf,Cappelli:1987xt,Kato:1987td}. The corresponding modular invariant partition functions are given in Table \ref{tab:WZW}. We now look for a partition function containing the characters 
\begin{eqnarray}
	\CZ_{\text{gauged}} \supset \chi_{\sqrt{4k+ 9} - 1} \bar\chi_{0}\;.
\end{eqnarray}
If it exists, the defect $\mathcal{L}_\lambda$ in the diagonal $SU(2)_k$ can be gauged away while preserving $SU(2)$ chiral algebra. Otherwise, $\mathcal{L}_\lambda$ cannot be gauged away, hence the corresponding continuous non-invertible symmetry is termed \emph{intrinsic} while preserving the $SU(2)$ chiral algebra.

Inspection of Table~\ref{tab:WZW} shows that the possible solutions are the partition functions corresponding to the diagrams $D_3,E_6,E_8$. These cases are all well-known, and the result of gauging the line connected to the non-local current to make it local is also known (see e.g.~\cite{Neupert_2016}):
\begin{enumerate}
    \item $D_3$ corresponds to $k=4$. $SU(2)_4$ has a current at the end of the line $\CL_4$, corresponding to the embedding $SU(2)_4\subset SU(3)_1$.
    \item  $E_6$ corresponds to $k=10$. $SU(2)_{10}$ has a current at the end of the line $\CL_6$, corresponding to the embedding $SU(2)_{10}\subset SO(5)_1$.
    \item  $E_8$ corresponds to $k=28$. $SU(2)_{28}$ has a current at the end of the line $\CL_{10}$, corresponding to the embedding $SU(2)_{28}\subset (G_2)_1$.
\end{enumerate}

In summary, we find that only when $k=4, 10, 28$, the continuous non-invertible  symmetry can be derived from an invertible symmetry by gauging while preserving an $SU(2)$ chiral algebra.
In all other cases, the non-invertible continuous symmetries are intrinsically non-invertible if we preserve the $SU(2)$ chiral algebra, and do not originate from invertible ones. The simplest example is $SU(2)_{18}$ with a current at the end of the line $\CL_8$.

\subsubsection{Probe from 3d TQFT}
\label{sec:probe3dtqft}

The above method uses the ADE classification, which is specific to $SU(2)_k$. To generalize to other groups, it is useful to present an alternative method of probing the intrinsic-ness. One way to determine possible discrete gaugings is to use the 3d TQFT \cite{Komargodski:2020mxz,Elitzur:1989nr}.

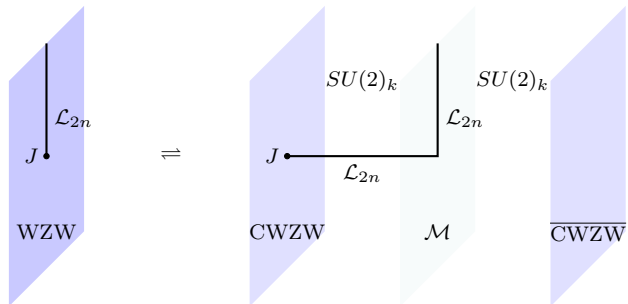
\begin{figure}
	\centering
 {\scriptsize
 \raisebox{-63pt}{\begin{tikzpicture}

			\fill[fill=blue!70, opacity=.3] (1-2,4) -- (1-2,1) -- (0-2,0) -- (0-2,3);
			 \draw (-0.5-1,1) node{$\text{WZW}$};

            \filldraw (-1.5,2) circle (1pt);
            \node[left] at (-1.5,2) {$J$};
            \draw[thick] (-1.5,2) -- (-1.5,3.5); 
             \node[right] at (-1.5, 2.5) {$\mathcal{L}_{2n}$};
	\end{tikzpicture}}
    \qquad 
    \raisebox{-10pt}{\begin{tikzpicture}
        \node[] at (0,2) {$\rightleftharpoons$};
    \end{tikzpicture}}
    \qquad 
	\raisebox{-63pt}{\begin{tikzpicture}
             \fill[fill=blue!40,opacity=.3] (3,4) -- (3,1) -- (2,0) -- (2,3) -- cycle;
             \fill[fill=teal!10,opacity=.3] (3-2,4) -- (3-2,1) -- (2-2,0) -- (2-2,3) -- cycle;
             
			\fill[fill=blue!40, opacity=.3] (1-2,4) -- (1-2,1) -- (0-2,0) -- (0-2,3);
			\draw (-0.5-1,1) node{$\text{CWZW}$};
		    \draw (2.5,1) node{ $\overline{\text{CWZW}}$};

            \node[] at (-0.5,3) {$SU(2)_k$};

            \node[] at (1.5,3) {$SU(2)_k$};

            \node[] at (0.5,1) {$\mathcal{M}$};

            \filldraw (-1.5,2) circle (1pt);
            \node[left] at (-1.5,2) {$J$};
            \draw[thick] (-1.5,2) -- (0.5,2)-- (0.5,3.5); 
            \node[below] at (-0.5, 2) {$\mathcal{L}_{2n}$};
            \node[right] at (0.5, 2.5) {$\mathcal{L}_{2n}$};
	\end{tikzpicture}}
 }
	\caption{
 3d TQFT construction of the diagonal $SU(2)_k$ WZW model and non-local conserved currents for $k=(n+2)(n-1)$. 
	}
	\label{fig:SU2k}
\end{figure}

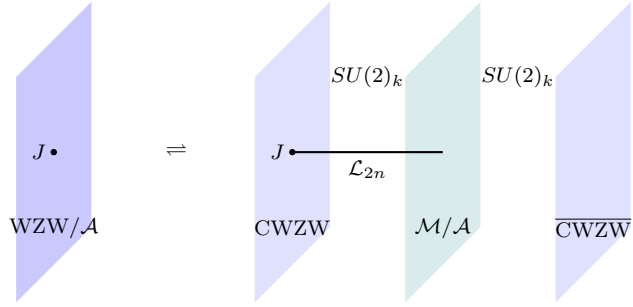
\begin{figure}[t]
	\centering
 {\scriptsize
 \raisebox{-63pt}{\begin{tikzpicture}
             
			\fill[fill=blue!70, opacity=.3] (1-2,4) -- (1-2,1) -- (0-2,0) -- (0-2,3);
			 \draw (-0.5-1,1) node{$\text{WZW}/\mathcal{A}$};

            \filldraw (-1.5,2) circle (1pt);
            \node[left] at (-1.5,2) {$J$};
	\end{tikzpicture}}
    \qquad 
    \raisebox{-10pt}{\begin{tikzpicture}
        \node[] at (0,2) {$\rightleftharpoons$};
    \end{tikzpicture}}
    \qquad 
	\raisebox{-63pt}{\begin{tikzpicture}
             \fill[fill=blue!40,opacity=.3] (3,4) -- (3,1) -- (2,0) -- (2,3) -- cycle;
             \fill[fill=teal!50,opacity=.3] (3-2,4) -- (3-2,1) -- (2-2,0) -- (2-2,3) -- cycle;
             
			\fill[fill=blue!40, opacity=.3] (1-2,4) -- (1-2,1) -- (0-2,0) -- (0-2,3);
			\draw (-0.5-1,1) node{$\text{CWZW}$};
		    \draw (2.5,1) node{ $\overline{\text{CWZW}}$};

            \node[] at (-0.5,3) {$SU(2)_k$};

            \node[] at (1.5,3) {$SU(2)_k$};

            \node[] at (0.5,1) {$\mathcal{M}/\mathcal{A}$};

            \filldraw (-1.5,2) circle (1pt);
            \node[left] at (-1.5,2) {$J$};
            \draw[thick] (-1.5,2) -- (0.5,2); 
            \node[below] at (-0.5, 2) {$\mathcal{L}_{2n}$};
	\end{tikzpicture}}
 }
	\caption{
 3d TQFT construction of diagonal $SU(2)_k$ WZW model gauged by a Frobenius algebra $\mathcal{A}$ and local conserved currents for $k=(n+2)(n-1)$. 
	}
	\label{fig:SU2kgauged}
\end{figure}

The diagonal $SU(2)_k$ WZW model can be expanded into a 3d sandwich, where the bulk is an $SU(2)_k$ Chern-Simons theory;  the left boundary is the chiral boundary supporting a chiral $SU(2)_k$ WZW model; the right boundary is the anti-chiral boundary supporting an anti-chiral $SU(2)_k$ WZW model. In the middle, there is an identity surface defect $\mathcal{M}$, indicating that the anyons on the left and right are trivially connected. The Verlinde line operator lives on this trivial surface defect $\mathcal{M}$ of the 3d TQFT. The current attached to $\mathcal{L}_{2n}$ (where $k=(n+2)(n-1)$) is expanded as the right hand side of Figure \ref{fig:SU2k}, where the line $\mathcal{L}_{2n}$ extends to the bulk, and along $\mathcal{M}$. 

Gauging a Frobenius algebra $\mathcal{A}$ amounts to swapping the identity defect $\mathcal{M}$ for some other defect $\mathcal{M}/\mathcal{A}$ of the 3d TQFT. The defect can be specified by a matrix $W_{\mu\nu}$ with nonnegative integer elements, obeying \cite{Lan:2014uaa}
\begin{eqnarray}\label{eq:WST}
    SW=WS\;,\quad TW=WT\;,
\end{eqnarray}
and also 
\begin{equation}\label{eq:Winequality}
    W_{\mu\nu}W_{\rho\sigma}\leq N_{\mu \rho}^\alpha W_{\alpha \beta}N_{\nu \sigma}^\beta\;.
\end{equation}
Here, $S$ and $T$ are the modular matrices reviewed in Section \ref{sec:su2k}. To see whether the non-local current becomes a local current after gauging, we ask whether the surface defect $\mathcal{M}/\mathcal{A}$ maps $\mathcal{L}_{2n}$ to a trivial line, see Figure \ref{fig:SU2kgauged}. In other words, we examine whether $W_{0,2n}\neq 0$. If so, the non-local current can be made local by gauging. 
Since \eqref{eq:WST} is a linear equation in $W$, one can easily solve it (say, in Mathematica), and examine whether $W_{0,2n}\neq 0$. We find that this is satisfied only when $n=2,3,5$, i.e.~$k=4,10,28$, as expected from the ADE classification. We present the $W$ matrices in Appendix \ref{app:SU2kWZW}.

Note that~\eqref{eq:Winequality} is a stability condition on the topological defect, which essentially requires that the defect is simple. For our purposes, this condition is inconsequential; in the examples we examined, all solutions of~\eqref{eq:WST} automatically satisfy~\eqref{eq:Winequality}.

\subsection{Current Algebra of Non-Local Conserved Current?}

An important feature of the local currents $J^a$ is the existence of a current algebra \eqref{eq:JJOPE}. Specifically, the singular parts of the OPE of two local currents can only include the identity operator and other currents. We now show that non-local currents generically don't obey such a simple OPE. This is to be expected, since one assumption in the derivation of a current algebra is mutual locality of the currents, which is generically not obeyed for non-local currents. We focus on highest-weight states of each $SU(2)$ representation in the following for simplicity. We also only describe the manipulations of topological lines briefly, for a detailed version of these calculations see \cite{Chang:2020imq}. Along the way we compute the general chiral OPE coefficient of our non-local currents. 

Let us start with the OPE of two chiral operators,
\begin{equation}\label{eq:ppope}
    \phi_{\mu}(z) \phi_{\nu}(w) \sim \sum_{\rho} c_{\mu\nu}^{\rho} \phi_{\rho}(w)(z-w)^{h_{\rho}-h_{\mu}-h_{\nu}}+\cdots
\end{equation}
Here $\phi_{\mu}$ is the highest weight state of an $SU(2)_k$ primary of spin $\mu/2$, and is also a chiral primary operator of conformal dimension $(h_\mu,0)$.  $\phi_\mu$ lives at the end of the Verlinde line $\mathcal{L}_\mu$. 
Diagrammatically, \eqref{eq:ppope} means
\begin{eqnarray}\label{eq:opepp}
    \raisebox{-40pt}{
    \begin{tikzpicture}
    \draw[thick] (0,-1) -- (0,1);
    \filldraw (0,-1) circle (1pt);
    \node[below] at (0,-1) {$\phi_{\mu}(z)$};
    \node[left] at (0,1) {$\mathcal{L}_\mu$};
    \end{tikzpicture}
    }
    \raisebox{-40pt}{
    \begin{tikzpicture}
    \draw[thick] (0,-1) -- (0,1);
    \filldraw (0,-1) circle (1pt);
    \node[below] at (0,-1) {$\phi_{\nu}(w)$};
    \node[right] at (0,1) {$\mathcal{L}_\nu$};
    \end{tikzpicture}
    }
    =\sum_{\rho}\frac{c_{\mu\nu}^{\rho}}{(z-w)^{h_\mu+ h_\nu-h_\rho}}
    \raisebox{-40pt}{
    \begin{tikzpicture}
    \draw[thick] (0,0.5) -- (-1,1);
    \draw[thick] (0,0.5) -- (1,1);
    \draw[thick] (0,-1) -- (0,0.5);
    \node[right] at (0,0) {$\mathcal{L}_\rho$};
    \node[right] at (1,1) {$\mathcal{L}_\nu$};
    \node[left] at (-1,1) {$\mathcal{L}_\mu$};
    \filldraw (0,-1) circle (1pt);
    \node[below] at (0,-1) {$\phi_{\rho}(w)$};
    
    \end{tikzpicture}
    }
\end{eqnarray}
Since the topological junctions between the Verlinde lines are all one dimensional, we suppress their labels for simplicity. The coefficient $c_{\mu\nu}^{\rho}$ is the fusion coefficient between the chiral operators \cite{Zamolodchikov:1986bd,Recknagel:2013uja}.

We can now read off correlators.  Using the OPE we find for two-point functions of chiral operators of spin $\mu/2$ \begin{eqnarray}
\raisebox{-20pt}{
    \begin{tikzpicture}
    \draw[thick] (-1,0) -- (1,0);
    \filldraw (-1,0) circle (1pt);
    \filldraw (1,0) circle (1pt);
    \node[above] at (0,0) {$\mathcal{L}_\mu$};
    \node[below] at (-1,0) {$\phi_{\mu}(z)$};
    \node[below] at (1,0) {$\phi_{\mu}(w)$};
    \end{tikzpicture}}
    = \sum_{\rho} 
    \frac{c_{\mu\mu}^\rho}{(z-w)^{2h_\mu - h_\rho}}
    \raisebox{-40pt}{
    \begin{tikzpicture}
    \draw[thick] (0,-0.5) -- (0,0.5);
    \draw[thick] (0,1) circle (15pt);
    \node[left] at (0,0.1) {$\mathcal{L}_\rho$};
    \node[left] at (0.5,1) {$\mathcal{L}_{\mu}$};
    \node[below] at (0,-0.5) {$\phi_{\mu}(w)$};
    \filldraw (0,-0.5) circle (1pt);
    \end{tikzpicture}
    }
\end{eqnarray}
The no-tadpole condition \cite{Chang:2018iay} sets $\rho=0$, hence
\begin{equation}
    \langle \phi_\mu(w)\phi_\mu(w) \rangle = \frac{c_{\mu\mu}^0 d_\mu}{(z-w)^{2h_\mu}}\;,
\end{equation}
where $d_\mu$ is the quantum dimension of the $\mathcal{L}_\mu$.

Let's move on to chiral 3-point functions. We want to compute 
\begin{equation}
\begin{tikzpicture}[baseline={(current bounding box.center)}]
\def\L{2.5}
\fill (0, 0) circle (1pt) node[above left] {$\phi_{\rho}(z_3)$};
\fill (\L, 0) circle (1pt) node[above right] {$\phi_{\nu}(z_2)$};
\fill ({\L/2}, {-sqrt(3)*\L/2}) circle (1pt) node[below] {$\phi_{\mu}(z_1)$};

\draw[thick] (0,0) -- ({1.5*\L/3}, {-sqrt(3)*\L/6});
\draw[thick] (\L, 0) -- ({1.5*\L/3}, {-sqrt(3)*\L/6});
\draw[thick] ({\L/2}, {-sqrt(3)*\L/2}) -- ({1.5*\L/3}, {-sqrt(3)*\L/6});

\node[anchor=north east] at ({1.5*\L/6}, {-sqrt(3)*\L/12}) {$\mathcal{L}_{\rho}$};
\node[anchor=north west] at ({1.5*\L/2}, {-sqrt(3)*\L/12}) {$\mathcal{L}_\nu$};
\node[anchor=west] at ({.5*\L}, {-2*sqrt(3)*\L/6}) {$\mathcal{L}_\mu$};

\end{tikzpicture}
\end{equation}
By first using the OPE between $\phi_\mu$ and $\phi_\nu$, and then with $\phi_\rho$, we find that the 3-point function is\footnote{This OPE only produces the specific limit where $z_1\to z_2$ and $z_3\to z_2$. To obtain a result independent of this particular limit, we use conformal invariance to fix the $z$ dependence. }
\begin{equation}\label{eq:chiral_OPE}
    \langle \phi_\mu(z_1)\phi_\nu(z_2)\phi_\rho(z_3) \rangle =  \frac{c_{\mu\nu}^\rho c_{\rho\rho}^0 \sqrt{d_\mu d_\nu d_\rho}}{z_{12}^{h_{\mu}+h_\nu-h_\rho}z_{23}^{h_{\nu}+ h_\rho-h_\mu}z_{13}^{h_\mu+h_\rho-h_\nu}}\;.
\end{equation}

Next we can also discuss non-chiral operators. For nonchiral (diagonal) operators we find the usual OPE
\begin{equation}
    \langle \phi_\mu(z_1,\bar z_1)\phi_\nu(z_2,\bar z_2)\phi_\rho(z_3,\bar z_3) \rangle =
    \frac{\CC_{\mu\nu\rho}}{|z_{12}|^{2(h_{\mu}+h_\nu-h_\rho)}|z_{23}|^{2(h_{\nu}+ h_\rho-h_\mu)}|z_{13}|^{2(h_\mu+h_\rho-h_\nu)}}\;.
\end{equation}
we can relate this to the OPE of non-chiral operators, using the manipulations of \cite{Chang:2020imq}. We find
\begin{eqnarray}
    \CC_{\mu\nu\rho}=(c_{\mu\nu}^\rho)^2(c_{\rho\rho}^0)^2\sqrt{d_\mu d_\nu d_\rho}\;.
\end{eqnarray}
In the normalization where $c_{\rho\rho}^0=1$ for any $\rho$, the relation is simplified, and we can isolate $c_{\mu\nu}^\rho$:
\begin{eqnarray}\label{eq:result_for_c}
    (c_{\mu\nu}^\rho)^2=\frac{\CC_{\mu\nu\rho}}{\sqrt{d_\mu d_\nu d_\rho}}\;.
\end{eqnarray}
The OPE coefficients $\CC_{\mu\nu\rho}$ are known, as we review in Appendix \ref{app:SU2_OPE}. 
Plugging this result into \eqref{eq:result_for_c} gives the chiral OPE coefficient $c_{\mu\nu}^{\rho}$. 

We can now discuss the (non-)existence of a current algebra. For $k\in\mathbb{K}$ there is a non-local primary operator $\phi_{2n}$ which is the non-local conserved current $J$, and a current algebra means that only $c_{2n ~2n}^{0}$ and $c_{2n ~2n}^{2n}$ are nonzero. Instead we find that
\begin{eqnarray}
    c_{2n ~2n}^{\mu}= 0  \;\; \text{iff}\;\;N_{2n ~2n}^\mu= 0\;.
\end{eqnarray}
In particular there is a current algebra only for $k=4$ (where the line $\mathcal{L}_{4}$ is invertible). In fact, the current algebra for $k=4$ is a $\bZ_2$ graded algebra.

We discuss the two special cases $k=10,28$, where the current is attached to a line which can be gauged while preserving the $SU(2)$ chiral algebra.
Although the line is gaugeable and the current can be freed, there is no current algebra, and additional operators which are not currents appear in the $JJ$ OPE. However, these additional operators are all projected out of the spectrum when performing the gauging to free $J$, and so once $J$ is free it does participate in a current algebra. These cases show that although our non-local currents don't obviously participate in a current algebra, it is possible that there exists some gauging which frees them and leads to a current algebra for the freed currents.

\subsection{Action of Conserved Charges}

Having justified the existence of continuous non-invertible symmetries in $SU(2)_k$ theories, we proceed to study their physical properties---their action on local operators. In this section we focus on the action of the conserved charge \eqref{eq:charge} on local operators as a preliminary step. 

It is convenient to use the 3d TQFT description of various operators. First, the local primary operator $\Phi_\lambda$ of the diagonal $SU(2)_k$ WZW model is obtained from compactification of a line segment in $SU(2)_k$ Chern-Simons, i.e.~$\Phi_\mu= \phi_\mu \bar{\phi}_\mu$, connected through a Wilson line $\mathcal{L}_{\mu}$ in the $SU(2)_k$ TQFT. The conserved charge $Q$ follows from the definition \eqref{eq:charge}. Again we focus on $J$ being the highest-weight state of $SU(2)$, so we can suppress the additional $SU(2)$ index of $J$. Consider the action of $Q$ on a local primary $\phi_\mu$:
\begin{equation}\label{eq:one_current}
\oint\frac{dz}{2\pi i}\raisebox{-9ex}{
    {\scriptsize \begin{tikzpicture}
             \fill[fill=blue!40,opacity=.3] (3,4) -- (3,1) -- (1.5,0) -- (1.5,3) -- cycle;

             \fill[fill=teal!10,opacity=.3] (3-2,4) -- (3-2,1) -- (1.5-2,0) -- (1.5-2,3) -- cycle;

			\fill[fill=blue!40, opacity=.3] (-1,4) -- (-1,1) -- (-2.5,0) -- (-2.5,3);

            \filldraw (-1.75,2) circle (1pt);
            \node[above] at (-1.75,2) {$\phi_\mu$};
            \draw[thick, dashed] (-1.75,2) ellipse [x radius=0.4, y radius=1];

            \draw[thick] (0.25,2) ellipse [x radius=0.4, y radius=1];

            \draw[thick] (0.25,3)-- (-1.75,3);
            
            \filldraw[red] (-1.75,3) circle (1pt);
            \node[red, above] at (-1.75,3) {$J(z)$};
            \node[above left] at (-0.1,2.5) {$\mathcal{B}$};
            
            \filldraw (2.25,2) circle (1pt);

            \filldraw (0.25,2) circle (1pt);
            
            \node[above] at (2.25,2) {$\bar{\phi}_\mu$};
            
            \draw[thick] (-1.75,2) -- (2.25,2); 

            \node[ above] at (1.5,2) {$\mathcal{L}_\mu$};

            \node[above] at (-1,3) {$\mathcal{L}_{2n}$};
	\end{tikzpicture}}}
\end{equation}
where the dashed line is the integration contour. The loop in the middle is the base loop $\mathcal{B}$, which has a topological junction with $\mathcal{L}_{2n}$. 

To evaluate this expression, we first take the OPE between $J(z)$ and $\phi_\mu(w)$ on the left boundary, using \eqref{eq:opepp}. This gives
\begin{equation}
    \oint\frac{dz}{2\pi i}\sum_{\mu'}\frac{c_{2n ~\mu}^{\mu'}}{z^{1+h_{\mu}-h_{\mu'}}}
     \raisebox{-9ex}{
    {\scriptsize \begin{tikzpicture}
             \fill[fill=blue!40,opacity=.3] (3,4) -- (3,1) -- (1.5,0) -- (1.5,3) -- cycle;

             \fill[fill=teal!10,opacity=.3] (3-2,4) -- (3-2,1) -- (1.5-2,0) -- (1.5-2,3) -- cycle;
             
			\fill[fill=blue!40, opacity=.3] (-1,4) -- (-1,1) -- (-2.5,0) -- (-2.5,3);

            \filldraw (-1.75,2) circle (1pt);
            \node[left] at (-1.75,2) {$\phi_{\mu'}$};

            \draw[thick] (0.25,2) ellipse [x radius=0.4, y radius=1];

            \draw[thick] (0.25,3).. controls (-0.75,3)  .. (-0.75,2);
 
            \node[above left] at (1.1,2.5) {$\mathcal{B}$};
            
            \filldraw (2.25,2) circle (1pt);

            \filldraw (0.25,2) circle (1pt);
            
            \node[right] at (2.25,2) {$\bar{\phi}_\mu$};
            
            \draw[thick] (-1.75,2) -- (2.25,2); 

            \node[ above] at (1.5,2) {$\mathcal{L}_\mu$};

            \node[above] at (-0.5,3) {$\mathcal{L}_{2n}$};

            \node[below] at (-1,2) {$\mathcal{L}_{\mu'}$};
	\end{tikzpicture}}}
\end{equation}
Here we are summing over all possible $\lambda'$ such that $\CL_{\mu'}\in\CL_{2n}\times \CL_{\mu}$.
Now, when we contract the $\mathcal{B}$ circle, we must end up with a local operator of conformal dimension $(h_\mu',h_\mu)$. But the only such local operator in the diagonal $SU(2)_k$ WZW model has $\mu=\mu'$.  So a nontrivial result is available only if $\mathcal{L}_{2n}\times \mathcal{L}_\mu\ni\mathcal{L}_\mu$, in which case we find
\begin{equation}
    \oint\frac{dz}{2\pi i}\frac{c_{2n ~\mu}^{\mu}}{z}
     \raisebox{-9ex}{
    {\scriptsize \begin{tikzpicture}
             \fill[fill=blue!40,opacity=.3] (3,4) -- (3,1) -- (1.5,0) -- (1.5,3) -- cycle;

             \fill[fill=teal!10,opacity=.3] (3-2,4) -- (3-2,1) -- (1.5-2,0) -- (1.5-2,3) -- cycle;

			\fill[fill=blue!40, opacity=.3] (-1,4) -- (-1,1) -- (-2.5,0) -- (-2.5,3);

            \filldraw (-1.75,2) circle (1pt);
            \node[left] at (-1.75,2) {$\phi_{\mu}$};

            \draw[thick] (0.25,2) ellipse [x radius=0.4, y radius=1];

            \draw[thick] (0.25,3).. controls (-0.75,3)  .. (-0.75,2);

            \node[above left] at (1.1,2.5) {$\mathcal{B}$};
            
            \filldraw (2.25,2) circle (1pt);

            \filldraw (0.25,2) circle (1pt);
            
            \node[right] at (2.25,2) {$\bar{\phi}_\mu$};
            
            \draw[thick] (-1.75,2) -- (2.25,2); 

            \node[ above] at (1.5,2) {$\mathcal{L}_\mu$};

            \node[above] at (-0.5,3) {$\mathcal{L}_{2n}$};

            \node[below] at (-1,2) {$\mathcal{L}_{\mu}$};
	\end{tikzpicture}}}
\end{equation}
We can now perform the $z$ integral trivially to obtain $c_{2n ~ \mu}^{\mu}$ times the evaluation of the figure above. Using the standard diagrammatic calculation of the topological lines (see for instance \cite{Aasen:2020jwb}), one can evaluate the figure. For instance, when the base line $\mathcal{B}$ is chosen to be the simple line $\mathcal{L}_\beta$ with $\beta$ in the range \eqref{eq:nbeta}, the result is 
\begin{eqnarray}\label{eq:action_of_Q}
    c_{2n ~\mu}^{\mu} \sum_{\eta} \sqrt{\frac{d_\eta^2 d_{2n}}{d_{\mu}^2}} (F^{\mu \eta \beta}_{2n})_{\beta \mu} 
    \raisebox{-9ex}{
    {\scriptsize \begin{tikzpicture}
             \fill[fill=blue!40,opacity=.3] (3,4) -- (3,1) -- (1.5,0) -- (1.5,3) -- cycle;

             \fill[fill=teal!10,opacity=.3] (3-2,4) -- (3-2,1) -- (1.5-2,0) -- (1.5-2,3) -- cycle;
             
			\fill[fill=blue!40, opacity=.3] (-1,4) -- (-1,1) -- (-2.5,0) -- (-2.5,3);

            \filldraw (-1.75,2) circle (1pt);
            \node[left] at (-1.75,2) {$\phi_\mu$};
            
            \filldraw (2.25,2) circle (1pt);

            \filldraw (0.25,2) circle (1pt);
            
            \node[right] at (2.25,2) {$\bar{\phi}_\mu$};
            
            \draw[thick] (-1.75,2) -- (2.25,2); 

            \node[ above] at (1.5,2) {$\mathcal{L}_\mu$};

	\end{tikzpicture}}}
\end{eqnarray}

This shows in particular that the non-invertible symmetry we have been discussing is non-trivial. It would have been logically possible that our defect $\mathcal N_\alpha$ was actually identical to the baseline $\mathcal B$, and that the higher orders $\mathcal N_\alpha\sim \mathcal B+i\alpha Q+\mathcal O(\alpha^2)$ did not act on anything. This is certainly topological, but entirely useless; it is reassuring to see that this is not the case.

We have only discussed the action of the charge $Q$, and not the full defect $\mathcal N_\alpha$. We leave a full discussion of the action of $\mathcal N_\alpha$ on the various operators of the theory to future work. For the time being, we comment that, in the invertible case, the action of $Q$ is enough to determine the Ward identities of the symmetry, and one does not need to know the full action of $U_\alpha$. So it is conceivable that~\eqref{eq:action_of_Q} is sufficient in order to establish the constraints our symmetry imposes on correlation functions, a problem that we also relegate to the future.

We finish this discussion by stressing that~\eqref{eq:action_of_Q} gives the action of $Q$ on a local operator, in the $SU(2)_k$ theory. In other CFTs, a similar formula holds, but it depends on the choice of baseline. More importantly, in some cases a choice of baseline does not lead to a non-zero charge: such is the case, for example, in the cosine symmetry, where the first non-trivial term in the $\alpha$ expansion of $\mathcal N_\alpha$ occurs at order $\alpha^2$. In cases like these, one should replace the initial configuration~\eqref{eq:one_current} by one with more currents, or consider an alternative base line.

\subsection{Other WZW Models}

To show that the existence of these non-local currents is not special to $SU(2)_k$ WZW models, we briefly discuss here more general examples. We will not provide a comprehensive list of the non-local currents as in the $SU(2)_k$ case, and instead only provide some additional examples where they exist and obey some similar properties. 

A $G_k$ WZW model has central charge $c=\frac{k\dim G}{k+g}$ with $g$ the dual Coxeter number. Its primaries are labeled by integrable highest weights $\lambda=(\lambda_1,...,\lambda_{r})$, where $r=\operatorname{rank}(G)$, subject to 
\begin{eqnarray}
    \sum_{i=1}^r a_i^\vee \lambda_i\leq k\;,
\end{eqnarray}
where $a_i^\vee$ are the comarks of $\mathfrak g$. The corresponding holomorphic dimension is
\begin{equation}
    h_\lambda=C(\lambda)/(2(k+g))\;,
\end{equation}
where $C(\lambda)=\sum_{ij}\lambda_i(\lambda_j+2)F_{ij}$ is the quadratic Casimir, with $F_{ij}=(\omega_i,\omega_j)$ the scalar product in the basis of fundamental weights. 

The operators that live at the end of a Verlinde line $\mathcal L_{\lambda}$ have weights $(h_\mu,h_\nu)\mod1$, for any $\mu,\nu$ such that $N^\nu_{\lambda\mu}\neq0$. Since we are after a holomorphic current $(1,0)$, and thanks to unitarity, we can restrict our attention to primaries $(h_\lambda,0)$, which automatically satisfy the fusion rules constraint. In other words, our task is to find an integrable representation $\lambda$ that satisfies
\begin{equation}\label{eq:current_eq}
h_{\lambda}=1\;.
\end{equation}

It turns out that it is simple to find families of solutions to this equation: fix some $\lambda$, and choose $k$ such that $h_\lambda=1$, namely $k=\frac12C(\lambda)-g$ (as long as this is a non-negative integer). Here we give some specific examples:
\begin{itemize}
    \item For $G=SU(N)$, one can choose $\lambda=(nN,0,\dots,0)$ and $k=\frac{1}{2}(n^2 (N-1) N+n (N-1)N-2N)$. Alternatively one can choose $\lambda=(0,n,0,\dots,0,2n)$ and  $k=3n^2+(2N-3)n-N$.
    \item For $G=Spin(N)$ one can choose $\lambda=(n+1,0,\dots,0)$ and $k=\frac{1}{2} (n^2+(n-1)N+3)$.
\end{itemize}
It is clear that such examples are ubiquitous in general WZW models, and are not specific to $SU(2)_k$. We remark that, unlike in the $SU(2)_k$ discussed earlier, the $SU(N)_k$ examples are not real representations, and therefore the line is not self-dual. Thus, one should keep track of its orientation when constructing the corresponding defects $\mathcal N_\alpha$. Furthermore, it is not clear whether a baseline satisfying $B\times\mathcal L\supset\mathcal B$ always exists; we checked by hand for multiple values of $N,n$, and we always found at least one simple line that does the job, but we don't have a general argument that it always exists.

As in the $SU(2)$ case, one can once again ask whether it is always possible to make such currents local while preserving $G$. We checked explicitly for the simple case of $G=SU(3)$ and found examples where this is not possible. Explicitly, the simplest such cases are $\lambda=(3,3)$ and $k=12$ and $\lambda=(6,0)$ and $k=15$.\footnote{An intuitive argument to show that these currents generically cannot be made local while preserving $G$ is that the condition $h_\lambda=1$ is much weaker than modular invariance, and therefore we expect many more non-local currents than available gaugings.}

\section{Continuous Non-Invertible Symmetries in $\mathcal{M}_m \times \overline{\mathcal{M}_m}$ and Defect Conformal Manifolds in $\mathcal{M}_m$}\label{sec:phantom_syms}

We now look for other simple examples of non-local currents. As we already mentioned in the beginning of Section \ref{sec:c=1} and showed in Appendix~\ref{app:min_mod}, a single minimal model does not host local or non-local currents. How about products of minimal models $\mathcal{M}_m \otimes \overline{\mathcal{M}_m}$? Indeed, by taking $m=3$ (i.e.~Ising $\times ~\overline{\text{Ising}}$), we find a theory with $c=1$, and which has several non-invertible symmetries as discussed in Section \ref{sec:c=1}. 
In this section we generalize this construction to $m>3$ and discuss some applications of these symmetries. Our notation for the minimal models appears in Appendix \ref{app:min_mod}. We will focus on unitary minimal models for simplicity.

\subsection{Non-Local Currents in Products of Minimal Models}
\label{sec:non-localcurrentminmodels}

We focus on the CFT defined by a product of a unitary minimal model $\mathcal{M}_m$ with its orientation reversal $\overline{\mathcal{M}_m}$, i.e.~
\begin{eqnarray}
    \mathcal{M}_m\otimes \overline{\mathcal{M}_m}\;.
\end{eqnarray}
The properties of $\mathcal{M}_m$ are reviewed in Appendix \ref{app:min_mod}. In $\mathcal{M}_m$, the primary chiral operator attached to $\mathcal{L}_{r,s}$ has conformal weight 
\begin{eqnarray}\label{eq:hrs}
    h_{r,s} = \frac{((m+1)r - ms)^2 -1}{4m(m+1)}\;,
\end{eqnarray}
where $1\leq r\leq m-1, 1\leq s\leq m$. 
Since $h_{r,s}= h_{m-r, m+1-s}$, we only consider the reduced parameter regime 
\begin{eqnarray}\label{eq:range}
    \begin{split}
        \text{odd }m: & \quad 1\leq r\leq \frac{m-1}{2}, 1\leq s\leq m\;,\\
        \text{even }m: & \quad 1\leq r\leq m-1, 1\leq s\leq \frac{m}{2}\;.
    \end{split}
\end{eqnarray}
The Verlinde lines are denoted as $\mathcal{L}_{r,s}$.

\begin{table}
    \centering
    \begin{tabular}{|c|c|}
    \hline
        $m$ & $[(r_1,s_1)_{h_{r_1,s_1}}, (r_2, s_2)_{h_{r_2,s_2}}]$\\
        \hline\hline
        3 & $[(1,3)_{\frac12}, (1,3)_{\frac12}]$ \\
        \hline
        8 & $[(3,2)_{\frac{5}{12}}, (5,4)_{\frac{7}{12}}]$\\
        \hline
        10 & $[(1,2)_{\frac{2}{11}}, (1,3)_{\frac{9}{11}}]$\\
        \hline
        11 & $[(2,1)_{\frac{7}{22}}, (3,5)_{\frac{15}{22}}]$\\
        \hline
        13 & $[(4,3)_{\frac{36}{91}}, (5,7)_{\frac{55}{91}}]$\\
        \hline
        15 & $[(1,3)_{\frac{7}{8}}, (4,5)_{\frac{1}{8}}]$\\
        \hline
        16 & $[(3,2)_{\frac{45}{136}}, (5,7)_{\frac{91}{136}}]$\\
        \hline
        18 & $[(5,4)_{\frac{22}{57}}, (7,9)_{\frac{35}{57}}]$\\
        \hline
        20 & $[(9,8)_{\frac{1}{2}}, (9,8)_{\frac{1}{2}}]$\\
        \hline
        21 & $[(1,3)_{\frac{10}{11}}, (8,9)_{\frac{1}{11}}], [(4,3)_{\frac{26}{77}}, (7,9)_{\frac{51}{77}}]$\\
        \hline
        23 & $[(3,5)_{\frac{77}{92}}, (4,5)_{\frac{15}{92}}], [(6,5)_{\frac{35}{92}}, (9,11)_{\frac{57}{92}}]$\\
        \hline
    \end{tabular}
    \caption{Minimal model $\mathcal{M}_m\otimes \overline{\mathcal{M}_m}$ with non-local currents at the end of Verlinde line $\mathcal{L}_{r_1,s_1}\otimes \bar{\mathcal{L}}_{r_2,s_2}$. }
    \label{tab:minsol}
\end{table}

Now we look for non-local currents in $\mathcal{M}_m\otimes \overline{\mathcal{M}_m}$. Following the same line of argument as in previous sections, it is enough to look for two pairs $(r_1,s_1)$ and $(r_2,s_2)$ such that 
\begin{eqnarray}\label{eq:hrs2}
    h_{r_1,s_1}+h_{r_2,s_2}=1
\end{eqnarray}
to ensure a $(1,0)$ operator living at the end of $\mathcal{L}^{L}_{r_1,s_1}\otimes \mathcal{L}^{R}_{r_2,s_2}$.\footnote{We emphasize that $\mathcal{L}^L_{r_1,s_1}$ and $\mathcal{L}^R_{r_2,s_2}$ act on different theories, to be distinguished from two lines acting in the same theory and hence can be simplified by using the fusion rule. } Here we use the superscripts $L$ and $R$ to distinguish the lines from the two copies. Using \eqref{eq:hrs}, the condition \eqref{eq:hrs2} amounts to
\begin{eqnarray}\label{eq:AB}
    A^2 + B^2 = (2m+1)^2 + 1\;,
\end{eqnarray}
with 
\begin{eqnarray}\label{eq:ABrs}
    A= (m+1)r - ms\;, \qquad B= (m+1)r' - ms'\;.
\end{eqnarray}
Using Bezout's lemma, one can show that for $1\leq A,B\leq 2m+1$, a solution to \eqref{eq:ABrs} exists for $r,s$ in the appropriate range as long as 
\begin{eqnarray}\label{eq:Aconstr}
    A\neq 0\mod m\;,\quad A\neq 0\mod m+1\;,\quad A\neq 2m+1\;,
\end{eqnarray}
and similarly for $B$. Thus a solution exists for $m$ such that a solution to \eqref{eq:AB} exists where $A,B$ obey \eqref{eq:Aconstr}.
We search for solutions for small $m$ in Mathematica, and enumerate them up to permutations\footnote{For instance, we find the solution $[(3,2)_{\frac{5}{12}}, (5,4)_{\frac{7}{12}}]$ for $m=8$, which implies that $[(5,4)_{\frac{7}{12}}, (3,2)_{\frac{5}{12}}]$ is also a solution by permutation. } in Table \ref{tab:minsol}.

Performing a similar computation to the one done in Section \ref{sec:su2rep} for $SU(2)_k$, we computed the $q$ expansion of the twisted partition function and confirmed that there is only a single current living at the end of the topological lines, as opposed to multiple currents as in the $SU(2)_k$ examples.

Since the central charge of $\mathcal{M}_m\otimes \overline{\mathcal{M}_m}$ is greater than 1, the classification of its topological operators is not available yet. However, if we focus on the topological operators commuting with two sets of Virasoro symmetries $Vir^L$ and $Vir^R$ from each copy of the minimal models, then their topological lines are simply $\mathcal{L}_{a_1,b_1}^L \otimes \mathcal{L}_{a_2,b_2}^R$. We therefore can ask if the line attached to the current $\mathcal{L}_{r_1,s_1}^L\otimes \mathcal{L}_{r_2,s_2}^R$ can be gauged away while preserving $Vir^L \otimes {Vir}^R$. This can be achieved using the same method as described in Section \ref{sec:probe3dtqft}. We do not solve this problem in this work.\footnote{The main obstruction is that the $S$ and $T$ matrices for the doubled theory is too large, hence solving them requires more advanced numerical treatment. }

\subsection{Defect Conformal Manifolds of $\mathcal{M}_m$ from Non-Local Currents in $\mathcal{M}_m \otimes \overline{\mathcal{M}_m}$}

One interesting application of the continuous non-invertible symmetry, and hence the non-local conserved current, in the double minimal model $\mathcal{M}_m \otimes \overline{\mathcal{M}_m}$ is that it implies the existence of a defect conformal manifold in a single minimal model \cite{Oshikawa:1996dj, Antinucci:2025uvj}.

It is well-known that a single minimal model is rigid in the sense that it does not belong to any conformal manifold. Moreover, it also does not have continuous families of topological defects---i.e.~continuous invertible or non-invertible symmetries.   Surprisingly, they often host continuous families of conformal defects, which form defect conformal manifolds (DCM). For instance the case $m=3$, i.e.~the Ising model, was discussed in \cite{Oshikawa_1997}. For higher $m$, however, the defect conformal manifolds have not been fully explored. For recent explorations of defect conformal manifolds in various models see \cite{Antinucci:2025uvj,Copetti:2025sym,Oshikawa_1997,Karch:2018uft}.

A defect conformal manifold can be obtained using a pinning field construction. We start with a conformal operator (which will be denoted as the base line), and deform it by an exactly marginal operator. In general it is difficult to prove that a deformation is exactly marginal.
However, as elucidated in \cite{Antinucci:2025uvj,Copetti:2025sym}, in certain situations, we are able to ensure the exact marginality by using a (non-invertible) symmetry based argument. 
The idea is to use the continuous non-invertible symmetry explored in this paper. We start with the trivial line in $\mathcal{M}_m$. Upon folding it yields a seed conformal boundary state $|0\rangle$ in $\mathcal{M}_m \otimes \overline{\mathcal{M}_m}$. One can generate a continuous family of conformal boundary conditions by fusing the seed state $|0\rangle$ with a continuous non-invertible symmetry found in Section \ref{sec:non-localcurrentminmodels}. This gives rise to a boundary conformal manifold (BCM). See recent exploration of BCM in \cite{Choi:2025ebk,Antinucci:2025uvj,Copetti:2025sym,Oshikawa_1997,Karch:2018uft,Herzog:2023dop}.

We next unfold $\mathcal{M}_m \otimes \overline{\mathcal{M}_m}$ with a conformal boundary labeled by a continuous parameter. Suppose the BCM of $\mathcal{M}_m \otimes \overline{\mathcal{M}_m}$ originates from a deformation by a non-local current
\begin{eqnarray}
    J_{1,0} = \psi^L_{h_{r_1,s_1},0} \xi^R_{h_{r_1,s_2},0}\;,
\end{eqnarray}
where $\psi^L_{h_{r_1,s_1},0}$ and $\xi^R_{h_{r_1,s_2},0}$ are non-local primary operators attached to topological lines $\mathcal{L}^L_{r_1,s_1}$ and $\mathcal{L}^R_{r_2,s_2}$ in $\mathcal{M}_m$ and $\overline{\mathcal{M}_m}$ respectively. 
Since in Section \ref{sec:defectconstruction} we already showed that $J_{1,0}$ is an exactly marginal operator, then the operator
\begin{eqnarray}
    \varepsilon_{h_{r_1,s_1}, h_{r_2,s_2}}= \psi_{h_{r_1,s_1},0} \bar{\xi}_{0, h_{r_1,s_2}}
\end{eqnarray}
obtained upon unfolding is also exactly marginal. Deforming the base line by $\varepsilon_{h_{r_1,s_1},h_{r_2,s_2}}$ gives rise to a defect conformal manifold.

So far, we demonstrated, in brief, the existence of defect conformal manifolds using continuous non-invertible symmetries, but without studying the properties of the DCM in detail. The defect conformal manifolds enjoy rich and interesting structures, such as the dependence of the choice of base line,  the $g$ function and reflection and transmission coefficients. We leave the study of these properties to the future.

\section{Comments on Higher Dimensions}\label{sec:higher_d}

We now make some general comments about how our construction can be extended to higher-dimensional examples.

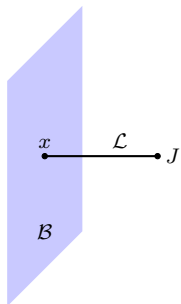
\begin{figure}
	\centering
 {\scriptsize
 \raisebox{-63pt}{\begin{tikzpicture}

			\fill[fill=blue!70, opacity=.3] (1-2,4) -- (1-2,1) -- (0-2,0) -- (0-2,3);
			 \draw (-0.5-1,1) node{$\mathcal{B}$};

            \filldraw (0,2) circle (1pt);
            \filldraw (-1.5,2) circle (1pt);
            \node[above] at (-1.5,2) {$x$};
            \node[right] at (0,2) {$J$};
            \draw[thick] (0,2) -- (-1.5,2); 
             \node[above] at (-0.5, 2) {$\mathcal{L}$};
	\end{tikzpicture}}
 }
	\caption{
 The non-local conserved current living at the end of a topological line $\mathcal{L}$. The $\mathcal{L}$ topologically terminates on a codimension 1 topological defect $\mathcal{B}$.  
	}
	\label{fig:Jonsurface}
\end{figure}

Given a non-local conserved current living at the end of a line $\mathcal{L}$, we can immediately construct a codimension-1 topological defect assuming $\mathcal{L}$ is gaugeable. To see this, we first look for a codimension 1 topological defect $\mathcal{B}$, i.e.~a base defect, which can topologically terminate $\mathcal{L}$, see Figure \ref{fig:Jonsurface}.  One natural candidate for $\mathcal{B}$ is a condensation defect of $\mathcal{L}$ (since we assumed there is no obstruction to $1-$gauging \cite{Roumpedakis:2022aik}). Shrinking $\mathcal{L}$ gives rise to a conserved current living on $\mathcal{B}$. Integrating this gives a continuous topological operator generating a continuous non-invertible symmetry.\footnote{Since $\CL$ is gaugeable, we can prove the topologicalness of the defect from integration by gauging back to an invertible symmetry. However, by generalizing the careful analysis in Section \ref{sec:Noether} to higher dimensions, one can prove topologicalness even without assuming $\mathcal{L}$ gaugeable. }

The non-invertible coset symmetry $(G\rtimes K)/K$, reviewed in Section \ref{sec:introgauging}, naturally fits into this construction. The line $\mathcal{L}$ is labeled by $\mathrm{Rep}(K)$, i.e.~the quantum symmetry after gauging $K$. The current $J$ is associated with $G$ symmetry. The base line is the condensation defect of $\mathcal{L}$. This non-invertible symmetry is non-intrinsic---it comes from an invertible symmetry $G\rtimes K$ by gauging. 

In fact, when the spacetime dimension $d$ is equal or higher than 4, a bosonic topological line $\mathcal{L}$ is always labeled by $\mathrm{Rep}(K)$ for some $K$ \cite{Lan:2018vjb,Balasubramanian:2024nei,Johnson-Freyd:2020twl,Johnson-Freyd:2021tbq}, and so $\mathcal{L}$ can always be gauged away. Therefore we conjecture that all the 0-form continuous non-invertible symmetries in $d\geq 4$ are non-intrinsic. This leaves room for interesting examples in $d=3$. It would be interesting to test this conjecture with more examples in $d\geq 4$, and explore new examples in $d=3$. 
One can also discuss higher form continuous non-invertible symmetries. We leave this to future works.

Finally, there are extensive ongoing discussions on whether the non-invertible ABJ symmetry \cite{Cordova:2022ieu,Choi:2022jqy,Chen:2022cyw} is continuous or a $\mathbb{Q}/\mathbb{Z}$ symmetry \cite{Karasik:2022kkq,Arbalestrier:2024oqg,GarciaEtxebarria:2022jky,Putrov:2022pua,Chen:2025buv}. If it is continuous, then one should be able to find a non-local conserved current. For instance, in \cite{Arbalestrier:2024oqg}, the topological operator for the non-invertible ABJ symmetry is
\begin{eqnarray}
    \mathcal{N}_\alpha = \int [\mathcal{D}c \mathcal{D}A \mathcal{D}\Phi]  \exp\left(i \frac{\alpha}{2}\int \star j\right) \mathcal{B}_\alpha
\end{eqnarray}
where the ``base defect" $\mathcal{B}_\alpha$ is
\begin{eqnarray}
    \mathcal{B}_\alpha= \exp\left(\frac{i}{2\pi} \int -\frac{\alpha}{4\pi}cdc + \Phi(dc + dA)\right)
\end{eqnarray}
Here, $c$ is an $\mathbb{R}$ valued 1-form, while $\Phi$ and $A$ are $U(1)$ valued 1-form. Notably, the parameter $\alpha$ is $U(1)$ valued, as opposed to $\mathbb{R}$ valued. One key difference here is that the choice of the base defect $\mathcal{B}_\alpha$ depends on $\alpha$, while the prescription described earlier in this section gives a parameter independent base defect. Hence it is not obvious how the non-invertible ABJ symmetry fits in our construction. It is likely is that our construction is not the most generic, and it would be interesting further generalize our construction to incorporate the non-invertible ABJ symmetry.

\section{Conclusions and Future Directions}\label{sec:conclusions}

In this paper we analyzed examples of continuous non-invertible symmetries in 1+1d CFTs. After reviewing the known non-intrinsic examples, we proved a general Noether's theorem, which states that every non-local current corresponds to a non-invertible symmetry and vice-versa. After reviewing the non-invertible symmetries in $c=1$ CFTs as a warm-up, we presented new examples in $G_k$ WZW models and products of minimal models. We proved that these symmetries are intrinsic if we demand that the gauging preserves some of the invertible symmetries. More generally, while we do not know of a gauging which makes these symmetries invertible, we could not prove that one does not exist. We also discussed how these symmetries act on local operators and some applications for them. 

This paper only begins the systematic study of continuous non-invertible symmetries. We now discuss various generalizations and open questions. 
\begin{enumerate}
    \item In this paper we mostly discuss 1+1d CFTs. It should be possible to extend the construction to general QFTs. The obstructions to this are mostly technical, since it is more complicated to look for currents (there are very few examples where one has a complete handle on the spectrum of twisted sector apart from free field theories), and it would be more complicated to check that the defect is topological. 
    One way to approach the question of general QFTs would be to look at relevant deformations of the theories discussed here and check whether the non-invertible symmetries we found are preserved by the RG flow \cite{Chang:2018iay}.  A first step would be to classify all deformations which preserve the topological line $\mathcal{B}$ that the current lives along, which is already a nontrivial condition which we found difficult to satisfy. 
    \item Our construction bears significant similarities to conserved charges which appear in the quantum group literature, see \cite{Gabai:2024puk,Gabai:2024qum} for a recent discussion.\footnote{The authors thank B. Gabai and A. Zhabin for interesting discussions on this topic.} Specifically, non-local currents have been used to construct some quantum group charges, see e.g.~\cite{Gabai:2024qum,BernardLeClair1991}. It would be very nice to understand better the connection between these two concepts, as they are clearly related. In particular, it would be nice to see if the currents we found can be used to construct quantum groups. For the sake of clarity, we compare some important properties between the constructions:
    \begin{itemize}
        \item We have shown that any non-local current defines a non-invertible symmetry, but it is not clear that any non-local current can be used to construct a quantum group. It would be nice to better understand what properties such a current should possess.
        \item The charges are constructed slightly differently in the two cases. In the quantum group case one integrates a current attached to a line, where the other end of the line is at some fixed point $\omega$ (and one eventually proves $\omega$-dependence). In the non-invertible symmetry case above, we showed that $J$ also lives \textit{along} some topological defect $\mathcal{B}$, and we choose to integrate the current along $\mathcal{B}$ as one would do for any operator living on a defect. Thus there are no ``external'' lines attached to the currents which we must take into account during integration. 
        \item The quantum group construction focuses on the charge, while the construction of non-invertible symmetries includes a charge as well as a full topological defect (with a well-defined defect Hilbert space). It would be nice to understand the quantum group analog of these objects.
    \end{itemize}  
     \item It would be nice to try to understand these topological lines using the framework of fusion categories that is used for finite symmetries. For example, can one define what a simple line is for continuous non-invertible symmetries? What constrains their fusion rules? 
     \item Interpreting our 1+1d CFT as a string worldsheet, it is natural to ask how the existence of a non-local current affects spacetime. While non-invertible symmetries are supposedly broken in spacetime at high enough order in the loop expansion \cite{Heckman:2024obe,Kaidi:2024wio} (see also \cite{Bachas:2012bj,Dijkgraaf:1989hb,Hamidi:1986vh}), it is possible that the currents themselves can lead to constraints in spacetime. 
     \item As commented at the end of Section \ref{sec:higher_d}, if the non-invertible ABJ symmetry is indeed a continuous symmetry, it does not fit nicely into our construction. This implies that either one needs to understand the non-invertible ABJ symmetry better to fit into our construction, or our construction of continuous non-invertible symmetry from the non-local current should be generalized. 

\item It would be very interesting to understand to what extent one can turn on background fields for continuous non-invertible symmetries. In the invertible case, being able to write $U_\alpha$ in terms of a conserved current $J$ is essential in coupling to a background: one simply adds $A\wedge J$ (plus, perhaps, higher-order terms) to the action. Is there an analogue of this for continuous non-invertible symmetries?
     
\end{enumerate}

\section*{Acknowledgements}

We thank Federico Ambrosino, Andrea Antinucci, Daniel Brennan, Yichul Choi, Christian Copetti,  Barak Gabai, Theodore Jacobson, Justin Kaidi, Ryohei Kobayashi,  Zohar Komargodski, Justin Kulp, Jacob McNamara, Miguel Montero, Fedor Popov, Rajath Radhakrishnan, Brandon C. Rayhaun, Ingo Runkel, Sahand Seifnashri, Shu-Heng Shao, Irene Valenzuela, Yifan Wang, Pengcheng Wei and  Aleksandr Zhabin for useful discussions. YZ thanks the KITP for hospitality during the program GenSym25 during the course of this work, and the program Symmetries in QFT and Particle Physics in Korea Advanced Institute of Science $\&$ Technology (KAIST) during the final stage of this work.  This research was supported in part by grant NSF PHY-2309135 to the Kavli Institute for Theoretical Physics (KITP). The work of Y.Z. is supported by starting funds from University of Chinese Academy of Sciences (UCAS) and from the Kavli Institute for Theoretical Sciences (KITS).

\appendix

\section{Minimal Models}
\label{app:min_mod}

In this appendix we consider the 1+1d minimal models $\mathcal{M}_{p,q}$ for coprime $p>q\geq 2$. The central charge is
\begin{equation} c=1-\frac{6 (p-q)^2}{p q}\;,\end{equation}
and the spectrum of holomorphic dimensions is
\begin{equation} \label{eq:min_mod_h}h_{r,s}=\frac{(p r-q s)^2-(p-q)^2}{4 p q} \;,\qquad 1\leq r\leq q-1\;,1\leq s\leq p-1\;,\end{equation}
and similarly for the antiholomorphic sector. 

For unitary minimal models, we set $p=m+1, q=m$, and denote the model as $\mathcal{M}_m$. The Verlinde lines $\mathcal{L}_{r,s}$ are also labeled by the same set of quantum numbers. The quantum dimension is
\begin{eqnarray}
    d_{r,s} = \frac{\sin\left( \frac{\pi r}{m} \right) \sin\left( \frac{\pi s}{m+1} \right)}{\sin\left( \frac{\pi}{m} \right) \sin\left( \frac{\pi}{m+1} \right)}\;.
\end{eqnarray}
Only $\mathcal{L}_{1,1}$ and $\mathcal{L}_{m-1,m}$ are invertible lines, the former being the identity line, while the latter being a $\bZ_2$ line.

\subsection{Absence of Non-Local Conserved Currents in Minimal Models}

Since the full spectrum of (local and twist) operators is known for the minimal models, one can easily search for non-local currents. The problem reduces to finding integers $r,s$ for which $h_{r,s}=1$, which requires
\begin{equation} q(s-1)=p (r+1)\;\;\text{  or  }\;\;q(s+1)=p (r-1) \;.\end{equation}
For the first condition, because $p$ and $q$ are coprime, $s-1\in p\bZ$, and $r+1\in q\bZ$. Moreover, for $r$ and $s$ to be within the range in \eqref{eq:min_mod_h}, the only possibility is $s=1, r= q-1$. But this would further imply $p=0$. Similarly, one get $q=0$ for the second condition. 
So non-local currents do not exist in unitary or non-unitary minimal models.

\subsection{Twisted Partition Function}

We focus on unitary minimal models where $p=m+1, q=m$. The Ising model corresponds to $m=3$. The character is 
\begin{eqnarray}
    \chi_{r,s}(\tau)= \mathcal{K}_{r,s}^m(\tau) - \mathcal{K}_{r,-s}^m(\tau)\;,
\end{eqnarray}
where
\begin{eqnarray}
    \mathcal{K}_{r,s}^{m}(\tau) = \frac{1}{\eta(q)} \sum_{n\in \bZ} q^{(2(m+1)m n + (m+1)r - m s)^2/4m(m+1)}\;.
\end{eqnarray}
The partition function is 
\begin{eqnarray}
    \mathcal{Z}(\tau) = \sum_{r,s}  \chi_{r,s}(\tau) \bar{\chi}_{r,s}(\bar{\tau})\;.
\end{eqnarray}
The range of summation depends on the parity of $m$. For odd $m$, we have $1\leq r\leq \frac{m-1}{2}, 1\leq s\leq m$. For even $m$, we have $1\leq r\leq m-1, 1\leq s \leq \frac{m}{2}$. The $\mathcal{L}_{r,s}$ twisted partition function is 
\begin{eqnarray}
    \mathcal{Z}_{r,s}(\tau) = \sum_{r',s', r'', s''} N_{(r,s)(r',s')}^{(r'',s'')} \chi_{r,s}(\tau) \bar{\chi}_{r,s}(\bar{\tau})\;.
\end{eqnarray}

\section{$SU(2)_k$ WZW Model}
\label{app:SU2kWZW}

In this appendix, we provide more details for Section \ref{sec:currentWZW}.

\subsection{Characters with Chemical Potential}

We consider the diagonal $SU(2)_k$ WZW model (we follow the notation of \cite{Eberhardt2019WZW}). The characters are labeled by an integer $0\leq \lambda \leq k$, i.e.~$\chi_\lambda$. They are defined as
\begin{eqnarray}\label{eq:characters}
	\chi_\lambda (\tau) = \text{Tr}_{\lambda}\left(q^{L_0-\frac{c}{24}}\right)= \frac{q^{\frac{(\lambda+1)^2}{4(k+2)}}}{[\eta(q)]^3} \sum_{n\in \bZ} [\lambda + 1 + 2n(k+2)] q^{n(\lambda + 1 + (k+2)n)}\;.
\end{eqnarray}
which transforms under modular S transformation as
\begin{eqnarray}
	\begin{split}
		\chi_\lambda(-1/\tau) = \sum_{\mu} S_{\lambda \mu} \chi_\mu(\tau), \quad \chi_\lambda (\tau+1) = e^{2\pi i h_\lambda} \chi_\lambda(\tau)\;.
	\end{split}
\end{eqnarray}
The $S_{\mu\nu}$ is defined to be \eqref{eq:modularS}. 
We can also define a character modified by a chemical potential for the Cartan of $SU(2)$:
\begin{equation}
    \chi_\lambda(z ; \tau)=\text{Tr}_{\lambda}\left(q^{L_0-\frac{c}{24}}y^{J_3}\right)\;,
\end{equation}
where $y=e^{2\pi i z}$. 
This is given by
\begin{equation}\label{eq:chemical_potential_character}
    \chi_{\ell}(z ; \tau)=\frac{\Theta_{2 \ell+1}^{(k+2)}(z ; \tau)-\Theta_{-2 \ell-1}^{(k+2)}(z ; \tau)}{\Theta_1^{(2)}(z ; \tau)-\Theta_{-1}^{(2)}(z ; \tau)}\;,
\end{equation}
where
\begin{equation}
\Theta_m^{(k)}(z ; \tau) \equiv \sum_{n \in \mathbb{Z}+\frac{m}{2 k}} q^{k n^2} y^{k n}\;.\end{equation}

The torus partition function of the CFT, with $SU(2)$ chemical potential turned on, is
\begin{eqnarray}\label{eq:diagpart}
	\CZ(\tau, z)= \sum_\lambda \chi_\lambda (z, \tau) \bar\chi_\lambda(\bar{z}, \bar\tau)\;.
\end{eqnarray}

\subsection{Twisted Sector Operators}\label{app:Z_twist}

The action of $\CL_\lambda$ on a conformal primary $\phi_{\mu, \mu}$ is 
\begin{eqnarray}
	\CL_\lambda \phi_{\mu, \mu} = \frac{S_{\lambda \mu}}{S_{0 \mu}} \phi_{\mu, \mu} \;.
\end{eqnarray}
This means that inserting $\CL_\lambda$ along the spatial direction gives the partition function 
\begin{eqnarray}
	\CZ^\lambda(\tau) = \sum_{\mu} \frac{S_{\lambda \mu}}{S_{0 \mu}} \chi_{\mu}(\tau) \bar{\chi}_\mu(\bar\tau)\;.
\end{eqnarray}
Under the modular $S$ transformation, we get the partition function with $\CL_\lambda$ inserted along the time direction, i.e.~as a defect
\begin{eqnarray}
	\CZ_{\lambda}(\tau, \bar\tau)= \sum_{\mu \nu \sigma} \frac{S_{\lambda \mu}}{S_{0 \mu}} S_{\mu \nu}\chi_{\nu}(\tau) S^*_{\mu\sigma}\bar{\chi}_\sigma(\bar\tau)= \sum_{\nu\sigma} N_{\lambda \nu}^\sigma \chi_\nu (\tau) \bar\chi_\sigma(\bar\tau)\;,
\end{eqnarray}
where in the second equality we used the Verlinde formula \eqref{eq:Verlindeformula}. 
From this expression, we find that the $\lambda$-twisted sector contains the primary operators of conformal weights $(h_\nu, h_\sigma)$ for $|\lambda-\nu|\leq \sigma \leq \min(\lambda+\nu, 2k - \lambda -\nu)$ and $\nu+\sigma+\lambda =0$  mod 2.

Similarly, the twisted partition function with chemical potential is 
\begin{eqnarray}\label{chemical_pot_partition_function}
	\CZ_{\lambda}(\tau, z)= \sum_{\nu\sigma} N_{\lambda \nu}^\sigma \chi_\nu (z;\tau) \bar\chi_\sigma(\bar z;\bar\tau)\;.
\end{eqnarray}

\subsection{OPE Coefficients}\label{app:SU2_OPE}

We review the OPE coefficients for three local (non-chiral) primaries of the diagonal $SU(2)_k$ WZW models. These were originally computed in \cite{Zamolodchikov:1986bd}, and a simple presentation is given in \cite[section 6.2]{Recknagel:2013uja}. Given three primaries $\lambda_1,\lambda_2,\lambda_3\in \{0,1,...,k\}$ one finds
\begin{equation}
    \tilde{\mathcal{C}}^{\lambda_3}_{\lambda_1,\lambda_2} = (J + 1)! \, P(J + 1) \, P(1)^{\frac{1}{2}} 
\prod_{\nu = 1}^{3} \frac{P(\hat{\lambda}_\nu) \, \hat{\lambda}_\nu!}{(\lambda_\nu + 1)^{\frac{1}{2}} (\lambda_\nu)! \, P(\lambda_\nu)^{\frac{1}{2}} \, P(\lambda_\nu + 1)^{\frac{1}{2}}}\;,
\end{equation}
where $ J = \frac{\lambda_1+\lambda_2+\lambda_3}{2} $ and $ \hat{\lambda}_\nu = J - \lambda_\nu $. We also define $P(\lambda) $ such that $P(0) = 1 $ and
\begin{equation}
P(\lambda) = \prod_{n=1}^{\lambda} \frac{\Gamma\bigl( \frac{n}{k+2} \bigr)}{\Gamma\bigl(1 - \frac{n}{k+2} \bigr)}\;.
\end{equation}
$\tilde{\mathcal{C}}^{\lambda_3}_{\lambda_1,\lambda_2}$ here is defined in the normalization where $\tilde{\mathcal{C}}_{\lambda\lambda}^0=1$. Instead we will be interested in $\mathcal{C}_{\lambda_1\lambda_2}^{\lambda_3}$ normalized such that $\mathcal{C}_{\lambda\lambda}^0=d_\lambda$. This is obtained by taking 
\begin{eqnarray}
    \mathcal{C}_{\lambda_1\lambda_2}^{\lambda_3}=\tilde{\mathcal{C}}_{\lambda_1\lambda_2}^{\lambda_3}\sqrt{d_{\lambda_1}d_{\lambda_2}d_{\lambda_3}}\;.
\end{eqnarray}
The OPE of $SU(2)$ and Virasoro descendants can be read off from these results, and we will not require them here.

\subsection{Solving Modular Equations}

To probe whether a non-local current can be made local by gauging, we find non-negative integer solutions to \eqref{eq:WST} and \eqref{eq:Winequality},\footnote{We also demand $W_{00}=1$. } and find whether there is a solution with $W_{0,2n}\neq 0$ for $k=(n+2)(n-1)$. If such a solution exists, then the current can be made local.

\subsubsection{$k=4, n=2$} 

The only non-trivial solution is
\begin{equation}
    W= \left(
\begin{smallmatrix}
 1 & \vphantom{0}\cdot & \vphantom{0}\cdot & \vphantom{0}\cdot & \textcolor{red}{1} \\
 \vphantom{0}\cdot & \vphantom{0}\cdot & \vphantom{0}\cdot & \vphantom{0}\cdot & \vphantom{0}\cdot \\
 \vphantom{0}\cdot & \vphantom{0}\cdot & 2 & \vphantom{0}\cdot & \vphantom{0}\cdot \\
 \vphantom{0}\cdot & \vphantom{0}\cdot & \vphantom{0}\cdot & \vphantom{0}\cdot & \vphantom{0}\cdot \\
 \textcolor{red}{1} & \vphantom{0}\cdot & \vphantom{0}\cdot & \vphantom{0}\cdot & 1 \\
\end{smallmatrix}
\right)\;,
\end{equation}
where the omitted entries are zero. Since $W_{04}=W_{40}=1\neq 0$ (marked in red), the non-local current can be made local by gauging.

\subsubsection{$k=10,n=3$}

There are two non-trivial solutions, 
\begin{equation}
    W_1= 
    \left(
\begin{smallmatrix}
 1 & \vphantom{0}\cdot & \vphantom{0}\cdot & \vphantom{0}\cdot & \vphantom{0}\cdot & \vphantom{0}\cdot & \vphantom{0}\cdot & \vphantom{0}\cdot & \vphantom{0}\cdot & \vphantom{0}\cdot & \vphantom{0}\cdot \\
 \vphantom{0}\cdot & \vphantom{0}\cdot & \vphantom{0}\cdot & \vphantom{0}\cdot & \vphantom{0}\cdot & \vphantom{0}\cdot & \vphantom{0}\cdot & \vphantom{0}\cdot & \vphantom{0}\cdot & 1 & \vphantom{0}\cdot \\
 \vphantom{0}\cdot & \vphantom{0}\cdot & 1 & \vphantom{0}\cdot & \vphantom{0}\cdot & \vphantom{0}\cdot & \vphantom{0}\cdot & \vphantom{0}\cdot & \vphantom{0}\cdot & \vphantom{0}\cdot & \vphantom{0}\cdot \\
 \vphantom{0}\cdot & \vphantom{0}\cdot & \vphantom{0}\cdot & \vphantom{0}\cdot & \vphantom{0}\cdot & \vphantom{0}\cdot & \vphantom{0}\cdot & 1 & \vphantom{0}\cdot & \vphantom{0}\cdot & \vphantom{0}\cdot \\
 \vphantom{0}\cdot & \vphantom{0}\cdot & \vphantom{0}\cdot & \vphantom{0}\cdot & 1 & \vphantom{0}\cdot & \vphantom{0}\cdot & \vphantom{0}\cdot & \vphantom{0}\cdot & \vphantom{0}\cdot & \vphantom{0}\cdot \\
 \vphantom{0}\cdot & \vphantom{0}\cdot & \vphantom{0}\cdot & \vphantom{0}\cdot & \vphantom{0}\cdot & 1 & \vphantom{0}\cdot & \vphantom{0}\cdot & \vphantom{0}\cdot & \vphantom{0}\cdot & \vphantom{0}\cdot \\
 \vphantom{0}\cdot & \vphantom{0}\cdot & \vphantom{0}\cdot & \vphantom{0}\cdot & \vphantom{0}\cdot & \vphantom{0}\cdot & 1 & \vphantom{0}\cdot & \vphantom{0}\cdot & \vphantom{0}\cdot & \vphantom{0}\cdot \\
 \vphantom{0}\cdot & \vphantom{0}\cdot & \vphantom{0}\cdot & 1 & \vphantom{0}\cdot & \vphantom{0}\cdot & \vphantom{0}\cdot & \vphantom{0}\cdot & \vphantom{0}\cdot & \vphantom{0}\cdot & \vphantom{0}\cdot \\
 \vphantom{0}\cdot & \vphantom{0}\cdot & \vphantom{0}\cdot & \vphantom{0}\cdot & \vphantom{0}\cdot & \vphantom{0}\cdot & \vphantom{0}\cdot & \vphantom{0}\cdot & 1 & \vphantom{0}\cdot & \vphantom{0}\cdot \\
 \vphantom{0}\cdot & 1 & \vphantom{0}\cdot & \vphantom{0}\cdot & \vphantom{0}\cdot & \vphantom{0}\cdot & \vphantom{0}\cdot & \vphantom{0}\cdot & \vphantom{0}\cdot & \vphantom{0}\cdot & \vphantom{0}\cdot \\
 \vphantom{0}\cdot & \vphantom{0}\cdot & \vphantom{0}\cdot & \vphantom{0}\cdot & \vphantom{0}\cdot & \vphantom{0}\cdot & \vphantom{0}\cdot & \vphantom{0}\cdot & \vphantom{0}\cdot & \vphantom{0}\cdot & 1 \\
\end{smallmatrix}
\right), \quad 
W_2= 
\left(
\begin{smallmatrix}
 1 & \vphantom{0}\cdot & \vphantom{0}\cdot & \vphantom{0}\cdot & \vphantom{0}\cdot & \vphantom{0}\cdot & \textcolor{red}{1} & \vphantom{0}\cdot & \vphantom{0}\cdot & \vphantom{0}\cdot & \vphantom{0}\cdot \\
 \vphantom{0}\cdot & \vphantom{0}\cdot & \vphantom{0}\cdot & \vphantom{0}\cdot & \vphantom{0}\cdot & \vphantom{0}\cdot & \vphantom{0}\cdot & \vphantom{0}\cdot & \vphantom{0}\cdot & \vphantom{0}\cdot & \vphantom{0}\cdot \\
 \vphantom{0}\cdot & \vphantom{0}\cdot & \vphantom{0}\cdot & \vphantom{0}\cdot & \vphantom{0}\cdot & \vphantom{0}\cdot & \vphantom{0}\cdot & \vphantom{0}\cdot & \vphantom{0}\cdot & \vphantom{0}\cdot & \vphantom{0}\cdot \\
 \vphantom{0}\cdot & \vphantom{0}\cdot & \vphantom{0}\cdot & 1 & \vphantom{0}\cdot & \vphantom{0}\cdot & \vphantom{0}\cdot & 1 & \vphantom{0}\cdot & \vphantom{0}\cdot & \vphantom{0}\cdot \\
 \vphantom{0}\cdot & \vphantom{0}\cdot & \vphantom{0}\cdot & \vphantom{0}\cdot & 1 & \vphantom{0}\cdot & \vphantom{0}\cdot & \vphantom{0}\cdot & \vphantom{0}\cdot & \vphantom{0}\cdot & 1 \\
 \vphantom{0}\cdot & \vphantom{0}\cdot & \vphantom{0}\cdot & \vphantom{0}\cdot & \vphantom{0}\cdot & \vphantom{0}\cdot & \vphantom{0}\cdot & \vphantom{0}\cdot & \vphantom{0}\cdot & \vphantom{0}\cdot & \vphantom{0}\cdot \\
 \textcolor{red}{1} & \vphantom{0}\cdot & \vphantom{0}\cdot & \vphantom{0}\cdot & \vphantom{0}\cdot & \vphantom{0}\cdot & 1 & \vphantom{0}\cdot & \vphantom{0}\cdot & \vphantom{0}\cdot & \vphantom{0}\cdot \\
 \vphantom{0}\cdot & \vphantom{0}\cdot & \vphantom{0}\cdot & 1 & \vphantom{0}\cdot & \vphantom{0}\cdot & \vphantom{0}\cdot & 1 & \vphantom{0}\cdot & \vphantom{0}\cdot & \vphantom{0}\cdot \\
 \vphantom{0}\cdot & \vphantom{0}\cdot & \vphantom{0}\cdot & \vphantom{0}\cdot & \vphantom{0}\cdot & \vphantom{0}\cdot & \vphantom{0}\cdot & \vphantom{0}\cdot & \vphantom{0}\cdot & \vphantom{0}\cdot & \vphantom{0}\cdot \\
 \vphantom{0}\cdot & \vphantom{0}\cdot & \vphantom{0}\cdot & \vphantom{0}\cdot & \vphantom{0}\cdot & \vphantom{0}\cdot & \vphantom{0}\cdot & \vphantom{0}\cdot & \vphantom{0}\cdot & \vphantom{0}\cdot & \vphantom{0}\cdot \\
 \vphantom{0}\cdot & \vphantom{0}\cdot & \vphantom{0}\cdot & \vphantom{0}\cdot & 1 & \vphantom{0}\cdot & \vphantom{0}\cdot & \vphantom{0}\cdot & \vphantom{0}\cdot & \vphantom{0}\cdot & 1 \\
\end{smallmatrix}
\right)\;.
\end{equation}
Since $W_{06}=W_{60}=1\neq 0$ (marked in red), the non-local current can be made local by gauging.

\subsubsection{$k=18, n=4$}

There is only one non-trivial solution, 
\begin{equation}
    W= 
    \left(
\begin{smallmatrix}
 1 & \vphantom{0}\cdot & \vphantom{0}\cdot & \vphantom{0}\cdot & \vphantom{0}\cdot & \vphantom{0}\cdot & \vphantom{0}\cdot & \vphantom{0}\cdot & \vphantom{0}\cdot & \vphantom{0}\cdot & \vphantom{0}\cdot & \vphantom{0}\cdot & \vphantom{0}\cdot & \vphantom{0}\cdot & \vphantom{0}\cdot & \vphantom{0}\cdot & \vphantom{0}\cdot & \vphantom{0}\cdot & \vphantom{0}\cdot \\
 \vphantom{0}\cdot & \vphantom{0}\cdot & \vphantom{0}\cdot & \vphantom{0}\cdot & \vphantom{0}\cdot & \vphantom{0}\cdot & \vphantom{0}\cdot & \vphantom{0}\cdot & \vphantom{0}\cdot & \vphantom{0}\cdot & \vphantom{0}\cdot & \vphantom{0}\cdot & \vphantom{0}\cdot & \vphantom{0}\cdot & \vphantom{0}\cdot & \vphantom{0}\cdot & \vphantom{0}\cdot & 1 & \vphantom{0}\cdot \\
 \vphantom{0}\cdot & \vphantom{0}\cdot & 1 & \vphantom{0}\cdot & \vphantom{0}\cdot & \vphantom{0}\cdot & \vphantom{0}\cdot & \vphantom{0}\cdot & \vphantom{0}\cdot & \vphantom{0}\cdot & \vphantom{0}\cdot & \vphantom{0}\cdot & \vphantom{0}\cdot & \vphantom{0}\cdot & \vphantom{0}\cdot & \vphantom{0}\cdot & \vphantom{0}\cdot & \vphantom{0}\cdot & \vphantom{0}\cdot \\
 \vphantom{0}\cdot & \vphantom{0}\cdot & \vphantom{0}\cdot & \vphantom{0}\cdot & \vphantom{0}\cdot & \vphantom{0}\cdot & \vphantom{0}\cdot & \vphantom{0}\cdot & \vphantom{0}\cdot & \vphantom{0}\cdot & \vphantom{0}\cdot & \vphantom{0}\cdot & \vphantom{0}\cdot & \vphantom{0}\cdot & \vphantom{0}\cdot & 1 & \vphantom{0}\cdot & \vphantom{0}\cdot & \vphantom{0}\cdot \\
 \vphantom{0}\cdot & \vphantom{0}\cdot & \vphantom{0}\cdot & \vphantom{0}\cdot & 1 & \vphantom{0}\cdot & \vphantom{0}\cdot & \vphantom{0}\cdot & \vphantom{0}\cdot & \vphantom{0}\cdot & \vphantom{0}\cdot & \vphantom{0}\cdot & \vphantom{0}\cdot & \vphantom{0}\cdot & \vphantom{0}\cdot & \vphantom{0}\cdot & \vphantom{0}\cdot & \vphantom{0}\cdot & \vphantom{0}\cdot \\
 \vphantom{0}\cdot & \vphantom{0}\cdot & \vphantom{0}\cdot & \vphantom{0}\cdot & \vphantom{0}\cdot & \vphantom{0}\cdot & \vphantom{0}\cdot & \vphantom{0}\cdot & \vphantom{0}\cdot & \vphantom{0}\cdot & \vphantom{0}\cdot & \vphantom{0}\cdot & \vphantom{0}\cdot & 1 & \vphantom{0}\cdot & \vphantom{0}\cdot & \vphantom{0}\cdot & \vphantom{0}\cdot & \vphantom{0}\cdot \\
 \vphantom{0}\cdot & \vphantom{0}\cdot & \vphantom{0}\cdot & \vphantom{0}\cdot & \vphantom{0}\cdot & \vphantom{0}\cdot & 1 & \vphantom{0}\cdot & \vphantom{0}\cdot & \vphantom{0}\cdot & \vphantom{0}\cdot & \vphantom{0}\cdot & \vphantom{0}\cdot & \vphantom{0}\cdot & \vphantom{0}\cdot & \vphantom{0}\cdot & \vphantom{0}\cdot & \vphantom{0}\cdot & \vphantom{0}\cdot \\
 \vphantom{0}\cdot & \vphantom{0}\cdot & \vphantom{0}\cdot & \vphantom{0}\cdot & \vphantom{0}\cdot & \vphantom{0}\cdot & \vphantom{0}\cdot & \vphantom{0}\cdot & \vphantom{0}\cdot & \vphantom{0}\cdot & \vphantom{0}\cdot & 1 & \vphantom{0}\cdot & \vphantom{0}\cdot & \vphantom{0}\cdot & \vphantom{0}\cdot & \vphantom{0}\cdot & \vphantom{0}\cdot & \vphantom{0}\cdot \\
 \vphantom{0}\cdot & \vphantom{0}\cdot & \vphantom{0}\cdot & \vphantom{0}\cdot & \vphantom{0}\cdot & \vphantom{0}\cdot & \vphantom{0}\cdot & \vphantom{0}\cdot & 1 & \vphantom{0}\cdot & \vphantom{0}\cdot & \vphantom{0}\cdot & \vphantom{0}\cdot & \vphantom{0}\cdot & \vphantom{0}\cdot & \vphantom{0}\cdot & \vphantom{0}\cdot & \vphantom{0}\cdot & \vphantom{0}\cdot \\
 \vphantom{0}\cdot & \vphantom{0}\cdot & \vphantom{0}\cdot & \vphantom{0}\cdot & \vphantom{0}\cdot & \vphantom{0}\cdot & \vphantom{0}\cdot & \vphantom{0}\cdot & \vphantom{0}\cdot & 1 & \vphantom{0}\cdot & \vphantom{0}\cdot & \vphantom{0}\cdot & \vphantom{0}\cdot & \vphantom{0}\cdot & \vphantom{0}\cdot & \vphantom{0}\cdot & \vphantom{0}\cdot & \vphantom{0}\cdot \\
 \vphantom{0}\cdot & \vphantom{0}\cdot & \vphantom{0}\cdot & \vphantom{0}\cdot & \vphantom{0}\cdot & \vphantom{0}\cdot & \vphantom{0}\cdot & \vphantom{0}\cdot & \vphantom{0}\cdot & \vphantom{0}\cdot & 1 & \vphantom{0}\cdot & \vphantom{0}\cdot & \vphantom{0}\cdot & \vphantom{0}\cdot & \vphantom{0}\cdot & \vphantom{0}\cdot & \vphantom{0}\cdot & \vphantom{0}\cdot \\
 \vphantom{0}\cdot & \vphantom{0}\cdot & \vphantom{0}\cdot & \vphantom{0}\cdot & \vphantom{0}\cdot & \vphantom{0}\cdot & \vphantom{0}\cdot & 1 & \vphantom{0}\cdot & \vphantom{0}\cdot & \vphantom{0}\cdot & \vphantom{0}\cdot & \vphantom{0}\cdot & \vphantom{0}\cdot & \vphantom{0}\cdot & \vphantom{0}\cdot & \vphantom{0}\cdot & \vphantom{0}\cdot & \vphantom{0}\cdot \\
 \vphantom{0}\cdot & \vphantom{0}\cdot & \vphantom{0}\cdot & \vphantom{0}\cdot & \vphantom{0}\cdot & \vphantom{0}\cdot & \vphantom{0}\cdot & \vphantom{0}\cdot & \vphantom{0}\cdot & \vphantom{0}\cdot & \vphantom{0}\cdot & \vphantom{0}\cdot & 1 & \vphantom{0}\cdot & \vphantom{0}\cdot & \vphantom{0}\cdot & \vphantom{0}\cdot & \vphantom{0}\cdot & \vphantom{0}\cdot \\
 \vphantom{0}\cdot & \vphantom{0}\cdot & \vphantom{0}\cdot & \vphantom{0}\cdot & \vphantom{0}\cdot & 1 & \vphantom{0}\cdot & \vphantom{0}\cdot & \vphantom{0}\cdot & \vphantom{0}\cdot & \vphantom{0}\cdot & \vphantom{0}\cdot & \vphantom{0}\cdot & \vphantom{0}\cdot & \vphantom{0}\cdot & \vphantom{0}\cdot & \vphantom{0}\cdot & \vphantom{0}\cdot & \vphantom{0}\cdot \\
 \vphantom{0}\cdot & \vphantom{0}\cdot & \vphantom{0}\cdot & \vphantom{0}\cdot & \vphantom{0}\cdot & \vphantom{0}\cdot & \vphantom{0}\cdot & \vphantom{0}\cdot & \vphantom{0}\cdot & \vphantom{0}\cdot & \vphantom{0}\cdot & \vphantom{0}\cdot & \vphantom{0}\cdot & \vphantom{0}\cdot & 1 & \vphantom{0}\cdot & \vphantom{0}\cdot & \vphantom{0}\cdot & \vphantom{0}\cdot \\
 \vphantom{0}\cdot & \vphantom{0}\cdot & \vphantom{0}\cdot & 1 & \vphantom{0}\cdot & \vphantom{0}\cdot & \vphantom{0}\cdot & \vphantom{0}\cdot & \vphantom{0}\cdot & \vphantom{0}\cdot & \vphantom{0}\cdot & \vphantom{0}\cdot & \vphantom{0}\cdot & \vphantom{0}\cdot & \vphantom{0}\cdot & \vphantom{0}\cdot & \vphantom{0}\cdot & \vphantom{0}\cdot & \vphantom{0}\cdot \\
 \vphantom{0}\cdot & \vphantom{0}\cdot & \vphantom{0}\cdot & \vphantom{0}\cdot & \vphantom{0}\cdot & \vphantom{0}\cdot & \vphantom{0}\cdot & \vphantom{0}\cdot & \vphantom{0}\cdot & \vphantom{0}\cdot & \vphantom{0}\cdot & \vphantom{0}\cdot & \vphantom{0}\cdot & \vphantom{0}\cdot & \vphantom{0}\cdot & \vphantom{0}\cdot & 1 & \vphantom{0}\cdot & \vphantom{0}\cdot \\
 \vphantom{0}\cdot & 1 & \vphantom{0}\cdot & \vphantom{0}\cdot & \vphantom{0}\cdot & \vphantom{0}\cdot & \vphantom{0}\cdot & \vphantom{0}\cdot & \vphantom{0}\cdot & \vphantom{0}\cdot & \vphantom{0}\cdot & \vphantom{0}\cdot & \vphantom{0}\cdot & \vphantom{0}\cdot & \vphantom{0}\cdot & \vphantom{0}\cdot & \vphantom{0}\cdot & \vphantom{0}\cdot & \vphantom{0}\cdot \\
 \vphantom{0}\cdot & \vphantom{0}\cdot & \vphantom{0}\cdot & \vphantom{0}\cdot & \vphantom{0}\cdot & \vphantom{0}\cdot & \vphantom{0}\cdot & \vphantom{0}\cdot & \vphantom{0}\cdot & \vphantom{0}\cdot & \vphantom{0}\cdot & \vphantom{0}\cdot & \vphantom{0}\cdot & \vphantom{0}\cdot & \vphantom{0}\cdot & \vphantom{0}\cdot & \vphantom{0}\cdot & \vphantom{0}\cdot & 1 \\
\end{smallmatrix}
\right)\;.
\end{equation}
Since $W_{80}=W_{08}=0$, the non-local current cannot be made local by gauging while preserving $SU(2)$ chiral algebra. 

\subsubsection{$k=28, n=5$}

There are two solutions, 
\begin{equation}
    W_1= 
    \left(
\begin{smallmatrix}
 1 & \vphantom{0}\cdot & \vphantom{0}\cdot & \vphantom{0}\cdot & \vphantom{0}\cdot & \vphantom{0}\cdot & \vphantom{0}\cdot & \vphantom{0}\cdot & \vphantom{0}\cdot & \vphantom{0}\cdot & \vphantom{0}\cdot & \vphantom{0}\cdot & \vphantom{0}\cdot & \vphantom{0}\cdot & \vphantom{0}\cdot & \vphantom{0}\cdot & \vphantom{0}\cdot & \vphantom{0}\cdot & \vphantom{0}\cdot & \vphantom{0}\cdot & \vphantom{0}\cdot & \vphantom{0}\cdot & \vphantom{0}\cdot & \vphantom{0}\cdot & \vphantom{0}\cdot & \vphantom{0}\cdot & \vphantom{0}\cdot & \vphantom{0}\cdot & 1 \\
 \vphantom{0}\cdot & \vphantom{0}\cdot & \vphantom{0}\cdot & \vphantom{0}\cdot & \vphantom{0}\cdot & \vphantom{0}\cdot & \vphantom{0}\cdot & \vphantom{0}\cdot & \vphantom{0}\cdot & \vphantom{0}\cdot & \vphantom{0}\cdot & \vphantom{0}\cdot & \vphantom{0}\cdot & \vphantom{0}\cdot & \vphantom{0}\cdot & \vphantom{0}\cdot & \vphantom{0}\cdot & \vphantom{0}\cdot & \vphantom{0}\cdot & \vphantom{0}\cdot & \vphantom{0}\cdot & \vphantom{0}\cdot & \vphantom{0}\cdot & \vphantom{0}\cdot & \vphantom{0}\cdot & \vphantom{0}\cdot & \vphantom{0}\cdot & \vphantom{0}\cdot & \vphantom{0}\cdot \\
 \vphantom{0}\cdot & \vphantom{0}\cdot & 1 & \vphantom{0}\cdot & \vphantom{0}\cdot & \vphantom{0}\cdot & \vphantom{0}\cdot & \vphantom{0}\cdot & \vphantom{0}\cdot & \vphantom{0}\cdot & \vphantom{0}\cdot & \vphantom{0}\cdot & \vphantom{0}\cdot & \vphantom{0}\cdot & \vphantom{0}\cdot & \vphantom{0}\cdot & \vphantom{0}\cdot & \vphantom{0}\cdot & \vphantom{0}\cdot & \vphantom{0}\cdot & \vphantom{0}\cdot & \vphantom{0}\cdot & \vphantom{0}\cdot & \vphantom{0}\cdot & \vphantom{0}\cdot & \vphantom{0}\cdot & 1 & \vphantom{0}\cdot & \vphantom{0}\cdot \\
 \vphantom{0}\cdot & \vphantom{0}\cdot & \vphantom{0}\cdot & \vphantom{0}\cdot & \vphantom{0}\cdot & \vphantom{0}\cdot & \vphantom{0}\cdot & \vphantom{0}\cdot & \vphantom{0}\cdot & \vphantom{0}\cdot & \vphantom{0}\cdot & \vphantom{0}\cdot & \vphantom{0}\cdot & \vphantom{0}\cdot & \vphantom{0}\cdot & \vphantom{0}\cdot & \vphantom{0}\cdot & \vphantom{0}\cdot & \vphantom{0}\cdot & \vphantom{0}\cdot & \vphantom{0}\cdot & \vphantom{0}\cdot & \vphantom{0}\cdot & \vphantom{0}\cdot & \vphantom{0}\cdot & \vphantom{0}\cdot & \vphantom{0}\cdot & \vphantom{0}\cdot & \vphantom{0}\cdot \\
 \vphantom{0}\cdot & \vphantom{0}\cdot & \vphantom{0}\cdot & \vphantom{0}\cdot & 1 & \vphantom{0}\cdot & \vphantom{0}\cdot & \vphantom{0}\cdot & \vphantom{0}\cdot & \vphantom{0}\cdot & \vphantom{0}\cdot & \vphantom{0}\cdot & \vphantom{0}\cdot & \vphantom{0}\cdot & \vphantom{0}\cdot & \vphantom{0}\cdot & \vphantom{0}\cdot & \vphantom{0}\cdot & \vphantom{0}\cdot & \vphantom{0}\cdot & \vphantom{0}\cdot & \vphantom{0}\cdot & \vphantom{0}\cdot & \vphantom{0}\cdot & 1 & \vphantom{0}\cdot & \vphantom{0}\cdot & \vphantom{0}\cdot & \vphantom{0}\cdot \\
 \vphantom{0}\cdot & \vphantom{0}\cdot & \vphantom{0}\cdot & \vphantom{0}\cdot & \vphantom{0}\cdot & \vphantom{0}\cdot & \vphantom{0}\cdot & \vphantom{0}\cdot & \vphantom{0}\cdot & \vphantom{0}\cdot & \vphantom{0}\cdot & \vphantom{0}\cdot & \vphantom{0}\cdot & \vphantom{0}\cdot & \vphantom{0}\cdot & \vphantom{0}\cdot & \vphantom{0}\cdot & \vphantom{0}\cdot & \vphantom{0}\cdot & \vphantom{0}\cdot & \vphantom{0}\cdot & \vphantom{0}\cdot & \vphantom{0}\cdot & \vphantom{0}\cdot & \vphantom{0}\cdot & \vphantom{0}\cdot & \vphantom{0}\cdot & \vphantom{0}\cdot & \vphantom{0}\cdot \\
 \vphantom{0}\cdot & \vphantom{0}\cdot & \vphantom{0}\cdot & \vphantom{0}\cdot & \vphantom{0}\cdot & \vphantom{0}\cdot & 1 & \vphantom{0}\cdot & \vphantom{0}\cdot & \vphantom{0}\cdot & \vphantom{0}\cdot & \vphantom{0}\cdot & \vphantom{0}\cdot & \vphantom{0}\cdot & \vphantom{0}\cdot & \vphantom{0}\cdot & \vphantom{0}\cdot & \vphantom{0}\cdot & \vphantom{0}\cdot & \vphantom{0}\cdot & \vphantom{0}\cdot & \vphantom{0}\cdot & 1 & \vphantom{0}\cdot & \vphantom{0}\cdot & \vphantom{0}\cdot & \vphantom{0}\cdot & \vphantom{0}\cdot & \vphantom{0}\cdot \\
 \vphantom{0}\cdot & \vphantom{0}\cdot & \vphantom{0}\cdot & \vphantom{0}\cdot & \vphantom{0}\cdot & \vphantom{0}\cdot & \vphantom{0}\cdot & \vphantom{0}\cdot & \vphantom{0}\cdot & \vphantom{0}\cdot & \vphantom{0}\cdot & \vphantom{0}\cdot & \vphantom{0}\cdot & \vphantom{0}\cdot & \vphantom{0}\cdot & \vphantom{0}\cdot & \vphantom{0}\cdot & \vphantom{0}\cdot & \vphantom{0}\cdot & \vphantom{0}\cdot & \vphantom{0}\cdot & \vphantom{0}\cdot & \vphantom{0}\cdot & \vphantom{0}\cdot & \vphantom{0}\cdot & \vphantom{0}\cdot & \vphantom{0}\cdot & \vphantom{0}\cdot & \vphantom{0}\cdot \\
 \vphantom{0}\cdot & \vphantom{0}\cdot & \vphantom{0}\cdot & \vphantom{0}\cdot & \vphantom{0}\cdot & \vphantom{0}\cdot & \vphantom{0}\cdot & \vphantom{0}\cdot & 1 & \vphantom{0}\cdot & \vphantom{0}\cdot & \vphantom{0}\cdot & \vphantom{0}\cdot & \vphantom{0}\cdot & \vphantom{0}\cdot & \vphantom{0}\cdot & \vphantom{0}\cdot & \vphantom{0}\cdot & \vphantom{0}\cdot & \vphantom{0}\cdot & 1 & \vphantom{0}\cdot & \vphantom{0}\cdot & \vphantom{0}\cdot & \vphantom{0}\cdot & \vphantom{0}\cdot & \vphantom{0}\cdot & \vphantom{0}\cdot & \vphantom{0}\cdot \\
 \vphantom{0}\cdot & \vphantom{0}\cdot & \vphantom{0}\cdot & \vphantom{0}\cdot & \vphantom{0}\cdot & \vphantom{0}\cdot & \vphantom{0}\cdot & \vphantom{0}\cdot & \vphantom{0}\cdot & \vphantom{0}\cdot & \vphantom{0}\cdot & \vphantom{0}\cdot & \vphantom{0}\cdot & \vphantom{0}\cdot & \vphantom{0}\cdot & \vphantom{0}\cdot & \vphantom{0}\cdot & \vphantom{0}\cdot & \vphantom{0}\cdot & \vphantom{0}\cdot & \vphantom{0}\cdot & \vphantom{0}\cdot & \vphantom{0}\cdot & \vphantom{0}\cdot & \vphantom{0}\cdot & \vphantom{0}\cdot & \vphantom{0}\cdot & \vphantom{0}\cdot & \vphantom{0}\cdot \\
 \vphantom{0}\cdot & \vphantom{0}\cdot & \vphantom{0}\cdot & \vphantom{0}\cdot & \vphantom{0}\cdot & \vphantom{0}\cdot & \vphantom{0}\cdot & \vphantom{0}\cdot & \vphantom{0}\cdot & \vphantom{0}\cdot & 1 & \vphantom{0}\cdot & \vphantom{0}\cdot & \vphantom{0}\cdot & \vphantom{0}\cdot & \vphantom{0}\cdot & \vphantom{0}\cdot & \vphantom{0}\cdot & 1 & \vphantom{0}\cdot & \vphantom{0}\cdot & \vphantom{0}\cdot & \vphantom{0}\cdot & \vphantom{0}\cdot & \vphantom{0}\cdot & \vphantom{0}\cdot & \vphantom{0}\cdot & \vphantom{0}\cdot & \vphantom{0}\cdot \\
 \vphantom{0}\cdot & \vphantom{0}\cdot & \vphantom{0}\cdot & \vphantom{0}\cdot & \vphantom{0}\cdot & \vphantom{0}\cdot & \vphantom{0}\cdot & \vphantom{0}\cdot & \vphantom{0}\cdot & \vphantom{0}\cdot & \vphantom{0}\cdot & \vphantom{0}\cdot & \vphantom{0}\cdot & \vphantom{0}\cdot & \vphantom{0}\cdot & \vphantom{0}\cdot & \vphantom{0}\cdot & \vphantom{0}\cdot & \vphantom{0}\cdot & \vphantom{0}\cdot & \vphantom{0}\cdot & \vphantom{0}\cdot & \vphantom{0}\cdot & \vphantom{0}\cdot & \vphantom{0}\cdot & \vphantom{0}\cdot & \vphantom{0}\cdot & \vphantom{0}\cdot & \vphantom{0}\cdot \\
 \vphantom{0}\cdot & \vphantom{0}\cdot & \vphantom{0}\cdot & \vphantom{0}\cdot & \vphantom{0}\cdot & \vphantom{0}\cdot & \vphantom{0}\cdot & \vphantom{0}\cdot & \vphantom{0}\cdot & \vphantom{0}\cdot & \vphantom{0}\cdot & \vphantom{0}\cdot & 1 & \vphantom{0}\cdot & \vphantom{0}\cdot & \vphantom{0}\cdot & 1 & \vphantom{0}\cdot & \vphantom{0}\cdot & \vphantom{0}\cdot & \vphantom{0}\cdot & \vphantom{0}\cdot & \vphantom{0}\cdot & \vphantom{0}\cdot & \vphantom{0}\cdot & \vphantom{0}\cdot & \vphantom{0}\cdot & \vphantom{0}\cdot & \vphantom{0}\cdot \\
 \vphantom{0}\cdot & \vphantom{0}\cdot & \vphantom{0}\cdot & \vphantom{0}\cdot & \vphantom{0}\cdot & \vphantom{0}\cdot & \vphantom{0}\cdot & \vphantom{0}\cdot & \vphantom{0}\cdot & \vphantom{0}\cdot & \vphantom{0}\cdot & \vphantom{0}\cdot & \vphantom{0}\cdot & \vphantom{0}\cdot & \vphantom{0}\cdot & \vphantom{0}\cdot & \vphantom{0}\cdot & \vphantom{0}\cdot & \vphantom{0}\cdot & \vphantom{0}\cdot & \vphantom{0}\cdot & \vphantom{0}\cdot & \vphantom{0}\cdot & \vphantom{0}\cdot & \vphantom{0}\cdot & \vphantom{0}\cdot & \vphantom{0}\cdot & \vphantom{0}\cdot & \vphantom{0}\cdot \\
 \vphantom{0}\cdot & \vphantom{0}\cdot & \vphantom{0}\cdot & \vphantom{0}\cdot & \vphantom{0}\cdot & \vphantom{0}\cdot & \vphantom{0}\cdot & \vphantom{0}\cdot & \vphantom{0}\cdot & \vphantom{0}\cdot & \vphantom{0}\cdot & \vphantom{0}\cdot & \vphantom{0}\cdot & \vphantom{0}\cdot & 2 & \vphantom{0}\cdot & \vphantom{0}\cdot & \vphantom{0}\cdot & \vphantom{0}\cdot & \vphantom{0}\cdot & \vphantom{0}\cdot & \vphantom{0}\cdot & \vphantom{0}\cdot & \vphantom{0}\cdot & \vphantom{0}\cdot & \vphantom{0}\cdot & \vphantom{0}\cdot & \vphantom{0}\cdot & \vphantom{0}\cdot \\
 \vphantom{0}\cdot & \vphantom{0}\cdot & \vphantom{0}\cdot & \vphantom{0}\cdot & \vphantom{0}\cdot & \vphantom{0}\cdot & \vphantom{0}\cdot & \vphantom{0}\cdot & \vphantom{0}\cdot & \vphantom{0}\cdot & \vphantom{0}\cdot & \vphantom{0}\cdot & \vphantom{0}\cdot & \vphantom{0}\cdot & \vphantom{0}\cdot & \vphantom{0}\cdot & \vphantom{0}\cdot & \vphantom{0}\cdot & \vphantom{0}\cdot & \vphantom{0}\cdot & \vphantom{0}\cdot & \vphantom{0}\cdot & \vphantom{0}\cdot & \vphantom{0}\cdot & \vphantom{0}\cdot & \vphantom{0}\cdot & \vphantom{0}\cdot & \vphantom{0}\cdot & \vphantom{0}\cdot \\
 \vphantom{0}\cdot & \vphantom{0}\cdot & \vphantom{0}\cdot & \vphantom{0}\cdot & \vphantom{0}\cdot & \vphantom{0}\cdot & \vphantom{0}\cdot & \vphantom{0}\cdot & \vphantom{0}\cdot & \vphantom{0}\cdot & \vphantom{0}\cdot & \vphantom{0}\cdot & 1 & \vphantom{0}\cdot & \vphantom{0}\cdot & \vphantom{0}\cdot & 1 & \vphantom{0}\cdot & \vphantom{0}\cdot & \vphantom{0}\cdot & \vphantom{0}\cdot & \vphantom{0}\cdot & \vphantom{0}\cdot & \vphantom{0}\cdot & \vphantom{0}\cdot & \vphantom{0}\cdot & \vphantom{0}\cdot & \vphantom{0}\cdot & \vphantom{0}\cdot \\
 \vphantom{0}\cdot & \vphantom{0}\cdot & \vphantom{0}\cdot & \vphantom{0}\cdot & \vphantom{0}\cdot & \vphantom{0}\cdot & \vphantom{0}\cdot & \vphantom{0}\cdot & \vphantom{0}\cdot & \vphantom{0}\cdot & \vphantom{0}\cdot & \vphantom{0}\cdot & \vphantom{0}\cdot & \vphantom{0}\cdot & \vphantom{0}\cdot & \vphantom{0}\cdot & \vphantom{0}\cdot & \vphantom{0}\cdot & \vphantom{0}\cdot & \vphantom{0}\cdot & \vphantom{0}\cdot & \vphantom{0}\cdot & \vphantom{0}\cdot & \vphantom{0}\cdot & \vphantom{0}\cdot & \vphantom{0}\cdot & \vphantom{0}\cdot & \vphantom{0}\cdot & \vphantom{0}\cdot \\
 \vphantom{0}\cdot & \vphantom{0}\cdot & \vphantom{0}\cdot & \vphantom{0}\cdot & \vphantom{0}\cdot & \vphantom{0}\cdot & \vphantom{0}\cdot & \vphantom{0}\cdot & \vphantom{0}\cdot & \vphantom{0}\cdot & 1 & \vphantom{0}\cdot & \vphantom{0}\cdot & \vphantom{0}\cdot & \vphantom{0}\cdot & \vphantom{0}\cdot & \vphantom{0}\cdot & \vphantom{0}\cdot & 1 & \vphantom{0}\cdot & \vphantom{0}\cdot & \vphantom{0}\cdot & \vphantom{0}\cdot & \vphantom{0}\cdot & \vphantom{0}\cdot & \vphantom{0}\cdot & \vphantom{0}\cdot & \vphantom{0}\cdot & \vphantom{0}\cdot \\
 \vphantom{0}\cdot & \vphantom{0}\cdot & \vphantom{0}\cdot & \vphantom{0}\cdot & \vphantom{0}\cdot & \vphantom{0}\cdot & \vphantom{0}\cdot & \vphantom{0}\cdot & \vphantom{0}\cdot & \vphantom{0}\cdot & \vphantom{0}\cdot & \vphantom{0}\cdot & \vphantom{0}\cdot & \vphantom{0}\cdot & \vphantom{0}\cdot & \vphantom{0}\cdot & \vphantom{0}\cdot & \vphantom{0}\cdot & \vphantom{0}\cdot & \vphantom{0}\cdot & \vphantom{0}\cdot & \vphantom{0}\cdot & \vphantom{0}\cdot & \vphantom{0}\cdot & \vphantom{0}\cdot & \vphantom{0}\cdot & \vphantom{0}\cdot & \vphantom{0}\cdot & \vphantom{0}\cdot \\
 \vphantom{0}\cdot & \vphantom{0}\cdot & \vphantom{0}\cdot & \vphantom{0}\cdot & \vphantom{0}\cdot & \vphantom{0}\cdot & \vphantom{0}\cdot & \vphantom{0}\cdot & 1 & \vphantom{0}\cdot & \vphantom{0}\cdot & \vphantom{0}\cdot & \vphantom{0}\cdot & \vphantom{0}\cdot & \vphantom{0}\cdot & \vphantom{0}\cdot & \vphantom{0}\cdot & \vphantom{0}\cdot & \vphantom{0}\cdot & \vphantom{0}\cdot & 1 & \vphantom{0}\cdot & \vphantom{0}\cdot & \vphantom{0}\cdot & \vphantom{0}\cdot & \vphantom{0}\cdot & \vphantom{0}\cdot & \vphantom{0}\cdot & \vphantom{0}\cdot \\
 \vphantom{0}\cdot & \vphantom{0}\cdot & \vphantom{0}\cdot & \vphantom{0}\cdot & \vphantom{0}\cdot & \vphantom{0}\cdot & \vphantom{0}\cdot & \vphantom{0}\cdot & \vphantom{0}\cdot & \vphantom{0}\cdot & \vphantom{0}\cdot & \vphantom{0}\cdot & \vphantom{0}\cdot & \vphantom{0}\cdot & \vphantom{0}\cdot & \vphantom{0}\cdot & \vphantom{0}\cdot & \vphantom{0}\cdot & \vphantom{0}\cdot & \vphantom{0}\cdot & \vphantom{0}\cdot & \vphantom{0}\cdot & \vphantom{0}\cdot & \vphantom{0}\cdot & \vphantom{0}\cdot & \vphantom{0}\cdot & \vphantom{0}\cdot & \vphantom{0}\cdot & \vphantom{0}\cdot \\
 \vphantom{0}\cdot & \vphantom{0}\cdot & \vphantom{0}\cdot & \vphantom{0}\cdot & \vphantom{0}\cdot & \vphantom{0}\cdot & 1 & \vphantom{0}\cdot & \vphantom{0}\cdot & \vphantom{0}\cdot & \vphantom{0}\cdot & \vphantom{0}\cdot & \vphantom{0}\cdot & \vphantom{0}\cdot & \vphantom{0}\cdot & \vphantom{0}\cdot & \vphantom{0}\cdot & \vphantom{0}\cdot & \vphantom{0}\cdot & \vphantom{0}\cdot & \vphantom{0}\cdot & \vphantom{0}\cdot & 1 & \vphantom{0}\cdot & \vphantom{0}\cdot & \vphantom{0}\cdot & \vphantom{0}\cdot & \vphantom{0}\cdot & \vphantom{0}\cdot \\
 \vphantom{0}\cdot & \vphantom{0}\cdot & \vphantom{0}\cdot & \vphantom{0}\cdot & \vphantom{0}\cdot & \vphantom{0}\cdot & \vphantom{0}\cdot & \vphantom{0}\cdot & \vphantom{0}\cdot & \vphantom{0}\cdot & \vphantom{0}\cdot & \vphantom{0}\cdot & \vphantom{0}\cdot & \vphantom{0}\cdot & \vphantom{0}\cdot & \vphantom{0}\cdot & \vphantom{0}\cdot & \vphantom{0}\cdot & \vphantom{0}\cdot & \vphantom{0}\cdot & \vphantom{0}\cdot & \vphantom{0}\cdot & \vphantom{0}\cdot & \vphantom{0}\cdot & \vphantom{0}\cdot & \vphantom{0}\cdot & \vphantom{0}\cdot & \vphantom{0}\cdot & \vphantom{0}\cdot \\
 \vphantom{0}\cdot & \vphantom{0}\cdot & \vphantom{0}\cdot & \vphantom{0}\cdot & 1 & \vphantom{0}\cdot & \vphantom{0}\cdot & \vphantom{0}\cdot & \vphantom{0}\cdot & \vphantom{0}\cdot & \vphantom{0}\cdot & \vphantom{0}\cdot & \vphantom{0}\cdot & \vphantom{0}\cdot & \vphantom{0}\cdot & \vphantom{0}\cdot & \vphantom{0}\cdot & \vphantom{0}\cdot & \vphantom{0}\cdot & \vphantom{0}\cdot & \vphantom{0}\cdot & \vphantom{0}\cdot & \vphantom{0}\cdot & \vphantom{0}\cdot & 1 & \vphantom{0}\cdot & \vphantom{0}\cdot & \vphantom{0}\cdot & \vphantom{0}\cdot \\
 \vphantom{0}\cdot & \vphantom{0}\cdot & \vphantom{0}\cdot & \vphantom{0}\cdot & \vphantom{0}\cdot & \vphantom{0}\cdot & \vphantom{0}\cdot & \vphantom{0}\cdot & \vphantom{0}\cdot & \vphantom{0}\cdot & \vphantom{0}\cdot & \vphantom{0}\cdot & \vphantom{0}\cdot & \vphantom{0}\cdot & \vphantom{0}\cdot & \vphantom{0}\cdot & \vphantom{0}\cdot & \vphantom{0}\cdot & \vphantom{0}\cdot & \vphantom{0}\cdot & \vphantom{0}\cdot & \vphantom{0}\cdot & \vphantom{0}\cdot & \vphantom{0}\cdot & \vphantom{0}\cdot & \vphantom{0}\cdot & \vphantom{0}\cdot & \vphantom{0}\cdot & \vphantom{0}\cdot \\
 \vphantom{0}\cdot & \vphantom{0}\cdot & 1 & \vphantom{0}\cdot & \vphantom{0}\cdot & \vphantom{0}\cdot & \vphantom{0}\cdot & \vphantom{0}\cdot & \vphantom{0}\cdot & \vphantom{0}\cdot & \vphantom{0}\cdot & \vphantom{0}\cdot & \vphantom{0}\cdot & \vphantom{0}\cdot & \vphantom{0}\cdot & \vphantom{0}\cdot & \vphantom{0}\cdot & \vphantom{0}\cdot & \vphantom{0}\cdot & \vphantom{0}\cdot & \vphantom{0}\cdot & \vphantom{0}\cdot & \vphantom{0}\cdot & \vphantom{0}\cdot & \vphantom{0}\cdot & \vphantom{0}\cdot & 1 & \vphantom{0}\cdot & \vphantom{0}\cdot \\
 \vphantom{0}\cdot & \vphantom{0}\cdot & \vphantom{0}\cdot & \vphantom{0}\cdot & \vphantom{0}\cdot & \vphantom{0}\cdot & \vphantom{0}\cdot & \vphantom{0}\cdot & \vphantom{0}\cdot & \vphantom{0}\cdot & \vphantom{0}\cdot & \vphantom{0}\cdot & \vphantom{0}\cdot & \vphantom{0}\cdot & \vphantom{0}\cdot & \vphantom{0}\cdot & \vphantom{0}\cdot & \vphantom{0}\cdot & \vphantom{0}\cdot & \vphantom{0}\cdot & \vphantom{0}\cdot & \vphantom{0}\cdot & \vphantom{0}\cdot & \vphantom{0}\cdot & \vphantom{0}\cdot & \vphantom{0}\cdot & \vphantom{0}\cdot & \vphantom{0}\cdot & \vphantom{0}\cdot \\
 1 & \vphantom{0}\cdot & \vphantom{0}\cdot & \vphantom{0}\cdot & \vphantom{0}\cdot & \vphantom{0}\cdot & \vphantom{0}\cdot & \vphantom{0}\cdot & \vphantom{0}\cdot & \vphantom{0}\cdot & \vphantom{0}\cdot & \vphantom{0}\cdot & \vphantom{0}\cdot & \vphantom{0}\cdot & \vphantom{0}\cdot & \vphantom{0}\cdot & \vphantom{0}\cdot & \vphantom{0}\cdot & \vphantom{0}\cdot & \vphantom{0}\cdot & \vphantom{0}\cdot & \vphantom{0}\cdot & \vphantom{0}\cdot & \vphantom{0}\cdot & \vphantom{0}\cdot & \vphantom{0}\cdot & \vphantom{0}\cdot & \vphantom{0}\cdot & 1 \\
\end{smallmatrix}
\right)
\end{equation}
\begin{equation}
W_2= 
\left(
\begin{smallmatrix}
 1 & \vphantom{0}\cdot & \vphantom{0}\cdot & \vphantom{0}\cdot & \vphantom{0}\cdot & \vphantom{0}\cdot & \vphantom{0}\cdot & \vphantom{0}\cdot & \vphantom{0}\cdot & \vphantom{0}\cdot & \textcolor{red}{1} & \vphantom{0}\cdot & \vphantom{0}\cdot & \vphantom{0}\cdot & \vphantom{0}\cdot & \vphantom{0}\cdot & \vphantom{0}\cdot & \vphantom{0}\cdot & 1 & \vphantom{0}\cdot & \vphantom{0}\cdot & \vphantom{0}\cdot & \vphantom{0}\cdot & \vphantom{0}\cdot & \vphantom{0}\cdot & \vphantom{0}\cdot & \vphantom{0}\cdot & \vphantom{0}\cdot & 1 \\
 \vphantom{0}\cdot & \vphantom{0}\cdot & \vphantom{0}\cdot & \vphantom{0}\cdot & \vphantom{0}\cdot & \vphantom{0}\cdot & \vphantom{0}\cdot & \vphantom{0}\cdot & \vphantom{0}\cdot & \vphantom{0}\cdot & \vphantom{0}\cdot & \vphantom{0}\cdot & \vphantom{0}\cdot & \vphantom{0}\cdot & \vphantom{0}\cdot & \vphantom{0}\cdot & \vphantom{0}\cdot & \vphantom{0}\cdot & \vphantom{0}\cdot & \vphantom{0}\cdot & \vphantom{0}\cdot & \vphantom{0}\cdot & \vphantom{0}\cdot & \vphantom{0}\cdot & \vphantom{0}\cdot & \vphantom{0}\cdot & \vphantom{0}\cdot & \vphantom{0}\cdot & \vphantom{0}\cdot \\
 \vphantom{0}\cdot & \vphantom{0}\cdot & \vphantom{0}\cdot & \vphantom{0}\cdot & \vphantom{0}\cdot & \vphantom{0}\cdot & \vphantom{0}\cdot & \vphantom{0}\cdot & \vphantom{0}\cdot & \vphantom{0}\cdot & \vphantom{0}\cdot & \vphantom{0}\cdot & \vphantom{0}\cdot & \vphantom{0}\cdot & \vphantom{0}\cdot & \vphantom{0}\cdot & \vphantom{0}\cdot & \vphantom{0}\cdot & \vphantom{0}\cdot & \vphantom{0}\cdot & \vphantom{0}\cdot & \vphantom{0}\cdot & \vphantom{0}\cdot & \vphantom{0}\cdot & \vphantom{0}\cdot & \vphantom{0}\cdot & \vphantom{0}\cdot & \vphantom{0}\cdot & \vphantom{0}\cdot \\
 \vphantom{0}\cdot & \vphantom{0}\cdot & \vphantom{0}\cdot & \vphantom{0}\cdot & \vphantom{0}\cdot & \vphantom{0}\cdot & \vphantom{0}\cdot & \vphantom{0}\cdot & \vphantom{0}\cdot & \vphantom{0}\cdot & \vphantom{0}\cdot & \vphantom{0}\cdot & \vphantom{0}\cdot & \vphantom{0}\cdot & \vphantom{0}\cdot & \vphantom{0}\cdot & \vphantom{0}\cdot & \vphantom{0}\cdot & \vphantom{0}\cdot & \vphantom{0}\cdot & \vphantom{0}\cdot & \vphantom{0}\cdot & \vphantom{0}\cdot & \vphantom{0}\cdot & \vphantom{0}\cdot & \vphantom{0}\cdot & \vphantom{0}\cdot & \vphantom{0}\cdot & \vphantom{0}\cdot \\
 \vphantom{0}\cdot & \vphantom{0}\cdot & \vphantom{0}\cdot & \vphantom{0}\cdot & \vphantom{0}\cdot & \vphantom{0}\cdot & \vphantom{0}\cdot & \vphantom{0}\cdot & \vphantom{0}\cdot & \vphantom{0}\cdot & \vphantom{0}\cdot & \vphantom{0}\cdot & \vphantom{0}\cdot & \vphantom{0}\cdot & \vphantom{0}\cdot & \vphantom{0}\cdot & \vphantom{0}\cdot & \vphantom{0}\cdot & \vphantom{0}\cdot & \vphantom{0}\cdot & \vphantom{0}\cdot & \vphantom{0}\cdot & \vphantom{0}\cdot & \vphantom{0}\cdot & \vphantom{0}\cdot & \vphantom{0}\cdot & \vphantom{0}\cdot & \vphantom{0}\cdot & \vphantom{0}\cdot \\
 \vphantom{0}\cdot & \vphantom{0}\cdot & \vphantom{0}\cdot & \vphantom{0}\cdot & \vphantom{0}\cdot & \vphantom{0}\cdot & \vphantom{0}\cdot & \vphantom{0}\cdot & \vphantom{0}\cdot & \vphantom{0}\cdot & \vphantom{0}\cdot & \vphantom{0}\cdot & \vphantom{0}\cdot & \vphantom{0}\cdot & \vphantom{0}\cdot & \vphantom{0}\cdot & \vphantom{0}\cdot & \vphantom{0}\cdot & \vphantom{0}\cdot & \vphantom{0}\cdot & \vphantom{0}\cdot & \vphantom{0}\cdot & \vphantom{0}\cdot & \vphantom{0}\cdot & \vphantom{0}\cdot & \vphantom{0}\cdot & \vphantom{0}\cdot & \vphantom{0}\cdot & \vphantom{0}\cdot \\
 \vphantom{0}\cdot & \vphantom{0}\cdot & \vphantom{0}\cdot & \vphantom{0}\cdot & \vphantom{0}\cdot & \vphantom{0}\cdot & 1 & \vphantom{0}\cdot & \vphantom{0}\cdot & \vphantom{0}\cdot & \vphantom{0}\cdot & \vphantom{0}\cdot & 1 & \vphantom{0}\cdot & \vphantom{0}\cdot & \vphantom{0}\cdot & 1 & \vphantom{0}\cdot & \vphantom{0}\cdot & \vphantom{0}\cdot & \vphantom{0}\cdot & \vphantom{0}\cdot & 1 & \vphantom{0}\cdot & \vphantom{0}\cdot & \vphantom{0}\cdot & \vphantom{0}\cdot & \vphantom{0}\cdot & \vphantom{0}\cdot \\
 \vphantom{0}\cdot & \vphantom{0}\cdot & \vphantom{0}\cdot & \vphantom{0}\cdot & \vphantom{0}\cdot & \vphantom{0}\cdot & \vphantom{0}\cdot & \vphantom{0}\cdot & \vphantom{0}\cdot & \vphantom{0}\cdot & \vphantom{0}\cdot & \vphantom{0}\cdot & \vphantom{0}\cdot & \vphantom{0}\cdot & \vphantom{0}\cdot & \vphantom{0}\cdot & \vphantom{0}\cdot & \vphantom{0}\cdot & \vphantom{0}\cdot & \vphantom{0}\cdot & \vphantom{0}\cdot & \vphantom{0}\cdot & \vphantom{0}\cdot & \vphantom{0}\cdot & \vphantom{0}\cdot & \vphantom{0}\cdot & \vphantom{0}\cdot & \vphantom{0}\cdot & \vphantom{0}\cdot \\
 \vphantom{0}\cdot & \vphantom{0}\cdot & \vphantom{0}\cdot & \vphantom{0}\cdot & \vphantom{0}\cdot & \vphantom{0}\cdot & \vphantom{0}\cdot & \vphantom{0}\cdot & \vphantom{0}\cdot & \vphantom{0}\cdot & \vphantom{0}\cdot & \vphantom{0}\cdot & \vphantom{0}\cdot & \vphantom{0}\cdot & \vphantom{0}\cdot & \vphantom{0}\cdot & \vphantom{0}\cdot & \vphantom{0}\cdot & \vphantom{0}\cdot & \vphantom{0}\cdot & \vphantom{0}\cdot & \vphantom{0}\cdot & \vphantom{0}\cdot & \vphantom{0}\cdot & \vphantom{0}\cdot & \vphantom{0}\cdot & \vphantom{0}\cdot & \vphantom{0}\cdot & \vphantom{0}\cdot \\
 \vphantom{0}\cdot & \vphantom{0}\cdot & \vphantom{0}\cdot & \vphantom{0}\cdot & \vphantom{0}\cdot & \vphantom{0}\cdot & \vphantom{0}\cdot & \vphantom{0}\cdot & \vphantom{0}\cdot & \vphantom{0}\cdot & \vphantom{0}\cdot & \vphantom{0}\cdot & \vphantom{0}\cdot & \vphantom{0}\cdot & \vphantom{0}\cdot & \vphantom{0}\cdot & \vphantom{0}\cdot & \vphantom{0}\cdot & \vphantom{0}\cdot & \vphantom{0}\cdot & \vphantom{0}\cdot & \vphantom{0}\cdot & \vphantom{0}\cdot & \vphantom{0}\cdot & \vphantom{0}\cdot & \vphantom{0}\cdot & \vphantom{0}\cdot & \vphantom{0}\cdot & \vphantom{0}\cdot \\
 \textcolor{red}{1} & \vphantom{0}\cdot & \vphantom{0}\cdot & \vphantom{0}\cdot & \vphantom{0}\cdot & \vphantom{0}\cdot & \vphantom{0}\cdot & \vphantom{0}\cdot & \vphantom{0}\cdot & \vphantom{0}\cdot & 1 & \vphantom{0}\cdot & \vphantom{0}\cdot & \vphantom{0}\cdot & \vphantom{0}\cdot & \vphantom{0}\cdot & \vphantom{0}\cdot & \vphantom{0}\cdot & 1 & \vphantom{0}\cdot & \vphantom{0}\cdot & \vphantom{0}\cdot & \vphantom{0}\cdot & \vphantom{0}\cdot & \vphantom{0}\cdot & \vphantom{0}\cdot & \vphantom{0}\cdot & \vphantom{0}\cdot & 1 \\
 \vphantom{0}\cdot & \vphantom{0}\cdot & \vphantom{0}\cdot & \vphantom{0}\cdot & \vphantom{0}\cdot & \vphantom{0}\cdot & \vphantom{0}\cdot & \vphantom{0}\cdot & \vphantom{0}\cdot & \vphantom{0}\cdot & \vphantom{0}\cdot & \vphantom{0}\cdot & \vphantom{0}\cdot & \vphantom{0}\cdot & \vphantom{0}\cdot & \vphantom{0}\cdot & \vphantom{0}\cdot & \vphantom{0}\cdot & \vphantom{0}\cdot & \vphantom{0}\cdot & \vphantom{0}\cdot & \vphantom{0}\cdot & \vphantom{0}\cdot & \vphantom{0}\cdot & \vphantom{0}\cdot & \vphantom{0}\cdot & \vphantom{0}\cdot & \vphantom{0}\cdot & \vphantom{0}\cdot \\
 \vphantom{0}\cdot & \vphantom{0}\cdot & \vphantom{0}\cdot & \vphantom{0}\cdot & \vphantom{0}\cdot & \vphantom{0}\cdot & 1 & \vphantom{0}\cdot & \vphantom{0}\cdot & \vphantom{0}\cdot & \vphantom{0}\cdot & \vphantom{0}\cdot & 1 & \vphantom{0}\cdot & \vphantom{0}\cdot & \vphantom{0}\cdot & 1 & \vphantom{0}\cdot & \vphantom{0}\cdot & \vphantom{0}\cdot & \vphantom{0}\cdot & \vphantom{0}\cdot & 1 & \vphantom{0}\cdot & \vphantom{0}\cdot & \vphantom{0}\cdot & \vphantom{0}\cdot & \vphantom{0}\cdot & \vphantom{0}\cdot \\
 \vphantom{0}\cdot & \vphantom{0}\cdot & \vphantom{0}\cdot & \vphantom{0}\cdot & \vphantom{0}\cdot & \vphantom{0}\cdot & \vphantom{0}\cdot & \vphantom{0}\cdot & \vphantom{0}\cdot & \vphantom{0}\cdot & \vphantom{0}\cdot & \vphantom{0}\cdot & \vphantom{0}\cdot & \vphantom{0}\cdot & \vphantom{0}\cdot & \vphantom{0}\cdot & \vphantom{0}\cdot & \vphantom{0}\cdot & \vphantom{0}\cdot & \vphantom{0}\cdot & \vphantom{0}\cdot & \vphantom{0}\cdot & \vphantom{0}\cdot & \vphantom{0}\cdot & \vphantom{0}\cdot & \vphantom{0}\cdot & \vphantom{0}\cdot & \vphantom{0}\cdot & \vphantom{0}\cdot \\
 \vphantom{0}\cdot & \vphantom{0}\cdot & \vphantom{0}\cdot & \vphantom{0}\cdot & \vphantom{0}\cdot & \vphantom{0}\cdot & \vphantom{0}\cdot & \vphantom{0}\cdot & \vphantom{0}\cdot & \vphantom{0}\cdot & \vphantom{0}\cdot & \vphantom{0}\cdot & \vphantom{0}\cdot & \vphantom{0}\cdot & \vphantom{0}\cdot & \vphantom{0}\cdot & \vphantom{0}\cdot & \vphantom{0}\cdot & \vphantom{0}\cdot & \vphantom{0}\cdot & \vphantom{0}\cdot & \vphantom{0}\cdot & \vphantom{0}\cdot & \vphantom{0}\cdot & \vphantom{0}\cdot & \vphantom{0}\cdot & \vphantom{0}\cdot & \vphantom{0}\cdot & \vphantom{0}\cdot \\
 \vphantom{0}\cdot & \vphantom{0}\cdot & \vphantom{0}\cdot & \vphantom{0}\cdot & \vphantom{0}\cdot & \vphantom{0}\cdot & \vphantom{0}\cdot & \vphantom{0}\cdot & \vphantom{0}\cdot & \vphantom{0}\cdot & \vphantom{0}\cdot & \vphantom{0}\cdot & \vphantom{0}\cdot & \vphantom{0}\cdot & \vphantom{0}\cdot & \vphantom{0}\cdot & \vphantom{0}\cdot & \vphantom{0}\cdot & \vphantom{0}\cdot & \vphantom{0}\cdot & \vphantom{0}\cdot & \vphantom{0}\cdot & \vphantom{0}\cdot & \vphantom{0}\cdot & \vphantom{0}\cdot & \vphantom{0}\cdot & \vphantom{0}\cdot & \vphantom{0}\cdot & \vphantom{0}\cdot \\
 \vphantom{0}\cdot & \vphantom{0}\cdot & \vphantom{0}\cdot & \vphantom{0}\cdot & \vphantom{0}\cdot & \vphantom{0}\cdot & 1 & \vphantom{0}\cdot & \vphantom{0}\cdot & \vphantom{0}\cdot & \vphantom{0}\cdot & \vphantom{0}\cdot & 1 & \vphantom{0}\cdot & \vphantom{0}\cdot & \vphantom{0}\cdot & 1 & \vphantom{0}\cdot & \vphantom{0}\cdot & \vphantom{0}\cdot & \vphantom{0}\cdot & \vphantom{0}\cdot & 1 & \vphantom{0}\cdot & \vphantom{0}\cdot & \vphantom{0}\cdot & \vphantom{0}\cdot & \vphantom{0}\cdot & \vphantom{0}\cdot \\
 \vphantom{0}\cdot & \vphantom{0}\cdot & \vphantom{0}\cdot & \vphantom{0}\cdot & \vphantom{0}\cdot & \vphantom{0}\cdot & \vphantom{0}\cdot & \vphantom{0}\cdot & \vphantom{0}\cdot & \vphantom{0}\cdot & \vphantom{0}\cdot & \vphantom{0}\cdot & \vphantom{0}\cdot & \vphantom{0}\cdot & \vphantom{0}\cdot & \vphantom{0}\cdot & \vphantom{0}\cdot & \vphantom{0}\cdot & \vphantom{0}\cdot & \vphantom{0}\cdot & \vphantom{0}\cdot & \vphantom{0}\cdot & \vphantom{0}\cdot & \vphantom{0}\cdot & \vphantom{0}\cdot & \vphantom{0}\cdot & \vphantom{0}\cdot & \vphantom{0}\cdot & \vphantom{0}\cdot \\
 1 & \vphantom{0}\cdot & \vphantom{0}\cdot & \vphantom{0}\cdot & \vphantom{0}\cdot & \vphantom{0}\cdot & \vphantom{0}\cdot & \vphantom{0}\cdot & \vphantom{0}\cdot & \vphantom{0}\cdot & 1 & \vphantom{0}\cdot & \vphantom{0}\cdot & \vphantom{0}\cdot & \vphantom{0}\cdot & \vphantom{0}\cdot & \vphantom{0}\cdot & \vphantom{0}\cdot & 1 & \vphantom{0}\cdot & \vphantom{0}\cdot & \vphantom{0}\cdot & \vphantom{0}\cdot & \vphantom{0}\cdot & \vphantom{0}\cdot & \vphantom{0}\cdot & \vphantom{0}\cdot & \vphantom{0}\cdot & 1 \\
 \vphantom{0}\cdot & \vphantom{0}\cdot & \vphantom{0}\cdot & \vphantom{0}\cdot & \vphantom{0}\cdot & \vphantom{0}\cdot & \vphantom{0}\cdot & \vphantom{0}\cdot & \vphantom{0}\cdot & \vphantom{0}\cdot & \vphantom{0}\cdot & \vphantom{0}\cdot & \vphantom{0}\cdot & \vphantom{0}\cdot & \vphantom{0}\cdot & \vphantom{0}\cdot & \vphantom{0}\cdot & \vphantom{0}\cdot & \vphantom{0}\cdot & \vphantom{0}\cdot & \vphantom{0}\cdot & \vphantom{0}\cdot & \vphantom{0}\cdot & \vphantom{0}\cdot & \vphantom{0}\cdot & \vphantom{0}\cdot & \vphantom{0}\cdot & \vphantom{0}\cdot & \vphantom{0}\cdot \\
 \vphantom{0}\cdot & \vphantom{0}\cdot & \vphantom{0}\cdot & \vphantom{0}\cdot & \vphantom{0}\cdot & \vphantom{0}\cdot & \vphantom{0}\cdot & \vphantom{0}\cdot & \vphantom{0}\cdot & \vphantom{0}\cdot & \vphantom{0}\cdot & \vphantom{0}\cdot & \vphantom{0}\cdot & \vphantom{0}\cdot & \vphantom{0}\cdot & \vphantom{0}\cdot & \vphantom{0}\cdot & \vphantom{0}\cdot & \vphantom{0}\cdot & \vphantom{0}\cdot & \vphantom{0}\cdot & \vphantom{0}\cdot & \vphantom{0}\cdot & \vphantom{0}\cdot & \vphantom{0}\cdot & \vphantom{0}\cdot & \vphantom{0}\cdot & \vphantom{0}\cdot & \vphantom{0}\cdot \\
 \vphantom{0}\cdot & \vphantom{0}\cdot & \vphantom{0}\cdot & \vphantom{0}\cdot & \vphantom{0}\cdot & \vphantom{0}\cdot & \vphantom{0}\cdot & \vphantom{0}\cdot & \vphantom{0}\cdot & \vphantom{0}\cdot & \vphantom{0}\cdot & \vphantom{0}\cdot & \vphantom{0}\cdot & \vphantom{0}\cdot & \vphantom{0}\cdot & \vphantom{0}\cdot & \vphantom{0}\cdot & \vphantom{0}\cdot & \vphantom{0}\cdot & \vphantom{0}\cdot & \vphantom{0}\cdot & \vphantom{0}\cdot & \vphantom{0}\cdot & \vphantom{0}\cdot & \vphantom{0}\cdot & \vphantom{0}\cdot & \vphantom{0}\cdot & \vphantom{0}\cdot & \vphantom{0}\cdot \\
 \vphantom{0}\cdot & \vphantom{0}\cdot & \vphantom{0}\cdot & \vphantom{0}\cdot & \vphantom{0}\cdot & \vphantom{0}\cdot & 1 & \vphantom{0}\cdot & \vphantom{0}\cdot & \vphantom{0}\cdot & \vphantom{0}\cdot & \vphantom{0}\cdot & 1 & \vphantom{0}\cdot & \vphantom{0}\cdot & \vphantom{0}\cdot & 1 & \vphantom{0}\cdot & \vphantom{0}\cdot & \vphantom{0}\cdot & \vphantom{0}\cdot & \vphantom{0}\cdot & 1 & \vphantom{0}\cdot & \vphantom{0}\cdot & \vphantom{0}\cdot & \vphantom{0}\cdot & \vphantom{0}\cdot & \vphantom{0}\cdot \\
 \vphantom{0}\cdot & \vphantom{0}\cdot & \vphantom{0}\cdot & \vphantom{0}\cdot & \vphantom{0}\cdot & \vphantom{0}\cdot & \vphantom{0}\cdot & \vphantom{0}\cdot & \vphantom{0}\cdot & \vphantom{0}\cdot & \vphantom{0}\cdot & \vphantom{0}\cdot & \vphantom{0}\cdot & \vphantom{0}\cdot & \vphantom{0}\cdot & \vphantom{0}\cdot & \vphantom{0}\cdot & \vphantom{0}\cdot & \vphantom{0}\cdot & \vphantom{0}\cdot & \vphantom{0}\cdot & \vphantom{0}\cdot & \vphantom{0}\cdot & \vphantom{0}\cdot & \vphantom{0}\cdot & \vphantom{0}\cdot & \vphantom{0}\cdot & \vphantom{0}\cdot & \vphantom{0}\cdot \\
 \vphantom{0}\cdot & \vphantom{0}\cdot & \vphantom{0}\cdot & \vphantom{0}\cdot & \vphantom{0}\cdot & \vphantom{0}\cdot & \vphantom{0}\cdot & \vphantom{0}\cdot & \vphantom{0}\cdot & \vphantom{0}\cdot & \vphantom{0}\cdot & \vphantom{0}\cdot & \vphantom{0}\cdot & \vphantom{0}\cdot & \vphantom{0}\cdot & \vphantom{0}\cdot & \vphantom{0}\cdot & \vphantom{0}\cdot & \vphantom{0}\cdot & \vphantom{0}\cdot & \vphantom{0}\cdot & \vphantom{0}\cdot & \vphantom{0}\cdot & \vphantom{0}\cdot & \vphantom{0}\cdot & \vphantom{0}\cdot & \vphantom{0}\cdot & \vphantom{0}\cdot & \vphantom{0}\cdot \\
 \vphantom{0}\cdot & \vphantom{0}\cdot & \vphantom{0}\cdot & \vphantom{0}\cdot & \vphantom{0}\cdot & \vphantom{0}\cdot & \vphantom{0}\cdot & \vphantom{0}\cdot & \vphantom{0}\cdot & \vphantom{0}\cdot & \vphantom{0}\cdot & \vphantom{0}\cdot & \vphantom{0}\cdot & \vphantom{0}\cdot & \vphantom{0}\cdot & \vphantom{0}\cdot & \vphantom{0}\cdot & \vphantom{0}\cdot & \vphantom{0}\cdot & \vphantom{0}\cdot & \vphantom{0}\cdot & \vphantom{0}\cdot & \vphantom{0}\cdot & \vphantom{0}\cdot & \vphantom{0}\cdot & \vphantom{0}\cdot & \vphantom{0}\cdot & \vphantom{0}\cdot & \vphantom{0}\cdot \\
 \vphantom{0}\cdot & \vphantom{0}\cdot & \vphantom{0}\cdot & \vphantom{0}\cdot & \vphantom{0}\cdot & \vphantom{0}\cdot & \vphantom{0}\cdot & \vphantom{0}\cdot & \vphantom{0}\cdot & \vphantom{0}\cdot & \vphantom{0}\cdot & \vphantom{0}\cdot & \vphantom{0}\cdot & \vphantom{0}\cdot & \vphantom{0}\cdot & \vphantom{0}\cdot & \vphantom{0}\cdot & \vphantom{0}\cdot & \vphantom{0}\cdot & \vphantom{0}\cdot & \vphantom{0}\cdot & \vphantom{0}\cdot & \vphantom{0}\cdot & \vphantom{0}\cdot & \vphantom{0}\cdot & \vphantom{0}\cdot & \vphantom{0}\cdot & \vphantom{0}\cdot & \vphantom{0}\cdot \\
 \vphantom{0}\cdot & \vphantom{0}\cdot & \vphantom{0}\cdot & \vphantom{0}\cdot & \vphantom{0}\cdot & \vphantom{0}\cdot & \vphantom{0}\cdot & \vphantom{0}\cdot & \vphantom{0}\cdot & \vphantom{0}\cdot & \vphantom{0}\cdot & \vphantom{0}\cdot & \vphantom{0}\cdot & \vphantom{0}\cdot & \vphantom{0}\cdot & \vphantom{0}\cdot & \vphantom{0}\cdot & \vphantom{0}\cdot & \vphantom{0}\cdot & \vphantom{0}\cdot & \vphantom{0}\cdot & \vphantom{0}\cdot & \vphantom{0}\cdot & \vphantom{0}\cdot & \vphantom{0}\cdot & \vphantom{0}\cdot & \vphantom{0}\cdot & \vphantom{0}\cdot & \vphantom{0}\cdot \\
 1 & \vphantom{0}\cdot & \vphantom{0}\cdot & \vphantom{0}\cdot & \vphantom{0}\cdot & \vphantom{0}\cdot & \vphantom{0}\cdot & \vphantom{0}\cdot & \vphantom{0}\cdot & \vphantom{0}\cdot & 1 & \vphantom{0}\cdot & \vphantom{0}\cdot & \vphantom{0}\cdot & \vphantom{0}\cdot & \vphantom{0}\cdot & \vphantom{0}\cdot & \vphantom{0}\cdot & 1 & \vphantom{0}\cdot & \vphantom{0}\cdot & \vphantom{0}\cdot & \vphantom{0}\cdot & \vphantom{0}\cdot & \vphantom{0}\cdot & \vphantom{0}\cdot & \vphantom{0}\cdot & \vphantom{0}\cdot & 1 \\
\end{smallmatrix}
\right)
\end{equation}
Since $W_{0,10}=W_{10,0}=1\neq 0$ (marked in red), the non-local current can be made local by gauging.

\bibliographystyle{JHEP}
\bibliography{ref}

\providecommand{\href}[2]{#2}\begingroup\raggedright\begin{thebibliography}{10}

\bibitem{Gaiotto:2014kfa}
D.~Gaiotto, A.~Kapustin, N.~Seiberg and B.~Willett, \emph{{Generalized Global Symmetries}}, \href{https://doi.org/10.1007/JHEP02(2015)172}{\emph{JHEP} {\bfseries 02} (2015) 172} [\href{https://arxiv.org/abs/1412.5148}{{\ttfamily 1412.5148}}].

\bibitem{Chang:2018iay}
C.-M.~Chang, Y.-H.~Lin, S.-H.~Shao, Y.~Wang and X.~Yin, \emph{{Topological Defect Lines and Renormalization Group Flows in Two Dimensions}}, \href{https://doi.org/10.1007/JHEP01(2019)026}{\emph{JHEP} {\bfseries 01} (2019) 026} [\href{https://arxiv.org/abs/1802.04445}{{\ttfamily 1802.04445}}].

\bibitem{Komargodski:2020mxz}
Z.~Komargodski, K.~Ohmori, K.~Roumpedakis and S.~Seifnashri, \emph{{Symmetries and strings of adjoint QCD$_{2}$}}, \href{https://doi.org/10.1007/JHEP03(2021)103}{\emph{JHEP} {\bfseries 03} (2021) 103} [\href{https://arxiv.org/abs/2008.07567}{{\ttfamily 2008.07567}}].

\bibitem{Thorngren:2019iar}
R.~Thorngren and Y.~Wang, \emph{{Fusion Category Symmetry I: Anomaly In-Flow and Gapped Phases}},  \href{https://arxiv.org/abs/1912.02817}{{\ttfamily 1912.02817}}.

\bibitem{Thorngren:2021yso}
R.~Thorngren and Y.~Wang, \emph{{Fusion Category Symmetry II: Categoriosities at $c$ = 1 and Beyond}},  \href{https://arxiv.org/abs/2106.12577}{{\ttfamily 2106.12577}}.

\bibitem{Bhardwaj:2017xup}
L.~Bhardwaj and Y.~Tachikawa, \emph{{On finite symmetries and their gauging in two dimensions}}, \href{https://doi.org/10.1007/JHEP03(2018)189}{\emph{JHEP} {\bfseries 03} (2018) 189} [\href{https://arxiv.org/abs/1704.02330}{{\ttfamily 1704.02330}}].

\bibitem{Kong:2020cie}
L.~Kong, T.~Lan, X.-G.~Wen, Z.-H.~Zhang and H.~Zheng, \emph{{Algebraic higher symmetry and categorical symmetry -- a holographic and entanglement view of symmetry}}, \href{https://doi.org/10.1103/PhysRevResearch.2.043086}{\emph{Phys. Rev. Res.} {\bfseries 2} (2020) 043086} [\href{https://arxiv.org/abs/2005.14178}{{\ttfamily 2005.14178}}].

\bibitem{Lin:2022dhv}
Y.-H.~Lin, M.~Okada, S.~Seifnashri and Y.~Tachikawa, \emph{{Asymptotic density of states in 2d CFTs with non-invertible symmetries}},  \href{https://arxiv.org/abs/2208.05495}{{\ttfamily 2208.05495}}.

\bibitem{Huang:2021zvu}
T.-C.~Huang, Y.-H.~Lin and S.~Seifnashri, \emph{{Construction of two-dimensional topological field theories with non-invertible symmetries}}, \href{https://doi.org/10.1007/JHEP12(2021)028}{\emph{JHEP} {\bfseries 12} (2021) 028} [\href{https://arxiv.org/abs/2110.02958}{{\ttfamily 2110.02958}}].

\bibitem{Lu:2022ver}
D.-C.~Lu and Z.~Sun, \emph{{On triality defects in 2d CFT}}, \href{https://doi.org/10.1007/JHEP02(2023)173}{\emph{JHEP} {\bfseries 02} (2023) 173} [\href{https://arxiv.org/abs/2208.06077}{{\ttfamily 2208.06077}}].

\bibitem{Aasen:2020jwb}
D.~Aasen, P.~Fendley and R.S.K.~Mong, \emph{{Topological Defects on the Lattice: Dualities and Degeneracies}},  \href{https://arxiv.org/abs/2008.08598}{{\ttfamily 2008.08598}}.

\bibitem{Lin:2023uvm}
Y.-H.~Lin and S.-H.~Shao, \emph{{Bootstrapping Non-invertible Symmetries}},  \href{https://arxiv.org/abs/2302.13900}{{\ttfamily 2302.13900}}.

\bibitem{Huang:2021nvb}
T.-C.~Huang, Y.-H.~Lin, K.~Ohmori, Y.~Tachikawa and M.~Tezuka, \emph{{Numerical Evidence for a Haagerup Conformal Field Theory}}, \href{https://doi.org/10.1103/PhysRevLett.128.231603}{\emph{Phys. Rev. Lett.} {\bfseries 128} (2022) 231603} [\href{https://arxiv.org/abs/2110.03008}{{\ttfamily 2110.03008}}].

\bibitem{Burbano:2021loy}
I.M.~Burbano, J.~Kulp and J.~Neuser, \emph{{Duality defects in E$_{8}$}}, \href{https://doi.org/10.1007/JHEP10(2022)187}{\emph{JHEP} {\bfseries 10} (2022) 186} [\href{https://arxiv.org/abs/2112.14323}{{\ttfamily 2112.14323}}].

\bibitem{Gaiotto:2020iye}
D.~Gaiotto and J.~Kulp, \emph{{Orbifold groupoids}}, \href{https://doi.org/10.1007/JHEP02(2021)132}{\emph{JHEP} {\bfseries 02} (2021) 132} [\href{https://arxiv.org/abs/2008.05960}{{\ttfamily 2008.05960}}].

\bibitem{Koide:2021zxj}
M.~Koide, Y.~Nagoya and S.~Yamaguchi, \emph{{Non-invertible topological defects in 4-dimensional $\mathbb {Z}_2$ pure lattice gauge theory}}, \href{https://doi.org/10.1093/ptep/ptab145}{\emph{PTEP} {\bfseries 2022} (2022) 013B03} [\href{https://arxiv.org/abs/2109.05992}{{\ttfamily 2109.05992}}].

\bibitem{Bhardwaj:2023ayw}
L.~Bhardwaj and S.~Schafer-Nameki, \emph{{Generalized Charges, Part II: Non-Invertible Symmetries and the Symmetry TFT}},  \href{https://arxiv.org/abs/2305.17159}{{\ttfamily 2305.17159}}.

\bibitem{Kaidi:2021xfk}
J.~Kaidi, K.~Ohmori and Y.~Zheng, \emph{{Kramers-Wannier-like Duality Defects in (3+1)D Gauge Theories}}, \href{https://doi.org/10.1103/PhysRevLett.128.111601}{\emph{Phys. Rev. Lett.} {\bfseries 128} (2022) 111601} [\href{https://arxiv.org/abs/2111.01141}{{\ttfamily 2111.01141}}].

\bibitem{Choi:2021kmx}
Y.~Choi, C.~Cordova, P.-S.~Hsin, H.T.~Lam and S.-H.~Shao, \emph{{Noninvertible duality defects in 3+1 dimensions}}, \href{https://doi.org/10.1103/PhysRevD.105.125016}{\emph{Phys. Rev. D} {\bfseries 105} (2022) 125016} [\href{https://arxiv.org/abs/2111.01139}{{\ttfamily 2111.01139}}].

\bibitem{Choi:2022fgx}
Y.~Choi, H.T.~Lam and S.-H.~Shao, \emph{{Non-invertible Gauss Law and Axions}},  \href{https://arxiv.org/abs/2212.04499}{{\ttfamily 2212.04499}}.

\bibitem{Choi:2022zal}
Y.~Choi, C.~Cordova, P.-S.~Hsin, H.T.~Lam and S.-H.~Shao, \emph{{Non-invertible Condensation, Duality, and Triality Defects in 3+1 Dimensions}},  \href{https://arxiv.org/abs/2204.09025}{{\ttfamily 2204.09025}}.

\bibitem{Inamura:2023qzl}
K.~Inamura and K.~Ohmori, \emph{{Fusion Surface Models: 2+1d Lattice Models from Fusion 2-Categories}},  \href{https://arxiv.org/abs/2305.05774}{{\ttfamily 2305.05774}}.

\bibitem{Bhardwaj:2023wzd}
L.~Bhardwaj and S.~Schafer-Nameki, \emph{{Generalized charges, part I: Invertible symmetries and higher representations}}, \href{https://doi.org/10.21468/SciPostPhys.16.4.093}{\emph{SciPost Phys.} {\bfseries 16} (2024) 093} [\href{https://arxiv.org/abs/2304.02660}{{\ttfamily 2304.02660}}].

\bibitem{Bhardwaj:2022kot}
L.~Bhardwaj, S.~Schafer-Nameki and A.~Tiwari, \emph{{Unifying Constructions of Non-Invertible Symmetries}},  \href{https://arxiv.org/abs/2212.06159}{{\ttfamily 2212.06159}}.

\bibitem{Bhardwaj:2022yxj}
L.~Bhardwaj, L.~Bottini, S.~Schafer-Nameki and A.~Tiwari, \emph{{Non-Invertible Higher-Categorical Symmetries}},  \href{https://arxiv.org/abs/2204.06564}{{\ttfamily 2204.06564}}.

\bibitem{Kaidi:2022cpf}
J.~Kaidi, K.~Ohmori and Y.~Zheng, \emph{{Symmetry TFTs for Non-Invertible Defects}},  \href{https://arxiv.org/abs/2209.11062}{{\ttfamily 2209.11062}}.

\bibitem{Bartsch:2022ytj}
T.~Bartsch, M.~Bullimore, A.E.V.~Ferrari and J.~Pearson, \emph{{Non-invertible Symmetries and Higher Representation Theory II}},  \href{https://arxiv.org/abs/2212.07393}{{\ttfamily 2212.07393}}.

\bibitem{Antinucci:2022vyk}
A.~Antinucci, F.~Benini, C.~Copetti, G.~Galati and G.~Rizi, \emph{{The holography of non-invertible self-duality symmetries}},  \href{https://arxiv.org/abs/2210.09146}{{\ttfamily 2210.09146}}.

\bibitem{Antinucci:2023ezl}
A.~Antinucci, F.~Benini, C.~Copetti, G.~Galati and G.~Rizi, \emph{{Anomalies of non-invertible self-duality symmetries: fractionalization and gauging}},  \href{https://arxiv.org/abs/2308.11707}{{\ttfamily 2308.11707}}.

\bibitem{Cordova:2023bja}
C.~Cordova, P.-S.~Hsin and C.~Zhang, \emph{{Anomalies of Non-Invertible Symmetries in (3+1)d}},  \href{https://arxiv.org/abs/2308.11706}{{\ttfamily 2308.11706}}.

\bibitem{Bhardwaj:2024xcx}
L.~Bhardwaj, T.~D{\'e}coppet, S.~Schafer-Nameki and M.~Yu, \emph{{Fusion 3-Categories for Duality Defects}},  \href{https://arxiv.org/abs/2408.13302}{{\ttfamily 2408.13302}}.

\bibitem{Bullimore:2024khm}
M.~Bullimore and J.J.~Pearson, \emph{{Towards All Categorical Symmetries in 2+1 Dimensions}},  \href{https://arxiv.org/abs/2408.13931}{{\ttfamily 2408.13931}}.

\bibitem{Heidenreich:2021xpr}
B.~Heidenreich, J.~McNamara, M.~Montero, M.~Reece, T.~Rudelius and I.~Valenzuela, \emph{{Non-invertible global symmetries and completeness of the spectrum}}, \href{https://doi.org/10.1007/JHEP09(2021)203}{\emph{JHEP} {\bfseries 09} (2021) 203} [\href{https://arxiv.org/abs/2104.07036}{{\ttfamily 2104.07036}}].

\bibitem{Jacobson:2024muj}
T.~Jacobson, \emph{{Gauging C on the Lattice}},  \href{https://arxiv.org/abs/2406.12075}{{\ttfamily 2406.12075}}.

\bibitem{Hsin:2024aqb}
P.-S.~Hsin, R.~Kobayashi and C.~Zhang, \emph{{Fractionalization of Coset Non-Invertible Symmetry and Exotic Hall Conductance}},  \href{https://arxiv.org/abs/2405.20401}{{\ttfamily 2405.20401}}.

\bibitem{Hsin:2025ria}
P.-S.~Hsin, R.~Kobayashi and C.~Zhang, \emph{{Anomalies of Coset Non-Invertible Symmetries}},  \href{https://arxiv.org/abs/2503.00105}{{\ttfamily 2503.00105}}.

\bibitem{Nguyen:2021yld}
M.~Nguyen, Y.~Tanizaki and M.~\"Unsal, \emph{{Semi-Abelian gauge theories, non-invertible symmetries, and string tensions beyond $N$-ality}}, \href{https://doi.org/10.1007/JHEP03(2021)238}{\emph{JHEP} {\bfseries 03} (2021) 238} [\href{https://arxiv.org/abs/2101.02227}{{\ttfamily 2101.02227}}].

\bibitem{Antinucci:2022eat}
A.~Antinucci, G.~Galati and G.~Rizi, \emph{{On Continuous 2-Category Symmetries and Yang-Mills Theory}},  \href{https://arxiv.org/abs/2206.05646}{{\ttfamily 2206.05646}}.

\bibitem{Fuchs:2007tx}
J.~Fuchs, M.R.~Gaberdiel, I.~Runkel and C.~Schweigert, \emph{{Topological defects for the free boson CFT}}, \href{https://doi.org/10.1088/1751-8113/40/37/016}{\emph{J. Phys. A} {\bfseries 40} (2007) 11403} [\href{https://arxiv.org/abs/0705.3129}{{\ttfamily 0705.3129}}].

\bibitem{GarciaEtxebarria:2022jky}
I.~Garc\'\i{}a~Etxebarria and N.~Iqbal, \emph{{A Goldstone theorem for continuous non-invertible symmetries}},  \href{https://arxiv.org/abs/2211.09570}{{\ttfamily 2211.09570}}.

\bibitem{Damia:2023gtc}
J.A.~Damia, R.~Argurio and S.~Chaudhuri, \emph{{When the moduli space is an orbifold: spontaneous breaking of continuous non-invertible symmetries}}, \href{https://doi.org/10.1007/JHEP03(2024)042}{\emph{JHEP} {\bfseries 03} (2024) 042} [\href{https://arxiv.org/abs/2309.06491}{{\ttfamily 2309.06491}}].

\bibitem{Antinucci:2025uvj}
A.~Antinucci, C.~Copetti, G.~Galati and G.~Rizi, \emph{{Defect Conformal Manifolds from Phantom (Non-Invertible) Symmetries}},  \href{https://arxiv.org/abs/2505.09668}{{\ttfamily 2505.09668}}.

\bibitem{Choi:2025ebk}
Y.~Choi, H.~Ha, D.~Kim, Y.~Kusuki, S.~Ohyama and S.~Ryu, \emph{{Higher Structures on Boundary Conformal Manifolds: Higher Berry Phase and Boundary Conformal Field Theory}},  \href{https://arxiv.org/abs/2507.12525}{{\ttfamily 2507.12525}}.

\bibitem{Seifnashri:2025fgd}
S.~Seifnashri, S.-H.~Shao and X.~Yang, \emph{{Gauging non-invertible symmetries on the lattice}},  \href{https://arxiv.org/abs/2503.02925}{{\ttfamily 2503.02925}}.

\bibitem{Kaidi:2022uux}
J.~Kaidi, G.~Zafrir and Y.~Zheng, \emph{{Non-invertible symmetries of $ \mathcal{N} $ = 4 SYM and twisted compactification}}, \href{https://doi.org/10.1007/JHEP08(2022)053}{\emph{JHEP} {\bfseries 08} (2022) 053} [\href{https://arxiv.org/abs/2205.01104}{{\ttfamily 2205.01104}}].

\bibitem{SSW:2021unpublished}
S.~Seifnashri, S.-H.~Shao and Y.~Wang, \emph{{On Continuous Non-invertible Symmetries}}, {\emph{unpublished} (2021) }.

\bibitem{Cao:2025qnc}
W.~Cao, Y.~Miao and M.~Yamazaki, \emph{{Global symmetries of quantum lattice models under non-invertible dualities}},  \href{https://arxiv.org/abs/2501.12514}{{\ttfamily 2501.12514}}.

\bibitem{Moore:1989yh}
G.W.~Moore and N.~Seiberg, \emph{{Taming the Conformal Zoo}}, \href{https://doi.org/10.1016/0370-2693(89)90897-6}{\emph{Phys. Lett. B} {\bfseries 220} (1989) 422}.

\bibitem{Vafa:1989ih}
C.~Vafa, \emph{{Quantum Symmetries of String Vacua}}, \href{https://doi.org/10.1142/S0217732389001842}{\emph{Mod. Phys. Lett. A} {\bfseries 4} (1989) 1615}.

\bibitem{Ginsparg:1988ui}
P.H.~Ginsparg, \emph{{APPLIED CONFORMAL FIELD THEORY}},  in \emph{{Les Houches Summer School in Theoretical Physics: Fields, Strings, Critical Phenomena}}, 9, 1988 [\href{https://arxiv.org/abs/hep-th/9108028}{{\ttfamily hep-th/9108028}}].

\bibitem{Ginsparg:1987eb}
P.H.~Ginsparg, \emph{{Curiosities at c = 1}}, \href{https://doi.org/10.1016/0550-3213(88)90249-0}{\emph{Nucl. Phys. B} {\bfseries 295} (1988) 153}.

\bibitem{hotatlam1}
H.T.~Lam, \emph{{Conformal Manifolds from Continuous SymTFTs}}, {\emph{Generalized Symmetries: High-Energy, Condensed Matter and Mathematics KITP conference} (2025) }.

\bibitem{hotatlam2}
H.T.~Lam, \emph{{Global Aspects of Exactly Marginal Current-Current Deformations}}, {\emph{CMSA Workshop on Symmetries and Gravity} (2025) }.

\bibitem{SSW:2021unpublished2}
S.-H.~Shao, \emph{{On Continuous Non-invertible Symmetries}}, {\emph{Talk at Simons Collaboration on Global Categorical Symmetries---SCGP Workshop} (2021) }.

\bibitem{Ambrosino:2025myh}
F.~Ambrosino, I.~Runkel and G.M.T.~Watts, \emph{{Non-local charges from perturbed defects via SymTFT in 2d CFT}},  \href{https://arxiv.org/abs/2504.05277}{{\ttfamily 2504.05277}}.

\bibitem{Lin:2019kpn}
Y.-H.~Lin and S.-H.~Shao, \emph{{Anomalies and Bounds on Charged Operators}}, \href{https://doi.org/10.1103/PhysRevD.100.025013}{\emph{Phys. Rev. D} {\bfseries 100} (2019) 025013} [\href{https://arxiv.org/abs/1904.04833}{{\ttfamily 1904.04833}}].

\bibitem{Lin:2021udi}
Y.-H.~Lin and S.-H.~Shao, \emph{{$\mathbb{Z}_N$ symmetries, anomalies, and the modular bootstrap}}, \href{https://doi.org/10.1103/PhysRevD.103.125001}{\emph{Phys. Rev. D} {\bfseries 103} (2021) 125001} [\href{https://arxiv.org/abs/2101.08343}{{\ttfamily 2101.08343}}].

\bibitem{Polchinski:1998rq}
J.~Polchinski, \emph{{String theory. Vol. 1: An introduction to the bosonic string}}, Cambridge Monographs on Mathematical Physics, Cambridge University Press (12, 2007), \href{https://doi.org/10.1017/CBO9780511816079}{10.1017/CBO9780511816079}.

\bibitem{Chang:2020imq}
C.-M.~Chang and Y.-H.~Lin, \emph{{Lorentzian dynamics and factorization beyond rationality}}, \href{https://doi.org/10.1007/JHEP10(2021)125}{\emph{JHEP} {\bfseries 10} (2021) 125} [\href{https://arxiv.org/abs/2012.01429}{{\ttfamily 2012.01429}}].

\bibitem{Argurio:2024ewp}
R.~Argurio, A.~Collinucci, G.~Galati, O.~Hulik and E.~Paznokas, \emph{{Non-Invertible T-duality at Any Radius via Non-Compact SymTFT}}, \href{https://doi.org/10.21468/SciPostPhys.18.3.089}{\emph{SciPost Phys.} {\bfseries 18} (2025) 089} [\href{https://arxiv.org/abs/2409.11822}{{\ttfamily 2409.11822}}].

\bibitem{Yu:2025iqf}
X.~Yu and H.Y.~Zhang, \emph{{von Neumann Subfactors and Non-invertible Symmetries}},  \href{https://arxiv.org/abs/2504.05374}{{\ttfamily 2504.05374}}.

\bibitem{francesco2012conformal}
P.~Francesco, P.~Mathieu and D.~S{\'e}n{\'e}chal, \emph{Conformal field theory}, Springer Science \& Business Media (2012).

\bibitem{Eberhardt2019WZW}
L.~Eberhardt, \emph{Wess--zumino--witten models},  Lecture Notes YRISW PhD School in Vienna, ETH Z{\"u}rich (Feb., 2019).

\bibitem{Choi:2023xjw}
Y.~Choi, B.C.~Rayhaun, Y.~Sanghavi and S.-H.~Shao, \emph{{Remarks on boundaries, anomalies, and noninvertible symmetries}}, \href{https://doi.org/10.1103/PhysRevD.108.125005}{\emph{Phys. Rev. D} {\bfseries 108} (2023) 125005} [\href{https://arxiv.org/abs/2305.09713}{{\ttfamily 2305.09713}}].

\bibitem{Cappelli:1986hf}
A.~Cappelli, C.~Itzykson and J.B.~Zuber, \emph{{Modular invariant partition functions in two dimensions}}, \href{https://doi.org/10.1016/0550-3213(87)90155-6}{\emph{Nucl. Phys. B} {\bfseries 280} (1987) 445}.

\bibitem{Cappelli:1987xt}
A.~Cappelli, C.~Itzykson and J.B.~Zuber, \emph{{The ADE Classification of Minimal and A1(1) Conformal Invariant Theories}}, \href{https://doi.org/10.1007/BF01221394}{\emph{Commun. Math. Phys.} {\bfseries 113} (1987) 1}.

\bibitem{Kato:1987td}
A.~Kato, \emph{{Classification of Modular Invariant Partition Functions in Two-dimensions}}, \href{https://doi.org/10.1142/S0217732387000732}{\emph{Mod. Phys. Lett. A} {\bfseries 2} (1987) 585}.

\bibitem{Neupert_2016}
T.~Neupert, H.~He, C.~von Keyserlingk, G.~Sierra and B.A.~Bernevig, \emph{Boson condensation in topologically ordered quantum liquids}, \href{https://doi.org/10.1103/physrevb.93.115103}{\emph{Physical Review B} {\bfseries 93} (2016) }.

\bibitem{Elitzur:1989nr}
S.~Elitzur, G.W.~Moore, A.~Schwimmer and N.~Seiberg, \emph{{Remarks on the Canonical Quantization of the Chern-Simons-Witten Theory}}, \href{https://doi.org/10.1016/0550-3213(89)90436-7}{\emph{Nucl. Phys. B} {\bfseries 326} (1989) 108}.

\bibitem{Lan:2014uaa}
T.~Lan, J.C.~Wang and X.-G.~Wen, \emph{{Gapped Domain Walls, Gapped Boundaries and Topological Degeneracy}}, \href{https://doi.org/10.1103/PhysRevLett.114.076402}{\emph{Phys. Rev. Lett.} {\bfseries 114} (2015) 076402} [\href{https://arxiv.org/abs/1408.6514}{{\ttfamily 1408.6514}}].

\bibitem{Zamolodchikov:1986bd}
A.B.~Zamolodchikov and V.A.~Fateev, \emph{{Operator Algebra and Correlation Functions in the Two-Dimensional Wess-Zumino SU(2) x SU(2) Chiral Model}}, {\emph{Sov. J. Nucl. Phys.} {\bfseries 43} (1986) 657}.

\bibitem{Recknagel:2013uja}
A.~Recknagel and V.~Schomerus, \emph{{Boundary Conformal Field Theory and the Worldsheet Approach to D-Branes}}, Cambridge Monographs on Mathematical Physics, Cambridge University Press (11, 2013), \href{https://doi.org/10.1017/CBO9780511806476}{10.1017/CBO9780511806476}.

\bibitem{Oshikawa:1996dj}
M.~Oshikawa and I.~Affleck, \emph{{Boundary conformal field theory approach to the critical two-dimensional Ising model with a defect line}}, \href{https://doi.org/10.1016/S0550-3213(97)00219-8}{\emph{Nucl. Phys. B} {\bfseries 495} (1997) 533} [\href{https://arxiv.org/abs/cond-mat/9612187}{{\ttfamily cond-mat/9612187}}].

\bibitem{Oshikawa_1997}
M.~Oshikawa, M.~Yamanaka and I.~Affleck, \emph{Magnetization plateaus in spin chains: “haldane gap” for half-integer spins}, \href{https://doi.org/10.1103/physrevlett.78.1984}{\emph{Physical Review Letters} {\bfseries 78} (1997) 1984–1987}.

\bibitem{Copetti:2025sym}
C.~Copetti, \emph{{'t Hooft Anomalies and Defect Conformal Manifolds: Topological Signatures from Modulated Effective Actions}},  \href{https://arxiv.org/abs/2507.15466}{{\ttfamily 2507.15466}}.

\bibitem{Karch:2018uft}
A.~Karch and Y.~Sato, \emph{{Conformal Manifolds with Boundaries or Defects}}, \href{https://doi.org/10.1007/JHEP07(2018)156}{\emph{JHEP} {\bfseries 07} (2018) 156} [\href{https://arxiv.org/abs/1805.10427}{{\ttfamily 1805.10427}}].

\bibitem{Herzog:2023dop}
C.P.~Herzog and V.~Schaub, \emph{{Tilting space of boundary conformal field theories}}, \href{https://doi.org/10.1103/PhysRevD.109.L061701}{\emph{Phys. Rev. D} {\bfseries 109} (2024) L061701} [\href{https://arxiv.org/abs/2301.10789}{{\ttfamily 2301.10789}}].

\bibitem{Roumpedakis:2022aik}
K.~Roumpedakis, S.~Seifnashri and S.-H.~Shao, \emph{{Higher Gauging and Non-invertible Condensation Defects}},  \href{https://arxiv.org/abs/2204.02407}{{\ttfamily 2204.02407}}.

\bibitem{Lan:2018vjb}
T.~Lan, L.~Kong and X.-G.~Wen, \emph{{Classification of (3+1)D Bosonic Topological Orders: The Case When Pointlike Excitations Are All Bosons}}, \href{https://doi.org/10.1103/PhysRevX.8.021074}{\emph{Phys. Rev. X} {\bfseries 8} (2018) 021074}.

\bibitem{Balasubramanian:2024nei}
M.~Balasubramanian, M.~Buican and R.~Radhakrishnan, \emph{{On the Classification of Bosonic and Fermionic One-Form Symmetries in $2+1$d and 't Hooft Anomaly Matching}},  \href{https://arxiv.org/abs/2408.00866}{{\ttfamily 2408.00866}}.

\bibitem{Johnson-Freyd:2020twl}
T.~Johnson-Freyd, \emph{{(3+1)D topological orders with only a $\mathbb{Z}_2$-charged particle}},  \href{https://arxiv.org/abs/2011.11165}{{\ttfamily 2011.11165}}.

\bibitem{Johnson-Freyd:2021tbq}
T.~Johnson-Freyd and M.~Yu, \emph{{Topological Orders in (4+1)-Dimensions}}, \href{https://doi.org/10.21468/SciPostPhys.13.3.068}{\emph{SciPost Phys.} {\bfseries 13} (2022) 068} [\href{https://arxiv.org/abs/2104.04534}{{\ttfamily 2104.04534}}].

\bibitem{Cordova:2022ieu}
C.~Cordova and K.~Ohmori, \emph{{Non-Invertible Chiral Symmetry and Exponential Hierarchies}},  \href{https://arxiv.org/abs/2205.06243}{{\ttfamily 2205.06243}}.

\bibitem{Choi:2022jqy}
Y.~Choi, H.T.~Lam and S.-H.~Shao, \emph{{Noninvertible Global Symmetries in the Standard Model}}, \href{https://doi.org/10.1103/PhysRevLett.129.161601}{\emph{Phys. Rev. Lett.} {\bfseries 129} (2022) 161601} [\href{https://arxiv.org/abs/2205.05086}{{\ttfamily 2205.05086}}].

\bibitem{Chen:2022cyw}
S.~Chen and Y.~Tanizaki, \emph{{Solitonic symmetry beyond homotopy: invertibility from bordism and non-invertibility from TQFT}},  \href{https://arxiv.org/abs/2210.13780}{{\ttfamily 2210.13780}}.

\bibitem{Karasik:2022kkq}
A.~Karasik, \emph{{On anomalies and gauging of U(1) non-invertible symmetries in 4d QED}},  \href{https://arxiv.org/abs/2211.05802}{{\ttfamily 2211.05802}}.

\bibitem{Arbalestrier:2024oqg}
A.~Arbalestrier, R.~Argurio and L.~Tizzano, \emph{{Noninvertible axial symmetry in QED comes full circle}}, \href{https://doi.org/10.1103/PhysRevD.110.105012}{\emph{Phys. Rev. D} {\bfseries 110} (2024) 105012} [\href{https://arxiv.org/abs/2405.06596}{{\ttfamily 2405.06596}}].

\bibitem{Putrov:2022pua}
P.~Putrov, \emph{{$\mathbb{Q}/\mathbb{Z}$ symmetry}},  \href{https://arxiv.org/abs/2208.12071}{{\ttfamily 2208.12071}}.

\bibitem{Chen:2025buv}
S.~Chen, A.~Cherman and M.~Neuzil, \emph{{Symmetry theta angles and topological Witten effects}},  \href{https://arxiv.org/abs/2507.00220}{{\ttfamily 2507.00220}}.

\bibitem{Gabai:2024puk}
B.~Gabai, V.~Gorbenko, J.~Qiao, B.~Zan and A.~Zhabin, \emph{{Quantum Groups as Global Symmetries}},  \href{https://arxiv.org/abs/2410.24142}{{\ttfamily 2410.24142}}.

\bibitem{Gabai:2024qum}
B.~Gabai, V.~Gorbenko, J.~Qiao, B.~Zan and A.~Zhabin, \emph{{Quantum Groups as Global Symmetries II. Coulomb Gas Construction}},  \href{https://arxiv.org/abs/2410.24143}{{\ttfamily 2410.24143}}.

\bibitem{BernardLeClair1991}
D.~Bernard and A.~LeClair, \emph{Quantum group symmetries and non-local currents in 2{D} {QFT}}, \href{https://doi.org/10.1007/BF02099173}{\emph{Communications in Mathematical Physics} {\bfseries 142} (1991) 99}.

\bibitem{Heckman:2024obe}
J.J.~Heckman, J.~McNamara, M.~Montero, A.~Sharon, C.~Vafa and I.~Valenzuela, \emph{{On the Fate of Stringy Non-Invertible Symmetries}},  \href{https://arxiv.org/abs/2402.00118}{{\ttfamily 2402.00118}}.

\bibitem{Kaidi:2024wio}
J.~Kaidi, Y.~Tachikawa and H.Y.~Zhang, \emph{{On a class of selection rules without group actions in field theory and string theory}}, \href{https://doi.org/10.21468/SciPostPhys.17.6.169}{\emph{SciPost Phys.} {\bfseries 17} (2024) 169} [\href{https://arxiv.org/abs/2402.00105}{{\ttfamily 2402.00105}}].

\bibitem{Bachas:2012bj}
C.~Bachas, I.~Brunner and D.~Roggenkamp, \emph{{A worldsheet extension of O(d,d:Z)}}, \href{https://doi.org/10.1007/JHEP10(2012)039}{\emph{JHEP} {\bfseries 10} (2012) 039} [\href{https://arxiv.org/abs/1205.4647}{{\ttfamily 1205.4647}}].

\bibitem{Dijkgraaf:1989hb}
R.~Dijkgraaf, C.~Vafa, E.P.~Verlinde and H.L.~Verlinde, \emph{{The Operator Algebra of Orbifold Models}}, \href{https://doi.org/10.1007/BF01238812}{\emph{Commun. Math. Phys.} {\bfseries 123} (1989) 485}.

\bibitem{Hamidi:1986vh}
S.~Hamidi and C.~Vafa, \emph{{Interactions on Orbifolds}}, \href{https://doi.org/10.1016/0550-3213(87)90006-X}{\emph{Nucl. Phys. B} {\bfseries 279} (1987) 465}.

\end{thebibliography}\endgroup

\end{document}